\documentclass[aps,prd,nofootinbib]{revtex4-2}
\usepackage{amsmath}
\usepackage{amsfonts}
\usepackage{amssymb}
\usepackage{makeidx}
\usepackage{graphicx,epsfig}
\usepackage{color}
\usepackage{subfigure}
\usepackage{ulem}
\usepackage{multirow}
\usepackage{epstopdf}
\usepackage{dcolumn}
\usepackage{comment}
\usepackage{caption}
\usepackage{subcaption}
\usepackage{epstopdf}
\usepackage{dcolumn}
\usepackage{slashed}
\usepackage{makecell}
\usepackage{url}
\usepackage{bm}
\usepackage[section]{placeins}
\usepackage[colorlinks=true, citecolor=blue, urlcolor = blue, linkcolor= red, bookmarks=true]{hyperref}
\newcommand{\beq}{\begin{equation}}
	\newcommand{\eeq}{\end{equation}}
\newcommand{\bea}{\begin{eqnarray}}
	\newcommand{\eea}{\end{eqnarray}}

\begin{document}
		\title {Associated Production of Charmonia-Bottomonia with Color-Octet Channels at the Z Factory, CEPC and FCC-ee}
	
	\author{ Xiao-Peng Wang $^{(a)}$}
	\author{ Yi-Jie Li $^{(a)}$}
	\author{Guang-Zhi Xu $^{(a)}$}
	\email{ xuguangzhi@lnu.edu.cn}
	\author{Kui-Yong Liu $^{(b,a)}$}
	\email{liukuiyong@lnu.edu.cn}
	\affiliation{ {\footnotesize (a)~School of Physics, Liaoning University, Shenyang 110036, China}\\
		{\footnotesize (b)~School of Physics and Electronic Technology, Liaoning Normal University, Dalian 116029, China}}	
	
	\date{\today}
	
	\begin{abstract}
 Within the nonrelativistic QCD (NRQCD) framework, we investigate the associated production of charmonia+bottomonia  at future super $Z$ factory and at the CEPC/FCC-ee. The color-octet(CO) channels in the $\gamma^*/Z^0$-propagated process are considered besides the color-singlet(CS) channels. We find that the contributions of the CO states to the total cross section are    dominant for almost all   processes   from the production threshold to the $Z^0$ mass.  Thus, the comparison between the theoretical results and future data may give   constraints to the CO matrix elements. In addition, we calculate the relativistic corrections to both the CS and CO channels, which   decrease the cross sections  significantly, with the $K$ factor  $\simeq0.5$. 
The predicted events of  ($J/\psi+\Upsilon$,  $\Upsilon+\eta_c$) production, with $J/\psi, \Upsilon$ reconstructed by lepton pair and $\eta_c$ reconstructed by hadronic decays, are (1,1) and (13,10) at the CEPC(2-yr) and FCC-ee(4-yr), respectively.

	\end{abstract}
	
	\maketitle
	 
  \section{Introduction}
	 
	 Since the discovery of the $J/\psi$ meson in 1974\cite{E598:1974sol,SLAC-SP-017:1974ind}, heavy quarkonium— characterized as a nonrelativistic bound state of a heavy quark-antiquark pair—has served as an ideal laboratory for exploring QCD and  provided crucial insights into the strong interaction. A foundational theoretical advance has been the development of the nonrelativistic QCD (NRQCD) effective field theory\cite{Bodwin:1994jh}, within which the production cross section can be decomposed into calculable short-distance coefficients (SDCs) and universal long-distance matrix elements (LDMEs).
	   Unlike the  early   color-singlet mechanism (CSM)\cite{Berger:1980ni,Baier:1981zz,Chang:1979nn,Kuhn:1979bb,Guberina:1980dc,Baier:1983va}, in which the heavy quark pair created  at short distance has the same color and angular momentum as the final quarkonium,  one pillar of NRQCD is the color-octet mechanism (COM), enabling the  quark pair  to exist in different Fock  state   from that of the   quarkonium. This mechanism resolved the infrared (IR) divergences occurring in P-wave heavy quarkonium decays\cite{Barbieri:1980yp,Bodwin:1992ye} and gave the proper prediction for   charmonium production at   Tevatron\cite{CDF:1992cmg,CDF:1993xyh,CDF:1997ykw,CDF:1997uzj,Braaten:1994vv,Braaten:1995cj,Kramer:2001hh}, marking a major triumph for NRQCD.
	 However, several challenges to the NRQCD approach remain, such as  the   polarisation puzzles, where   NRQCD predictions are inconsistent with experimental measurements\cite{Beneke:1995yb,Cho:1994ih,Braaten:1999qk,CDF:2007msx,ALICE:2011gej,LHCb:2013izl,CMS:2013gbz}, and  the problem of the universality of LDMEs, where several theoretical groups have obtained different values\cite{Butenschoen:2011yh,Chao:2012iv,Gong:2012ug,Bodwin:2014gia,Zhang:2014ybe}.  For further details on the current status of heavy quarkonium physics, we refer the readers to Refs. \cite{Brambilla:2010cs,Andronic:2015wma,Lansberg:2019adr,Chapon:2020heu,Boer:2024ylx}.

	  Compared with the hadron colliders, the  production mechanism in $e^+e^-$ colliders is simpler and the uncertainties in the theoretical calculations are smaller. Consequently, extensive work has been based on 
	  $e^+e^-$
	  annihilation. One prominent area is the paired production of heavy charmonium,  e.g., the large discrepancy between leading-order (LO) calculation\cite{Liu:2002wq,Braaten:2002fi}  and measurement  \cite{Belle:2004abn,BaBar:2005nic} for $J/\psi+\eta_c$ production has been accounted for  through the inclusion of   QCD correction\cite{Zhang:2005cha,Gong:2007db,Dong:2012xx,Huang:2022dfw,Feng:2019zmt} and relativistic correction\cite{He:2007te,Bodwin:2007ga}.    However, due to the low interaction energy of the existing experiments,  processes involving bottomonium, such as the associated  production  of charmonium and bottomonium,  have not been accessible.
	 The future Super Z factory is a large experimental facility based on $e^+e^-$ colliders, such as  CEPC  (  three operation modes $\sqrt{s}\sim91.2 GeV,  160 GeV$ and $240 GeV$,  corresponding to Z factory, Higgs factory, and the WW threshold, respectively)\cite{CEPCStudyGroup:2018ghi} and  FCC-ee ($\sqrt{s}: 90\sim365~ GeV$)\cite{Agapov:2022bhm}, and operates at the $Z^0$  resonance peak with a design luminosity up to $\mathcal{L}\simeq10^{34\sim36} cm^{-2}s^{-1}$\cite{Chang:2010am}. Like the Super Z factory, the GigaZ is a project  based on the ILC ($\sqrt{s}\sim250 GeV$  with a luminosity $\mathcal{L}\simeq0.7\times10^{34} cm^{-2}s^{-1}$)\cite{Erler:2000jg,ILC:2007bjz}. These   facilities are expected to achieve this goal and enable the study of a wide range of processes, such as the production of   $B_c$ meson\cite{Yang:2011ps, Berezhnoy:2016etd,Zheng:2017xgj,Chen:2020dtu,Zhan:2022etq},  doubly heavy baryon\cite{Jiang:2012jt,Luo:2022jxq,Zhan:2023vwp,Zhan:2023jfm,Niu:2023ojf,Ma:2025ito},  doubly heavy tetraquark\cite{Niu:2024ghc,Jiang:2024lsr,Niu:2025gcj},   fully heavy tetraquark state\cite{Liang:2025wbt}, and rare $Z^0$ decay\cite{dEnterria:2023wjq}.
	  
	The exclusive production of double heavy quarkonium in $e^+e^-$ annihilation at Z factory has been studied at LO\cite{Hagiwara:2003cw,Chen:2013mjb,Liao_2023,liaoqili2,liaoqili}. The next-to-leading order (NLO) QCD corrections to double heavy quarkonium \cite{Berezhnoy:2021tqb, Belov:2023hpc} and double  $B_c$ meson \cite{Berezhnoy:2016etd} production have also been calculated, showing that the QCD corrections are essential. In contrast, the one-loop QCD corrections are negligible for the associated S-wave charmonium-bottomonium production (generally, $10^{-6}<\sigma_{QCD}/\sigma_{EW}<0.1$)\cite{Belov:2021ftc}.	  
	 The contributions from the CO component are substantial in semi-exclusive processes \cite{Sun:2013liv}. 
    In our recent work, we demonstrated that the COM plays a significant or even dominant role in the production of double heavy quarkonia around the Z factory energy region \cite{Wang:2025sbx}, where the gluon fragmentation process is of key importance. By contrast, the associated production of charmonium and bottomonium, which also serves as an effective channel to probing the production mechanisms of heavy quarkonia\cite{Belov:2021ftc}, proceeds solely via the fragmentation mode at the tree-level diagram. It is therefore of great interest to explore the contributions from the COM to these processes, as a continuation of our previous studies. Consistent with our previous investigations, in this paper we study the charmonium-bottomonium associated production involving double $S$-wave states as well as $S$-wave plus $P$-wave states, with the specific processes including: $J/\psi+\Upsilon, J/\psi+\eta_b,  J/\psi+\chi_{bJ}, J/\psi+h_b,   \eta_c+h_{b}, \eta_c+\chi_{bJ}, \eta_c+\eta_b, \eta_{c}+\Upsilon, h_c+\Upsilon, \chi_{cJ}+\Upsilon, h_c+\eta_{b}, \chi_{cJ}+\eta_b$.

The rest of this paper is organized as follows. In Section \ref{solution}, we present the theoretical framework and computational method. Section \ref{inputldmes} details the input parameters and the LDMEs used in this work. In Section \ref{results}, we analyze all possible production channels, present the results for (differential) cross sections, and discuss event generation as well as associated uncertainties. Finally, Section \ref{summary} concludes the paper.
	
\section{Formula and Method} \label{solution}

		\begin{widetext}	
		\begin{figure}
			\begin{tabular}{c c }
				\includegraphics[width=0.5\textwidth]{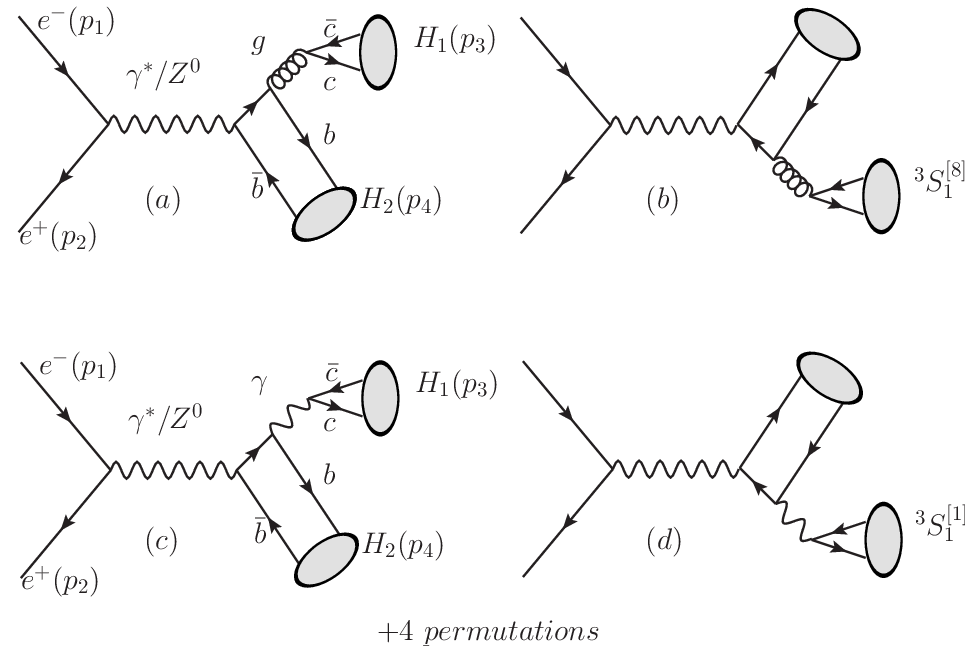}
			\end{tabular}
			\caption{Feynman diagrams for $e^-(p_1)+e^+(p_2)\rightarrow \gamma^*/Z^0\rightarrow H_1(p_3)+H_2(p_4)$ at the tree level. $H_1, H_2$ is charmonia and bottomonia, respectively. The permutation diagrams can be obtained by reversing the quark line.  (a) and (b) diagrams belong to CO (QCD) channels, (c) and (d) diagrams belong to CS (EW) channels.  }
			\label{feynmandia}
		\end{figure}
		\FloatBarrier	
	\end{widetext}
 
According to the NRQCD framework, the production cross sections are factorized into a sum of products of SDCs and LDMEs\cite{Bodwin:1994jh},
	\begin{equation}
		\label{eq:factorization} \hat{\sigma}(e^++e^-{\rightarrow}H_1+H_2)=\sum_{mn}\frac{F_{mn}}{m_{Q1}^{d_m-4}m_{Q2}^{d_n-4}}\langle0|\mathcal{O}_m^{H_1}|0\rangle\langle0|\mathcal{O}_n^{H_2}|0\rangle.
	\end{equation}
	The SDC $F_{mn}$ describes the production probability of intermediate quarks before hadronization. The LDME $\langle0|\mathcal{O}_{m,n}^{H}|0\rangle$ denotes the transition probability of a  $Q\bar{Q}$ pair to form the final bound state $H$. 
    The factor  $m_Q^{d_{m,n}-4}$ is introduced to make	$F_{mn}$ dimensionless.
 The heavy quarkonium states have a Fock state expansion formalism up-to the order of $\mathcal{O}(v^2)$ :
	\bea
|H\rangle=\mathcal{O}(1)|Q\bar{Q}(^{2S+1}L_J^{[1]})\rangle+\mathcal{O}(v)|Q\bar{Q}(^{2S+1}(L\pm1)_{J}^{[8]})g\rangle\cr
+\mathcal{O}(v^2)|Q\bar{Q}(^{2(S\pm1)+1}L_{J}^{[8]})g\rangle
+\mathcal{O}(v^2)|Q\bar{Q}(^{2S+1}L_{J}^{[1,8]})gg\rangle+...
\eea
In Table~\ref{tab:quarkonia_nrqcd_simplified}, the specific CS/CO intermediate states for the heavy quarkonium are listed.

    \begin{table}[htbp]
      \centering
      \caption{The color singlet and octet Fock states of the charmonium and bottomonium considered in this work.}
      \label{tab:quarkonia_nrqcd_simplified}
      \renewcommand\arraystretch{1.6} 
      \begin{tabular}{l|p{4.5cm}|p{6.5cm}}
        \toprule
        Particle & Color Singlet State & Color Octet States \\
        $J/\psi$/$\Upsilon$ & $^3S_1^{[1]}$ & $^3P_J^{[8]}, {}^1S_0^{[8]}, {}^3S_1^{[8]}$ \\
        $\eta_c$/$\eta_b$   & $^1S_0^{[1]}$ & $^1P_1^{[8]}, {}^3S_1^{[8]}, {}^1S_0^{[8]}$ \\
        $\chi_{cJ}$/$\chi_{bJ}$ & $^3P_J^{[1]}$ & $^3S_1^{[8]}$ \\
        $h_c$/$h_b$         & $^1P_1^{[1]}$ & $^1S_0^{[8]}$ \\
        \hline
      \end{tabular}
    \end{table}
    
    We also calculate the contributions from relativistic corrections in this work, for which the matrix elements of $\langle v^2\rangle$, defined by the ratios of the NLO LDME in $v^2$ to the LO LDME, are introduced below.
	\bea
	\langle v^2\rangle\equiv\frac{\langle0|\mathcal{P}^H(^{2s+1}L_J)|0\rangle}{m_Q^2\langle0|\mathcal{O}^H(^{2s+1}L_J)|0\rangle}
	\eea
    which adhere to the velocity power scaling rules and are of the order of $v^2$ as follows,
	\bea
	v^2=\langle v^2\rangle[1+\mathcal{O}(v^4)]
	\eea
    	
    In contrast to double charmonium or double bottomonium production \cite{Wang:2025sbx}, as elaborated in Ref.~\cite{Belov:2021ftc}, tree-level QCD diagrams cannot generate an intermediate CS $c\bar{c}$ state accompanied by a CS $b\bar{b}$ state. Namely, the lowest-order QCD contributions within CSM to this process are loop processes. LO electric-weak (EW) processes in CS states are $\mathcal{O}(\alpha^4)$. LO QCD processes in CO states are $\mathcal{O}(\alpha_s^2\alpha^2)$. Both types of processes involve mediated fragmentation mechanisms. For the EW processes, the $Z^0$-fragmentation channel is neglected, as the $Z^0$-propagator is suppressed when the propagator momentum runs at meson mass scale. The corresponding Feynman diagrams are shown in Fig.~\ref{feynmandia}, where the initial and final momenta are defined. The t-channel EW diagrams (two bosons exchange process) are not presented here which are forbidden or give small contributions, except for $J/\psi$ plus $\Upsilon$ production via two-photon exchange. Additionally, contributions from W-boson-involved processes (eight single-W-boson exchange s-channel diagrams, two double-W-boson exchange s-channel diagrams, and one double-W-boson exchange t-channel diagram) are also neglected, owing to the small magnitude of  $~|V_{bc}|(\simeq0.046)$ and W-propagator suppression.
    The overall colorless nature of the final state requires that both recoiled $Q\bar{Q}$ pairs must be either in the CS state ($c\bar{c}_1+b\bar{b}_1$) or in the CO state ($c\bar{c}_8+b\bar{b}_8$). As noted in our previous work \cite{Wang:2025sbx}, the CO channels actually correspond to inclusive processes. This particular CO configuration with close recoiled $Q\bar{Q}$ pairs induces factorization breakdown due to nonperturbative effects which are suppressed by inverse powers of the collision energy and thus usually negligible.
 
	From Fig. \ref{feynmandia}, the amplitudes can be written as follows,
	\bea
	i\mathcal{M}=\sum_{n}\bar{v}(p_2)\mathcal{L}_{\mu}u(p_1)\mathcal{D}^{\mu\nu}\mathcal{A}^{(n)}_{\nu},
	\eea
	the vertex $\mathcal{L}^{\mu}$ and the propagator $D_{\mu\nu}$ are:	
	\bea
	\mathcal{L}^\mu&=&-ie\gamma^\mu, ~~\frac{-ie}{4\cos\theta_W \sin\theta_W}\gamma^\mu(1-4\sin^2\theta_W-\gamma^5)\cr
	\mathcal{D}_{\mu\nu}&=&\frac{-ig_{\mu\nu}}{p^2}, ~~\frac{-ig_{\mu\nu}}{p^2-m_Z^2+im_Z\Gamma_Z}
	\eea
    where, the first (second) term corresponds to the process propagated by $\gamma^*$ ($Z^0$).  $e = \sqrt{4\pi\alpha}$ is the electromagnetic coupling constant,  $\theta_W$ is the Weinberg angle. $p=p_1+p_2$ is the four momentum of the propagator. $m_Z$ and $\Gamma_Z$ are the mass and the decay width of $Z^0$ boson.
	 
    The hadronic part of the amplitude $\mathcal{A}^{(n)}_{\nu}(n=a,b,c,...)$ is expressed as the product of traces in spinor space, with diagrams (a) and (d) provided as illustrative examples below.
	\bea	\mathcal{A}_\nu^{(a)}&=&-g^2Tr\bigg[\gamma^{\rho}\mathbb{P}_1\bigg]Tr\bigg[\mathbb{P}_2\gamma^{\rho}\frac{(\slashed{p}_{31}+\slashed{p}_{32}+\slashed{p}_{41})+m_{b}}{[(-p_{31}-p_{32}-p_{41})^2-m_{b}^2](-p_{31}-p_{32})^2}\mathcal{V}_b^{\nu}\bigg]~\cr
	\mathcal{A}_\nu^{(d)}&=&-Tr\bigg[\gamma^{\rho}\mathbb{P}_2\bigg]Tr\bigg[\mathbb{P}_1\mathcal{V}_c^{\nu}\frac{(-\slashed{p}_{32}-\slashed{p}_{41}-\slashed{p}_{42})+m_{c}}{[(p_{32}+p_{41}+p_{42})^2-m_{c}^2](p_{41}+p_{42})^2}\gamma^{\rho}\bigg]~
	\eea
    where $p_{31}=\frac{1}{2}p_3+q_3$, $p_{32}=\frac{1}{2}p_3-q_3$ denote the constituent quark momenta of quarkonium $H_1$ (with $q_3$ its internal relative momentum), $p_{41}=\frac{1}{2}p_4+q_4$, $p_{42}=\frac{1}{2}p_4-q_4$ denote those of $H_2$ (with $q_4$ its internal relative momentum). The vertexs $\mathcal{V}^{\mu}$ for $\gamma^*/Z^0$ propagation are defined as,
	\bea
	\mathcal{V}_c^\mu&=&-iee_c\gamma^\mu, ~~\frac{-ie}{4\cos\theta_W \sin\theta_W}\gamma^\mu(-1+4|e_c|\sin^2\theta_W+\gamma^5)\cr
	\mathcal{V}_b^\mu&=&-iee_b\gamma^\mu, ~~\frac{-ie}{4\cos\theta_W \sin\theta_W}\gamma^\mu(1-4|e_b|\sin^2\theta_W-\gamma^5)
	\eea
	 with $~e_c=2/3,~e_b=-1/3$. The projector operators $\mathbb{P}_{1,2}$ in the amplitudes have the following forms for spin-singlet and spin-triplet states,
	\bea
	\label{projectors}
\mathbb{P}(00)&\equiv&\sum_{s_Q s_{\bar{Q}}}\sum_{ kl}\langle s_Q s_{\bar{Q}};00\rangle \langle3,k;\bar{3},l|1,8\rangle v_l(p_{\bar{Q}})\bar{u}_k(p_{Q})=(-\slashed{p}_{\bar{Q}}+m_Q)\gamma_5\frac{\slashed{p}_Q+\slashed{p}_{\bar{Q}}+2E_q}{4\sqrt{2}E_q(E_q+m_Q)}(\slashed{p}_Q+m_Q)\otimes\mathbb{\pi}_{1,8}\\
\mathbb{P}(1s_z)&\equiv&\sum_{s_Q s_{\bar{Q}}}\sum_{ kl}\langle s_Q s_{\bar{Q}};1s_z\rangle \langle3,k;\bar{3},l|1,8\rangle v_l(p_{\bar{Q}})\bar{u}_k(p_{Q})=(-\slashed{p}_{\bar{Q}}+m_Q)\slashed{\varepsilon}(s_z) \frac{\slashed{p}_Q+\slashed{p}_{\bar{Q}}+2E_q}{4\sqrt{2}E_q(E_q+m_Q)}(\slashed{p}_Q+m_Q)\otimes\mathbb{\pi}_{1,8}
	\eea
	where $\varepsilon(s_z)$ is the spin polarization vector, $E_q=\sqrt{m_Q^2+\vec{q}^2}$ is the energy of $Q$ or $\bar{Q}$ in the meson's rest frame. In this frame, the 4-dimensional total meson momentum is expressed as $(2E_q,\vec{0})$, and the 4-dimensional relative momentum of $Q$ and $\bar{Q}$ in a pair is $(0,\vec{q})$. 
The invariant magnitude of $\vec{q}$ associated with the effective velocity $v$ is given by, specifically, $\vec{q}_3^2=m_c^2v_{c\bar{c}}^2,~\vec{q}_4^2=m_b^2v_{b\bar{b}}^2$.	
    Color projection operators $\mathbb{\pi}_{1,8}$ (subscripts $1,8$ denote the CS,CO states,respectively) assume the form of $\sum_{ij}\frac{\delta_{ij}}{\sqrt{N_c}}$ for CS case, and $\sqrt{2}T^a_{ij}~(a=1,2,\dots,8)$ for CO case.	For the CS channel, the total color factors of each diagram are \(N_c\), while for the CO channel, it is \(\frac{\delta_{ab}}{2}\) with $a,b$ being the color indexes of final states.  
    
    The amplitudes are expanded in terms of powers of $q_3,q_4$ as, 
    \bea
    \mathcal{A}(q_3,q_4)&=&\mathcal{A}_{0,0}+q_3^{\alpha_1}\mathcal{A}_{\alpha_1,0}+q_4^{\beta_1}\mathcal{A}_{0,\beta_1}\cr
    &+&\frac{1}{2}q_3^{\alpha_1}q_3^{\alpha_2}\mathcal{A}_{\alpha_1\alpha_2,0}+\frac{1}{2}q_4^{\beta_1}q_4^{\beta_2}\mathcal{A}_{0,\beta_1\beta_2}+\ldots.
    \eea
    The covariant Lorentz indices of the amplitude are omitted herein. Where derivative terms of all orders for the amplitude $\mathcal{A}$ with respect to $q_3,q_4$​ are defined as follows.
    \bea
    \mathcal{A}_{\alpha_1\cdots\alpha_m,\beta_1\cdots\beta_n}=\sqrt{\frac{m_c}{E_c}}\sqrt{\frac{m_b}{E_b}}[\frac{\partial^{m+n}\mathcal{A}(q_3,q_4)}{\partial q_3^{\alpha_1}\cdots\partial q_3^{\alpha_m}\partial q_4^{\beta_1}\cdots\partial q_4^{\beta_n}  }\Big|_{q_3=q_4=0}]
    \eea
    The factor $\sqrt{\frac{m_Q}{E_Q}}$ comes from relativistic normalization. Derivative operator changes the orbital quantum number of the amplitude by $\Delta L=\pm1$, therefore the LO $S$-wave and $P$-wave amplitudes in the expansion correspond to the zero-order term and the first-order derivative term, respectively. Furthermore, by virtue of the conservation of charge conjugation quantum number, the odd-order derivative terms in the expansion vanish for the $S$-wave, while the even-order derivative terms do so for the $P$-wave. We also account for the $\mathcal{O}(v^2)$ relativistic corrections which are essential in many production processes\cite{Braaten:2002fi,He:2007te,Jia:2009np,Li:2013csa,Li:2013nna,Xu:2012am,Xu:2014zra}, as done in the previous work \cite{Wang:2025sbx}. After integrating the freedoms of the relative momentum angle with the $S$ or $P$ state wave functions, the replacements are obtained,
    \bea
    \begin{aligned}
    & q_i^{\mu}q_i^{\nu} \rightarrow \frac{|\vec{q}_i|^2}{3}\Pi_i^{\mu\nu},~~\mathrm{for}\ \mathrm{S}\mathrm{-wave}\ \mathrm{states} \\
    & \left.
      \begin{aligned}
      & q_i^{\mu} \rightarrow |\vec{q}_i|\epsilon^{\mu}(L_z) \\
      & q_i^{\mu}q_i^{\nu}q_i^{\xi} \rightarrow \frac{|\vec{q}_i|^3}{5}\Pi_i^{\mu\nu\xi}=\frac{|\vec{q}_i|^3}{5}\left[\Pi_i^{\mu\nu}\epsilon^\xi(L_z)+\Pi_i^{\nu\xi}\epsilon^\mu(L_z)+\Pi_i^{\xi\mu}\epsilon^\nu(L_z)\right]
      \end{aligned}
      \right\} ~~\mathrm{for}\ \mathrm{P}\mathrm{-wave}\ \mathrm{states}
    \end{aligned}
    \eea
    with $\Pi_i^{\mu\nu}=-g^{\mu\nu}+\frac{p_i^{\mu}p^{\nu}_i}{p_i^2}$, $\epsilon^\rho(L_z)$ is the orbital polarization vector of the $P$-wave states. .

		It is convenient to introduce the Lorentz invariant Mandelstam variables, which are defined as,
	\bea
	s=(p_1+p_2)^2 ,
	t=(p_1-p_3)^2 ,
	u=(p_1-p_4)^2.
	\label{mandel}
	\eea
	with the relationship $s+t+u=M_{H_1}^2+M_{H_2}^2$. The variable $s$ is indepent of $|\vec{q}|^2$, while
 $t$ and $u$ are $|\vec{q}|^2$ dependent. Let $t_0, u_0$ be defined as the values of $t, u$ when $|\vec{q_3}|^2 = |\vec{q_4}|^2 = 0$. The following expansions are derived\cite{Li:2013csa},   
 \bea
 t&=&t_0+Ft(2m_c,2m_b)|\vec{q_3}|^2+Ft(2m_b,2m_c)|\vec{q_4}|^2+\mathcal{O}(v^4)\\
  u&=&u_0+Fu(2m_c,2m_b)|\vec{q_3}|^2+Fu(2m_b,2m_c)|\vec{q_4}|^2+\mathcal{O}(v^4)
 \eea
with
\bea
Ft(x,y)&=&\frac{4[(t_0+u_0)(t_0+y^2)-2y^2(t_0+x^2)]}{s[s-2(x^2+y^2)]+(x^2-y^2)^2}\\
Fu(x,y)&=&\frac{4[(t_0+u_0)(u_0+y^2)-2y^2(u_0+x^2)]}{s[s-2(x^2+y^2)]+(x^2-y^2)^2}
\eea

 \section{Input parameters and LDMEs}
 \label{inputldmes}
 We will take the following input parameters and LDMEs in the calculation, which are the same as those used in our previous work\cite{Wang:2025sbx}, 
 \bea
 \label{input}
 ~\alpha_s &=&0.26, ~\alpha=1/137,~m_c=1.5(1.75)~GeV,\cr
  ~m_b&=&4.7(4.94)~GeV , 
 m_Z=91.1876~ GeV,\cr
 ~\Gamma_Z&=&2.4952~GeV,~\sin^2\theta_w=0.2312,~v^2_{c\bar{c}}=0.23,~v^2_{b\bar{b}}=0.1,
 \eea
 where~$\alpha_s$~ is the strong coupling constant. ~$\alpha$~ is the EW fine structure constant. ~ $\theta_W$~ is the Weinberg angle. ~$m_c$,$~m_b$, and~$m_Z$ are the masses of charm quark, bottom quark, and~$Z^0$~ boson, respectively.~ $\Gamma_Z$ is the total decay width of ~$Z^0$~boson. $v^2_{Q\bar{Q}}$ is the square of the relative velocity between $Q$ and $\bar{Q}$ in heavy quarkonium, and its value is estimated via the G-K relation\cite{Gremm:1997dq}. Quark mass values enclosed in the brackets are for $P$-wave heavy quarkonia ($\chi_{QJ}(J=0,1,2)$ and $h_Q$).

 The LDMEs of charmonium are listed as follows.
 \bea
 \langle\mathcal{O}^{J/\psi}[^3S_1^{[1]}]\rangle&=&1.2~GeV^3\cr
 \langle\mathcal{O}^{J/\psi}[^1S_0^{[8]}]\rangle&=&0.0180\pm0.0087~GeV^3\cr
 \langle\mathcal{O}^{J/\psi}[^3S_1^{[8]}]\rangle&=&0.0013\pm0.0013~GeV^3\cr
 \langle\mathcal{O}^{J/\psi}[^3P_0^{[8]}]\rangle&=&(0.0180\pm0.0087)m_c^2~GeV^3 \nonumber
 \eea
 \bea
 \langle\mathcal{O}^{\eta_c}[^1S_0^{[1]}]\rangle&=&\frac{1}{3}\times1.2~GeV^3\cr
 \langle\mathcal{O}^{\eta_c}[^1S_0^{[8]}]\rangle&=&\frac{1}{3}\times(0.0013\pm0.0013)~GeV^3\cr
 \langle\mathcal{O}^{\eta_c}[^3S_1^{[8]}]\rangle&=&0.0180\pm0.0087~GeV^3\cr
 \langle\mathcal{O}^{\eta_c}[^1P_1^{[8]}]\rangle&=&	3\times(0.0180\pm0.0087)m_c^2~GeV^3\nonumber
 \eea
 \bea
 \langle\mathcal{O}^{h_c}[^1P_1^{[1]}]\rangle&=&	3\times0.054m_c^2~GeV^3\cr
 \langle\mathcal{O}^{h_c}[^1S_0^{[8]}]\rangle&=&	3\times(0.00187\pm0.00025)~GeV^3\nonumber
 \eea
 \bea
 \langle\mathcal{O}^{\chi_{c0}}[^3P_0^{[1]}]\rangle&=&0.054m_c^2~GeV^3\cr
 \langle\mathcal{O}^{\chi_{c0}}[^3S_1^{[8]}]\rangle&=&0.00187\pm0.00025~GeV^3\cr
 \langle\mathcal{O}^{\chi_{c1}}[^3P_1^{[1]}]\rangle&=&3\times0.054m_c^2~GeV^3\cr
 \langle\mathcal{O}^{\chi_{c1}}[^3S_1^{[8]}]\rangle&=&3\times(0.00187\pm0.00025)~GeV^3\cr
 \langle\mathcal{O}^{\chi_{c2}}[^3P_2^{[1]}]\rangle&=&5\times0.054m_c^2~GeV^3\cr
 \langle\mathcal{O}^{\chi_{c2}}[^3S_1^{[8]}]\rangle&=&5\times(0.00187\pm0.00025)~GeV^3\nonumber
 \label{ldmescc}
 \eea
 For bottomonium, the values of those are,
 \bea
 \langle\mathcal{O}^{\Upsilon}[^3S_1^{[1]}]\rangle&=&10.9~GeV^3\cr
 \langle\mathcal{O}^{\Upsilon}[^1S_0^{[8]}]\rangle&=&(0.0121\pm0.0400)~GeV^3\cr
 \langle\mathcal{O}^{\Upsilon}[^3S_1^{[8]}]\rangle&=&(0.0477\pm0.0334)~GeV^3\cr
 \langle\mathcal{O}^{\Upsilon}[^3P_0^{[8]}]\rangle&=&5\times(0.0121\pm0.0400)m_b^2~GeV^3 	 \nonumber
 \eea
 \bea
 \langle\mathcal{O}^{\chi_{b0}}[^3P_0^{[1]}]\rangle&=&0.1m_b^2~GeV^3\cr
 \langle\mathcal{O}^{\chi_{b0}}[^3S_1^{[8]}]\rangle&=&0.1008~GeV^3\cr
 \langle\mathcal{O}^{\chi_{b1}}[^3P_1^{[1]}]\rangle&=&3\times0.1m_b^2~GeV^3\cr
 \langle\mathcal{O}^{\chi_{b1}}[^3S_1^{[8]}]\rangle&=&3\times0.1008~GeV^3\cr
 \langle\mathcal{O}^{\chi_{b2}}[^3P_2^{[1]}]\rangle&=&5\times0.1m_b^2~GeV^3\cr
 \langle\mathcal{O}^{\chi_{b2}}[^3S_1^{[8]}]\rangle&=&5\times0.1008~GeV^3\nonumber
 \eea
 \bea
 \langle\mathcal{O}^{\eta_{b}}[^1S_0^{[1]}]\rangle&=&\frac{1}{3}\langle\mathcal{O}^{\Upsilon}[^3S_1^{[1]}]\rangle=3.633~GeV^3\cr
 \langle\mathcal{O}^{\eta_{b}}[^1S_0^{[8]}]\rangle&=&\frac{1}{3}\langle\mathcal{O}^{\Upsilon}[^3S_1^{[8]}]\rangle=(0.0159\pm0.0111)~GeV^3\cr
 \langle\mathcal{O}^{\eta_{b}}[^3S_1^{[8]}]\rangle&=&\langle\mathcal{O}^{\Upsilon}[^1S_0^{[8]}]\rangle=(0.0121\pm0.0400)~GeV^3\cr
 \langle\mathcal{O}^{\eta_{b}}[^1P_1^{[8]}]\rangle&=&3\times\langle\mathcal{O}^{\Upsilon}[^3P_0^{[8]}]\rangle=3\times5\times(0.0121\pm0.0400)m_b^2~GeV^3\nonumber
 \eea
 \bea
 \langle\mathcal{O}^{h_{b}}[^1P_1^{[1]}]\rangle&=&3\times\langle\mathcal{O}^{\chi_{b0}}[^3P_0^{[1]}]\rangle=3\times0.1m_b^2 ~GeV^3\cr
 \langle\mathcal{O}^{h_{b}}[^1S_0^{[8]}]\rangle&=&3\times\langle\mathcal{O}^{\chi_{b0}}[^3S_1^{[8]}]\rangle=3\times0.1008~ GeV^3\nonumber
 \eea
 
 The CS LDMEs are calculated using the potential model\cite{Cho:1995vh,Cho:1995ce,Eichten:1994gt},
and the CO ones are fitted to the hadron colliders data at LO in $\alpha_s$\cite{Sharma:2012dy}. Explicitly, they possess the heavy quark spin symmetry (HQSS) \cite{Bodwin:1994jh}.
 \bea
 \langle\mathcal{O}^{\chi_{QJ}}[^3P_J^{[1]}]\rangle&=&	(2J+1)\langle\mathcal{O}^{\chi_{Q0}}[^3P_0^{[1]}]\rangle\cr
 \langle\mathcal{O}^{\chi_{QJ}}[^3S_1^{[8]}]\rangle&=&	(2J+1)\langle\mathcal{O}^{\chi_{Q0}}[^3S_1^{[8]}]\rangle\cr
 \langle\mathcal{O}^{\eta_{Q}}[^1S_0^{[1]}/^1S_0^{[8]}]\rangle&=&\frac{1}{3}		\langle\mathcal{O}^{J/\psi~\mathrm{or}~\Upsilon}[^3S_1^{[1]}/^3S_1^{[8]}]\rangle	,
 \langle\mathcal{O}^{\eta_{Q}}[^3S_1^{[8]}]\rangle=	\langle\mathcal{O}^{J/\psi~\mathrm{or}~\Upsilon}[^1S_0^{[8]}]\rangle\cr
 \langle\mathcal{O}^{\eta_{Q}}[^1P_1^{[8]}]\rangle&=&	3\langle\mathcal{O}^{J/\psi~\mathrm{or}~\Upsilon}[^3P_0^{[8]}]\rangle,
 \langle\mathcal{O}^{h_{Q}}[^1P_1^{[1]}/^1S_0^{[8]}]\rangle=	3\langle\mathcal{O}^{\chi_{Q0}}[^3P_0^{[1]}/^3S_1^{[8]}]\rangle
 \label{hqss}
 \eea
 
%

 	\section{Results and discussion}\label{results}

 \subsection{Cross sections and CO contributions}\label{channels}


In the present calculation, we only consider the t-channel process of double-photon fragmentation for $J/\psi+\Upsilon$ production. For the other two t-channel processes with non-zero contributions in the final states ($J/\psi+\chi_{b1}$ and $\chi_{c1}+\Upsilon$), their contributions are negligible due to the suppression of the $Z^0$-propagator.
The production channels and corresponding cross sections at the $Z^0$ pole are shown in Table. \ref{channelsanalysis}, and we can compare this table with the corresponding one for the double charmonium/bottomonium production processes. It is observed that the CO contributions are dominant (basically $\geq95\%$) for almost all processes, except for the $J/\psi+\Upsilon, J/\psi+\eta_b, \Upsilon+h_c$ production. Unlike the double charmonium/bottomonium production\cite{Wang:2025sbx}, which has non-fragmentation diagrams in CO channels, only fragmentation ones ($g\rightarrow{}^3S_1^{[8]}$) exist for charmonium+bottomonium production. Specifically, for the production of $\eta_c+\chi_{bJ}(J=0,2), \eta_b+\chi_{cJ}(J=0,2), \eta_c+h_b, \eta_b+h_c$, and $\eta_c+\eta_b$, the CS channels have no contribution because neither photon nor Z-boson can fragment into $\eta_Q$, $h_Q$ and $\chi_{QJ}(J=0,2)$.

 	One could check these results with the fragmentation function of the gluon into $^3S_1^{[8]}$ state at $\mathcal{O}(\alpha_s)$
 in the high energy limit ($s\gg M_Q^2$),
 \bea\label{eqfragment}
 \sigma(e^+e^- \to H_1+ g \to H_1 + H_2 + X) = \sigma(e^+e^- \to H_1+ g) \, \int_0^1 dz \, D_{g \to H_2}(z),
 \eea
 where $z=2E_2/\sqrt{s}$. At LO of $\mathcal{O}(\alpha_s)$, the fragmentation function is given by
 $D_{g \to H_2(^3S_1^{[8]})}(z)  = \frac{\pi\alpha_s\delta(1 - z)}{3(N_c^2 - 1)}\frac{\langle\mathcal{O}^{H_2}[^3S_1^{[8]}]\rangle}{m_c^3}$\cite{Ma:2013yla}.

 At the tree level, the gluon fragmentation to $^3S_1^{[8]}$ state is most important, but for the higher-order diagrams, as discussed in Ref. \cite{Wang:2025sbx}, the gluon fragmentation into other intermediate states (e.g., $^3P_J^{[8]},^1S_0^{[1]}$) also yields significant contributions, which may weaken the importance of contributions from the $^3S_1^{[8]}$ channels, and a definitive conclusion therefore requires more comprehensive calculations that include higher-order QCD corrections.

		\begin{table}
		\caption{Production channels  and corresponding cross sections (units:$\times10^{-4} fb$) at $Z^0$ pole in $\mathcal{O}(v^0)$. The percentage in the brackets is the proportion of CO. The negligible results($<0.1\times 10^{-4} fb$) are not shown.
		}
		\begin{tabular}{|c| |cc| |cccc||cc||cc|}
			\hline
			$H_1$&\multicolumn{10}{c|}{ $H_2$}\\
			\hline
			\hline
			$J/\psi$& $\Upsilon$ &\makecell{370.5 \\(1.5 \%)}& $\chi_{bJ}$&\makecell{4.1 $(J=0)$\\( 96.2\%)}& \makecell{ 12.8$(J=1)$\\(92.1 \%)}& \makecell{20.0 $(J=2)$\\( 98.3\%)}&$\eta_b$&\makecell{21.8 \\( 14.2\%)} &$h_b$& \makecell{6.2 \\(83.3 \%)}\\
			\hline
			\hline
				$^3S_1^{[1]}$     &$^3S_1^{[1]}$& 364.9& $^3P_J^{[1]}$ & 0.2 &1.0 &0.3 & $^1S_0^{[1]}$  & 18.7 &$^1P_1^{[1]}$ & 1.0\\			    
			\hline
			$^3S_1^{[8]}$     &	$^3S_1^{[8]}$  & 0.7 & $^3S_1^{[8]}$ &1.3 &4.0 & 6.6&  $^3S_1^{[8]}$ &0.2  &  $^1S_0^{[8]}$  &5.2 \\		
			~&$^1S_0^{[8]}$&0.2  & &  &  &  & $^1S_0^{[8]}$  & 0.3  &   &  \\		
			~ &  $^3P_J^{[8]}$&3.3   &~ &~&~ &~& $^1P_1^{[8]}$& 2.3& &\\	
			\hline
			 $^1S_0^{[8]}$   &     $^3S_1^{[8]}$       &    0.4          & $^3S_1^{[8]}$ & 0.7&2.0 &3.3 &$^3S_1^{[8]}$ &  &  & \\		
			\hline	
		     $^3P_J^{[8]}$  & $^3S_1^{[8]}$ &1.1      &  $^3S_1^{[8]}$&1.9 &5.8  & 9.7& $^3S_1^{[8]}$ & 0.3& &  \\	
			\hline
			\hline
			$\eta_c$&$\eta_b$ & \makecell{38.4 \\( 100.0\%)}& $\chi_{bJ}$&\makecell{ 22.6$(J=0)$\\( 100.0\%)}& \makecell{ 67.8$(J=1)$\\( 100.0 \%)}& \makecell{ 113.0$(J=2)$\\(100.0 \%)}&$h_b$&\makecell{ 71.4\\( 100.0\%)} &$\Upsilon$&\makecell{ 61.5\\( 99.1\%)} \\
			\hline
			\hline
			 $^1S_0^{[1]}$ &  $^1S_0^{[1]}$  &     &  $^3P_J^{[1]}$  &  & &  & $^1P_1^{[1]}$ & &$^3S_1^{[1]}$ & 0.6\\	
			 \hline
			$^3S_1^{[8]}$ &$^3S_1^{[8]}$  & 2.3  &$^3S_1^{[8]}$  &18.4 &55.2 & 91.9&$^1S_0^{[8]}$ &71.4 &$^3S_1^{[8]}$  & 9.2\\
			 ~ & $^1S_0^{[8]}$ & 3.9   &   &  & &  & & &$^1S_0^{[8]}$ & 3.0\\	
			  ~ & $^1P_1^{[8]}$ & 31.5  &   &  & &  & & &$^3P_J^{[8]}$ & 46.3\\	
			\hline
		 $^1S_0^{[8]}$  & $^3S_1^{[8]}$   &   & $^3S_1^{[8]}$  & & & &   & & $^3S_1^{[8]}$ &  \\	
		\hline
			$^1P_1^{[8]}$       & $^3S_1^{[8]}$ & 0.6&$^3S_1^{[8]}$  &4.2 &12.6 &20.9 & & & $^3S_1^{[8]}$ & 2.3 \\
			\hline
			\hline
			$\eta_b$&$h_c$ &\makecell{ \\( 100.0\%)} & $\chi_{cJ}$&\makecell{ 2.5$(J=0)$\\( 100.0\%)}& \makecell{ 7.4  $(J=1)$\\( 100.0\%)}& \makecell{ 12.4$(J=2)$\\( 100.0\%)}& &   &   & \\
			\hline
				 \hline
				  $^1S_0^{[1]}$   & $^1P_1^{[1]}$ &     &  $^3P_J^{[1]}$&  &  &  & & & &\\	
				  \hline
		 $^3S_1^{[8]}$   & $^1S_0^{[8]}$ &    &  $^3S_1^{[8]}$& 0.2 & 0.5&0.8 & & & &\\	
			\hline
		   $^1S_0^{[8]}$    & &  &   $^3S_1^{[8]}$ & 0.3&0.8 & 1.3& & &    & ~  \\		
		   \hline
		      $^1P_1^{[8]}$    & &  &   $^3S_1^{[8]}$ & 2.1&6.2 & 10.3& & &    & ~  \\		
			\hline
			\hline
			$\Upsilon$& $h_c$&\makecell{ 0.5\\(18.6 \%)} & $\chi_{cJ}$&\makecell{ 3.9$(J=0)$\\( 99.5\%)}& \makecell{ 11.7$(J=1)$\\( 98.9\%)}& \makecell{19.3 $(J=2)$\\( 99.8 \%)}& & & & \\
			\hline
			\hline
			 $^3S_1^{[1]}$ &   $^1P_1^{[1]}$ & 0.4 & $^3P_J^{[1]}$  &  &  0.1&  &  &  & & \\	
			 \hline
			  $^3S_1^{[8]}$ &   $^1S_0^{[8]}$ & 0.1  & $^3S_1^{[8]}$    &  0.6& 1.9 &3.1   &  &  & & \\	
			\hline
		$^1S_0^{[8]}$  &  &   &$^3S_1^{[8]}$   &0.2& 0.6&  1.0& &  & & \\	
			\hline
		$^3P_J^{[8]}$ 	 &  & & $^3S_1^{[8]}$  & 3.0&9.1 &15.1 & & & & \\	
			\hline			
		\end{tabular}
		\label{channelsanalysis}
	\end{table}
	\FloatBarrier

\begin{widetext}
		\begin{figure*}[htbp]
			\begin{tabular}{c c c}
				\includegraphics[width=0.333\textwidth]{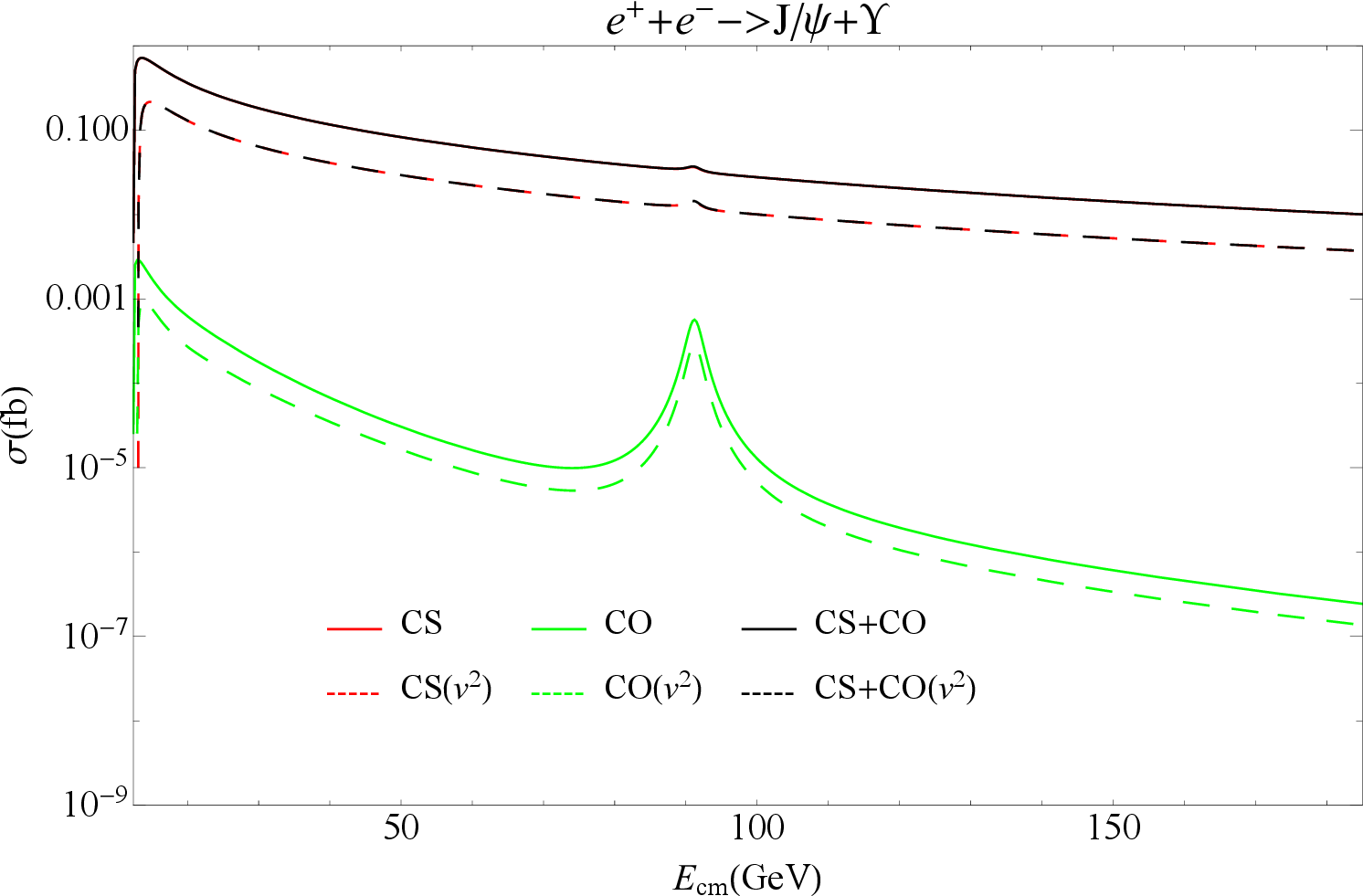}
				\includegraphics[width=0.333\textwidth]{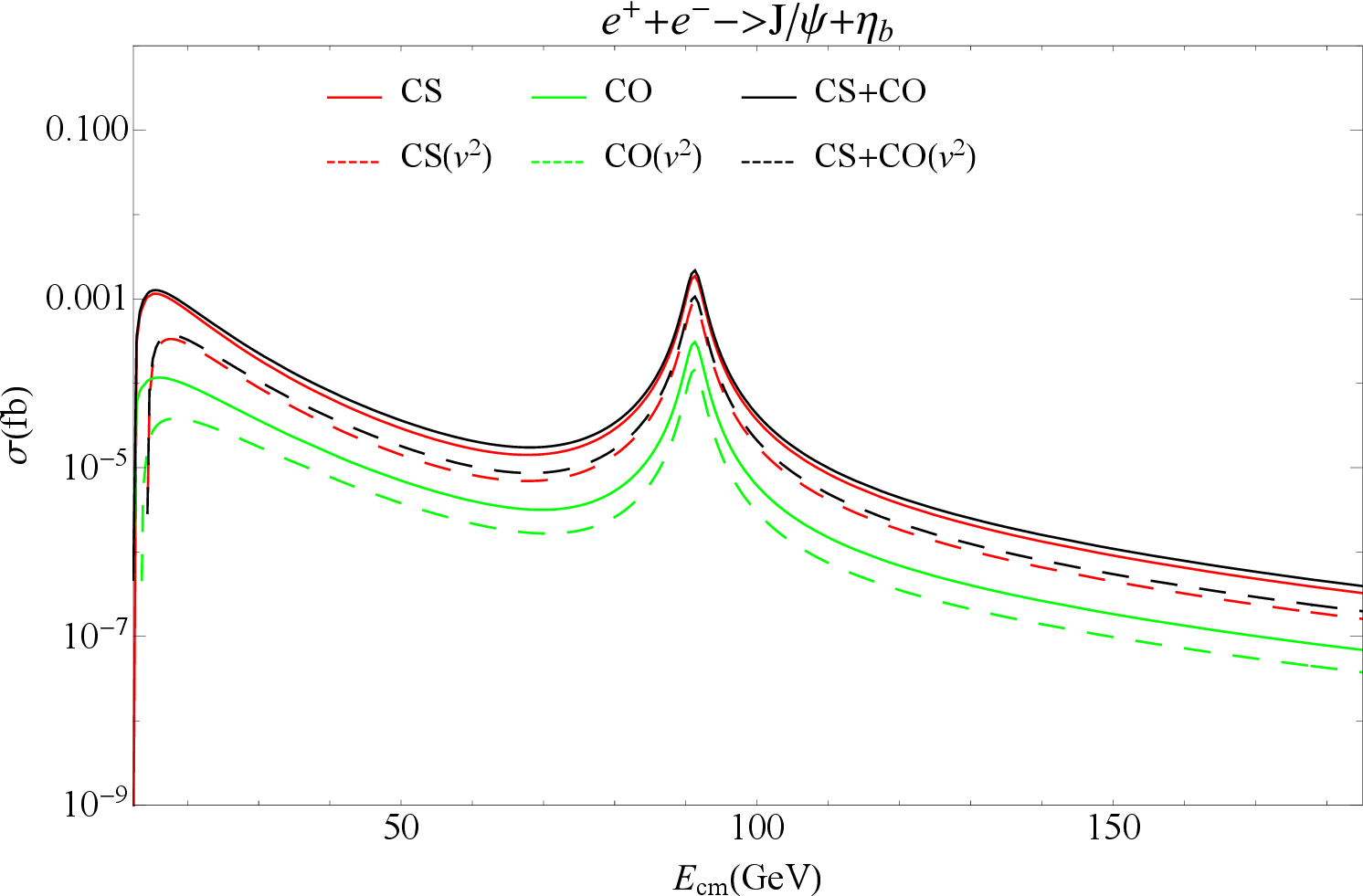}
				\includegraphics[width=0.333\textwidth]{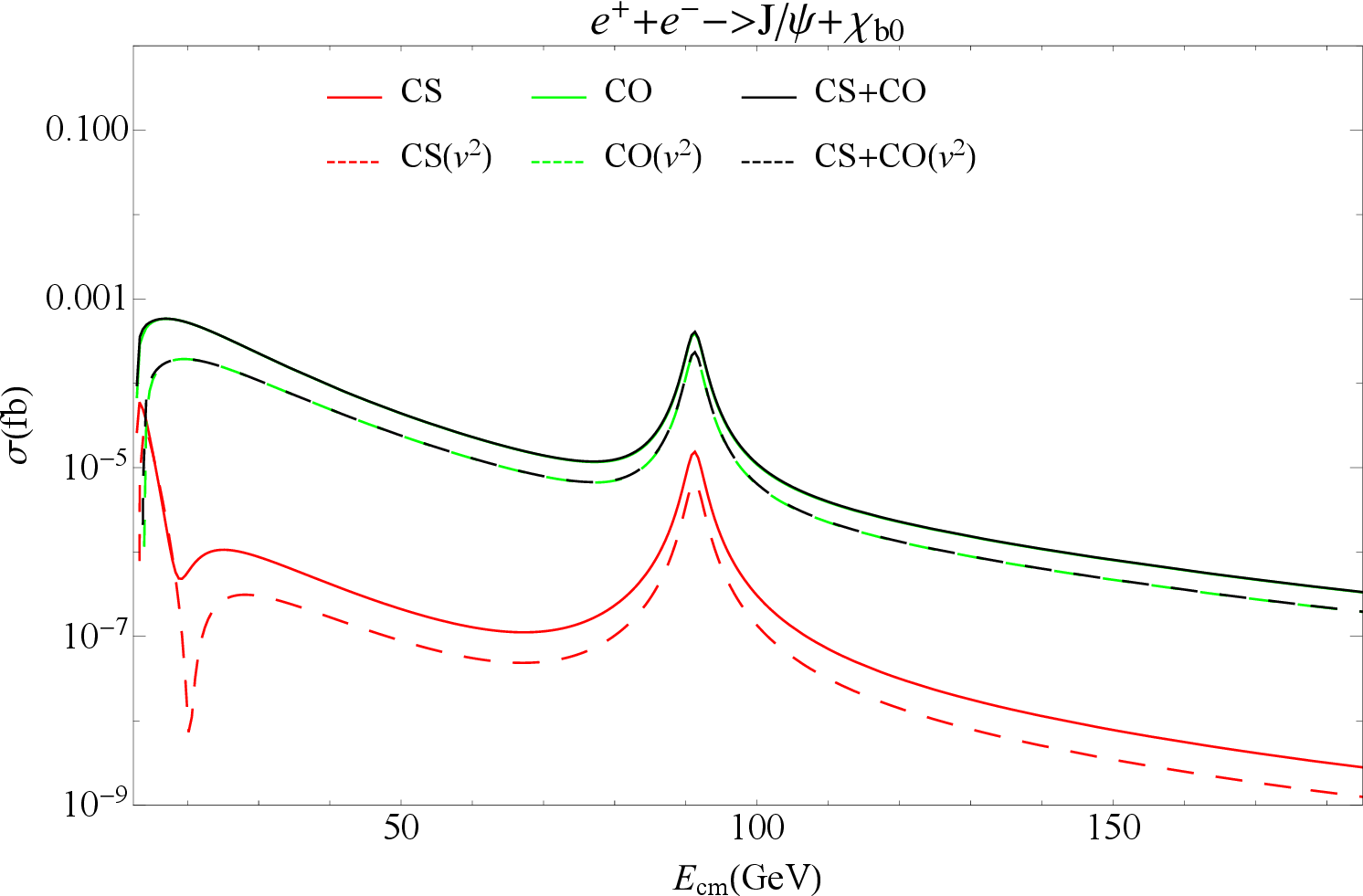}
			\end{tabular}
			\begin{tabular}{c c c}						
				\includegraphics[width=0.333\textwidth]{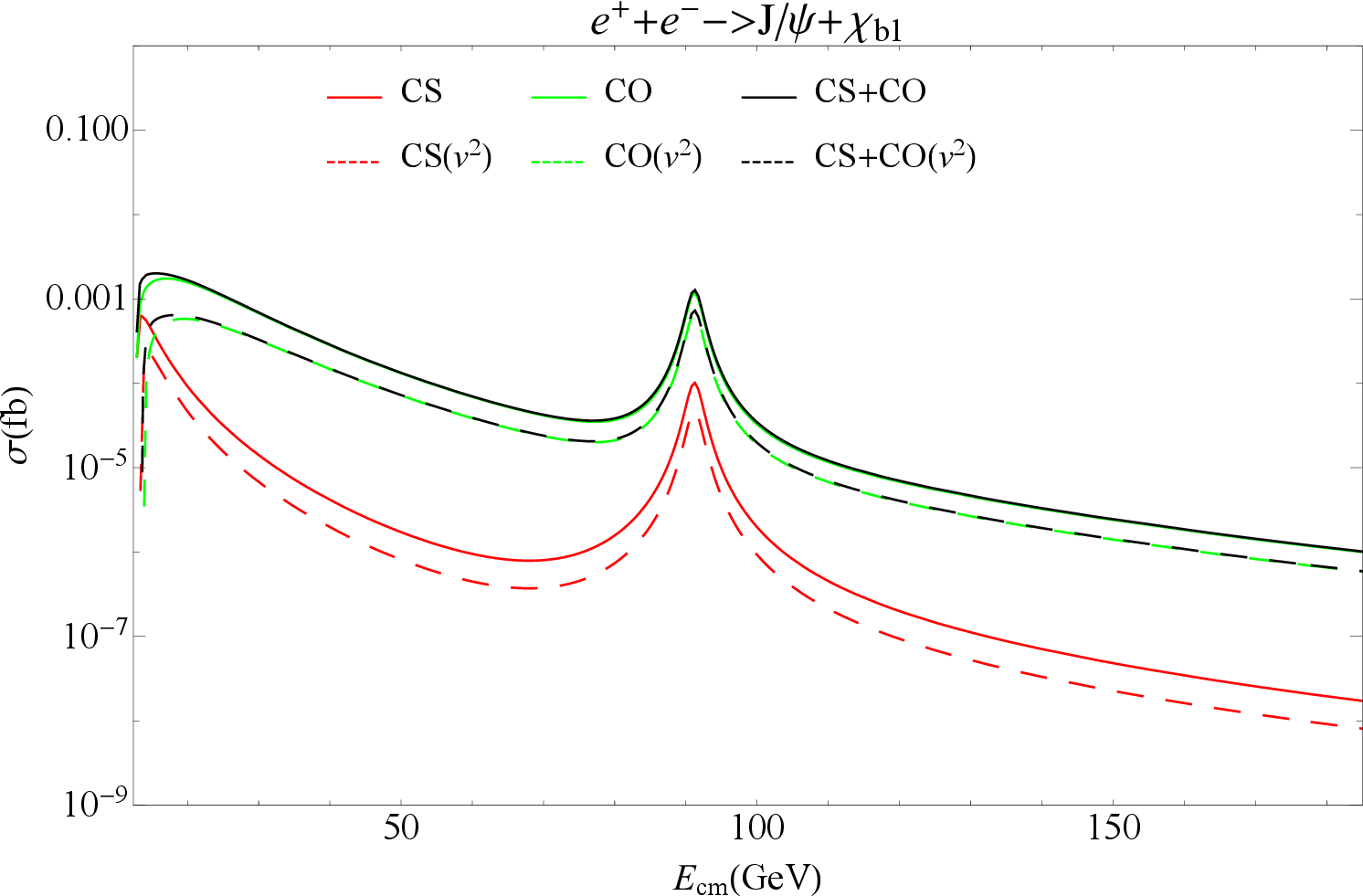}
				\includegraphics[width=0.333\textwidth]{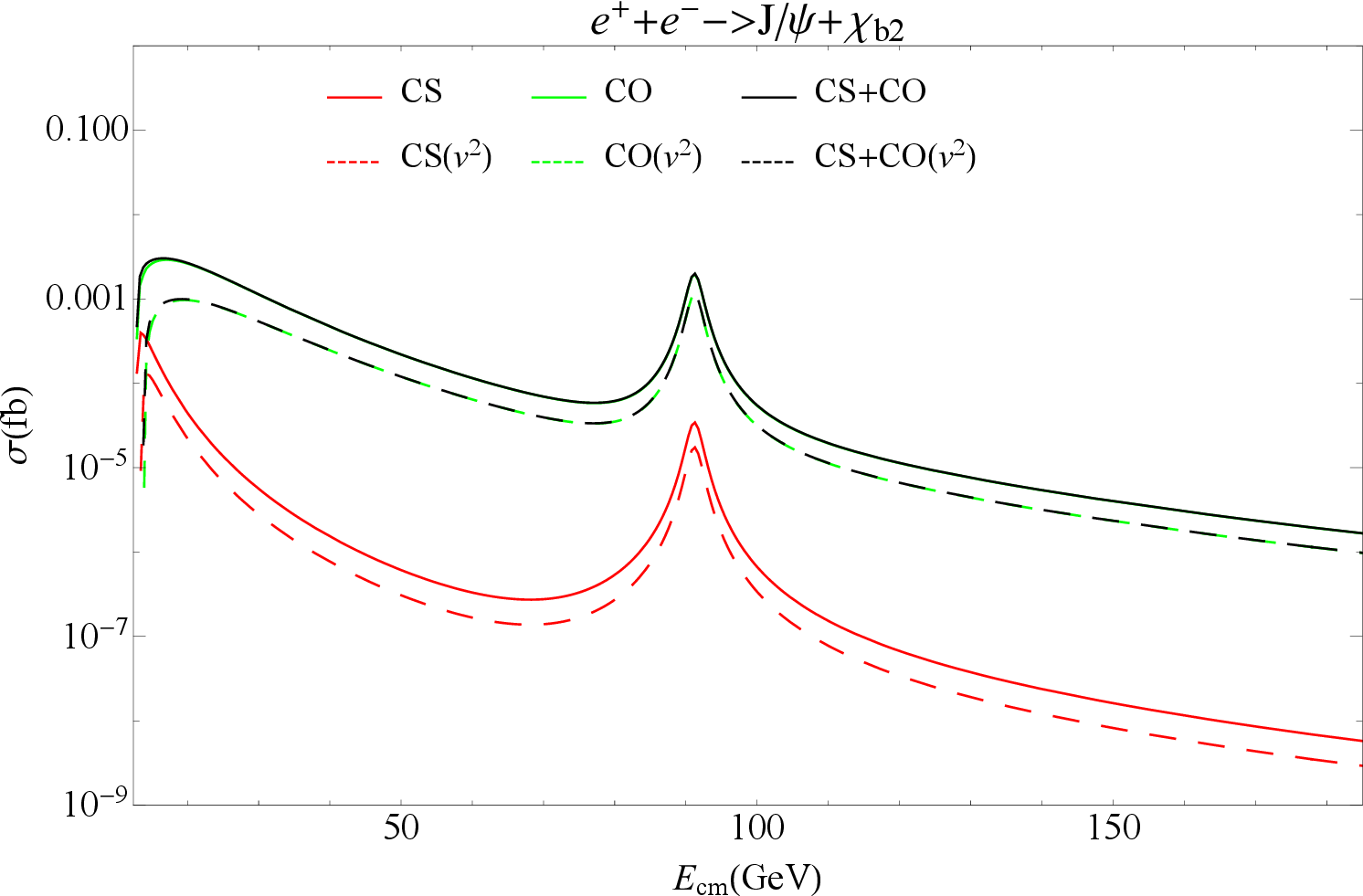}
				\includegraphics[width=0.333\textwidth]{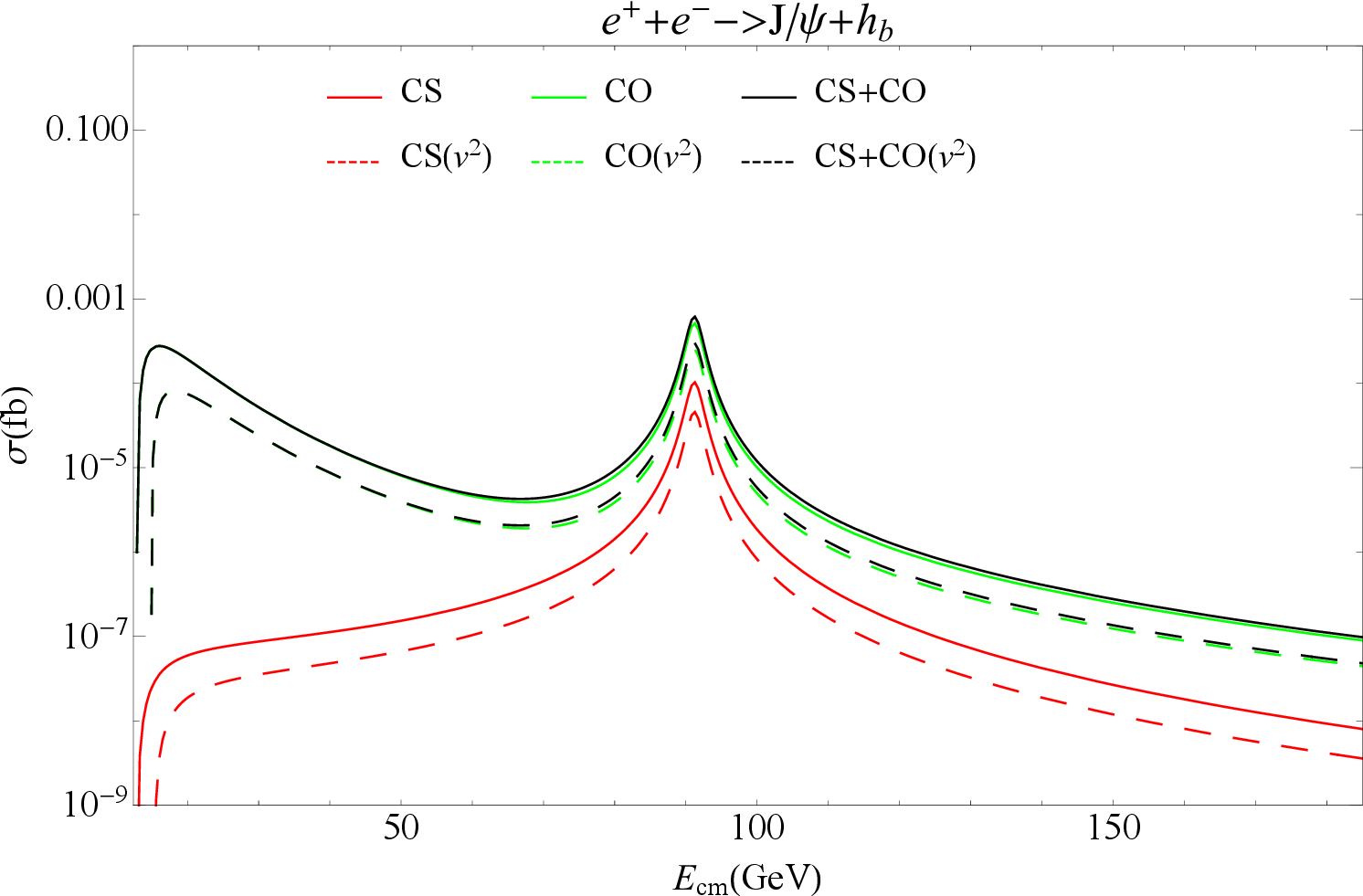}
			\end{tabular}
			
			\begin{tabular}{c c c }
				
				\includegraphics[width=0.333\textwidth]{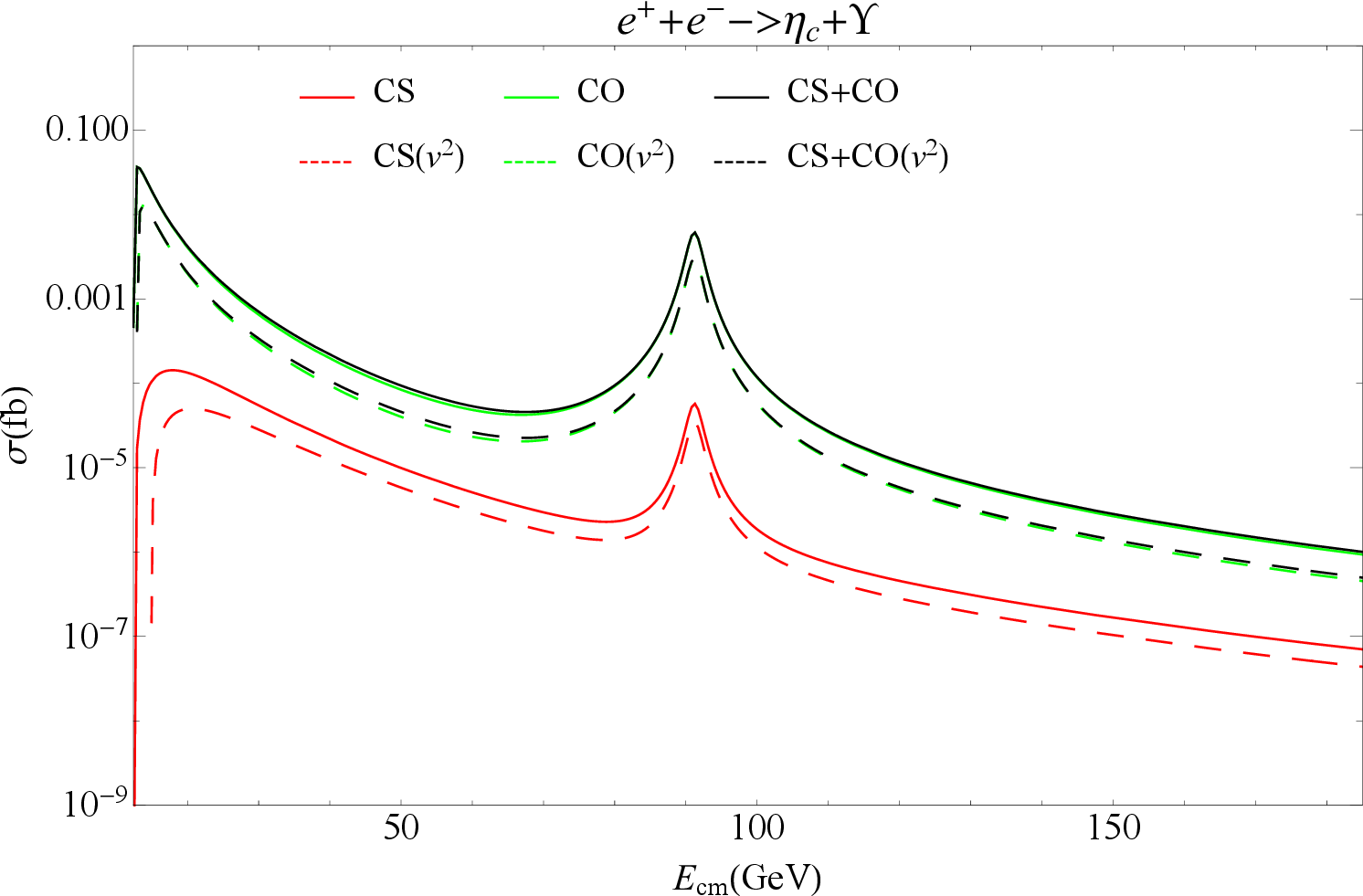}
				\includegraphics[width=0.333\textwidth]{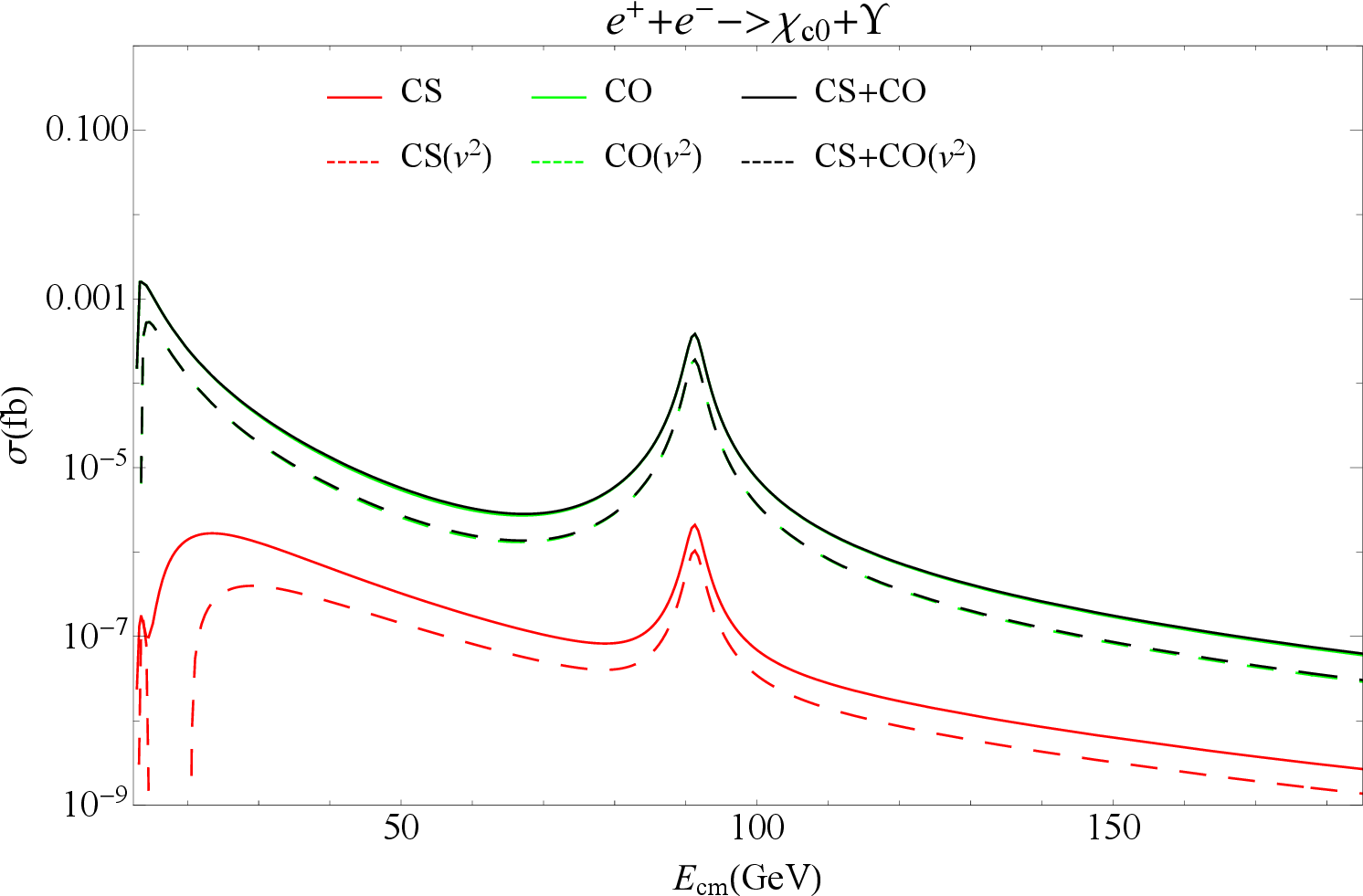}
				\includegraphics[width=0.333\textwidth]{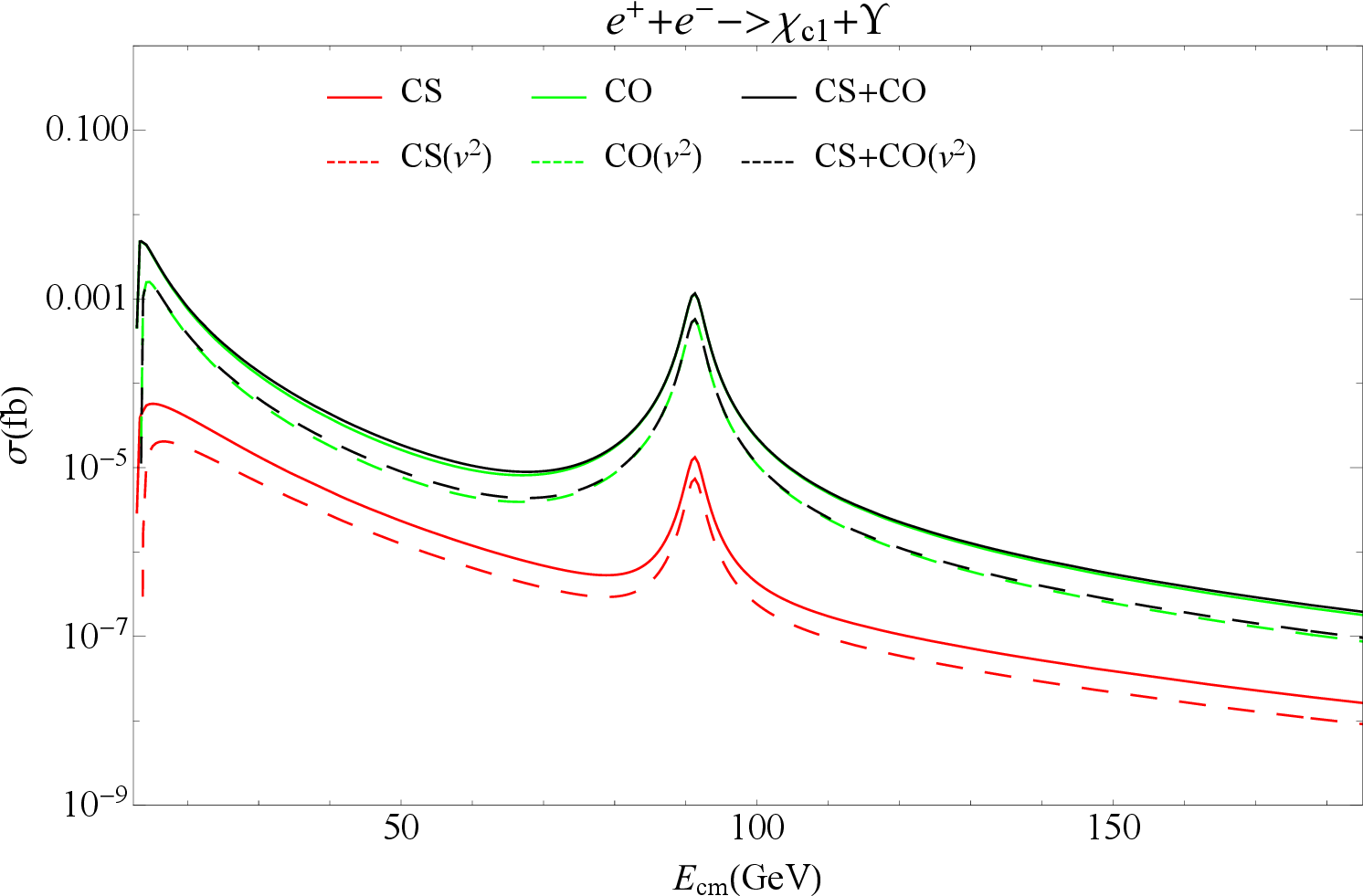}
			\end{tabular}
			\begin{tabular}{c c c }
				\includegraphics[width=0.333\textwidth]{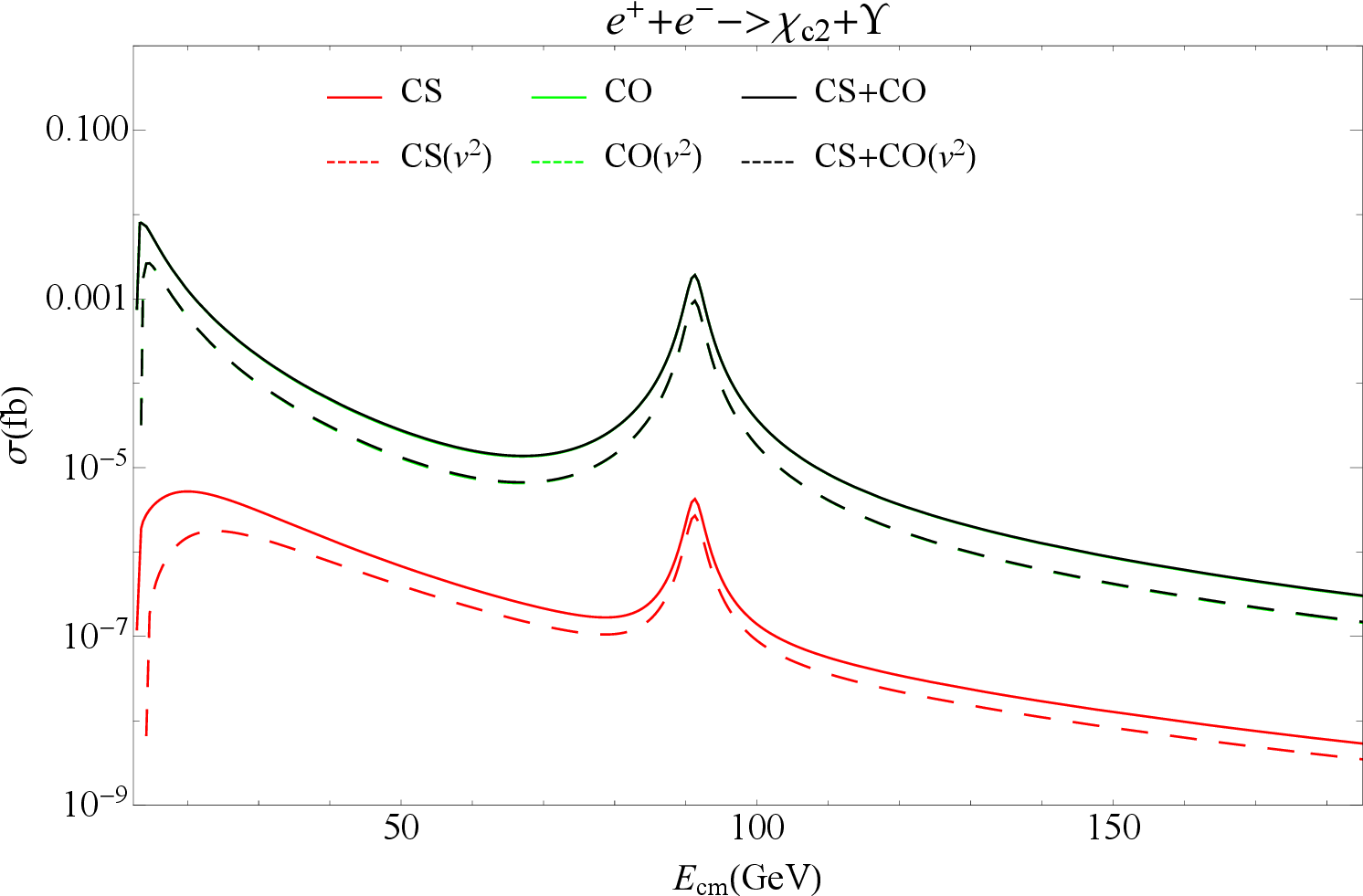}
				\includegraphics[width=0.333\textwidth]{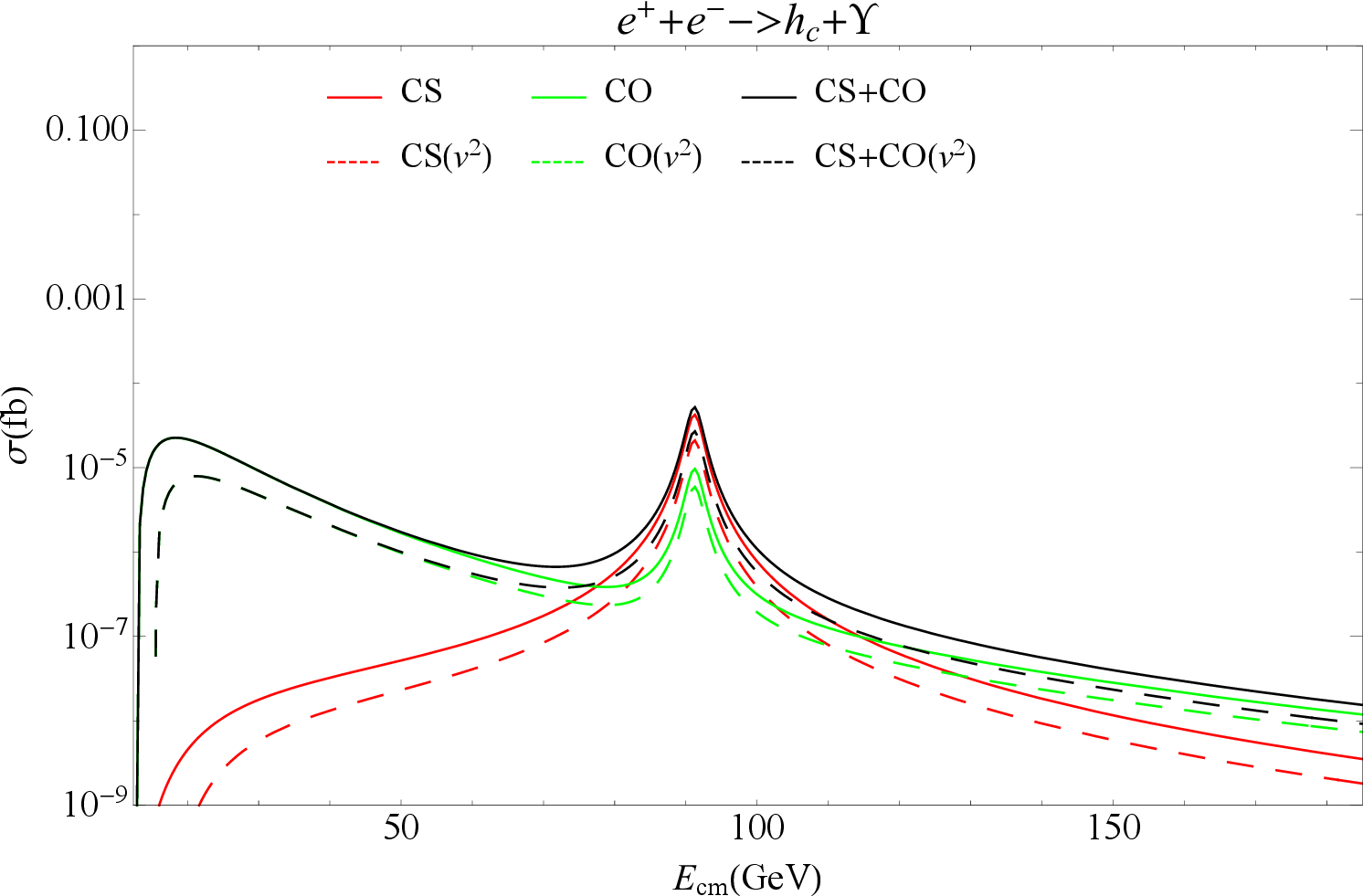}
				\includegraphics[width=0.333\textwidth]{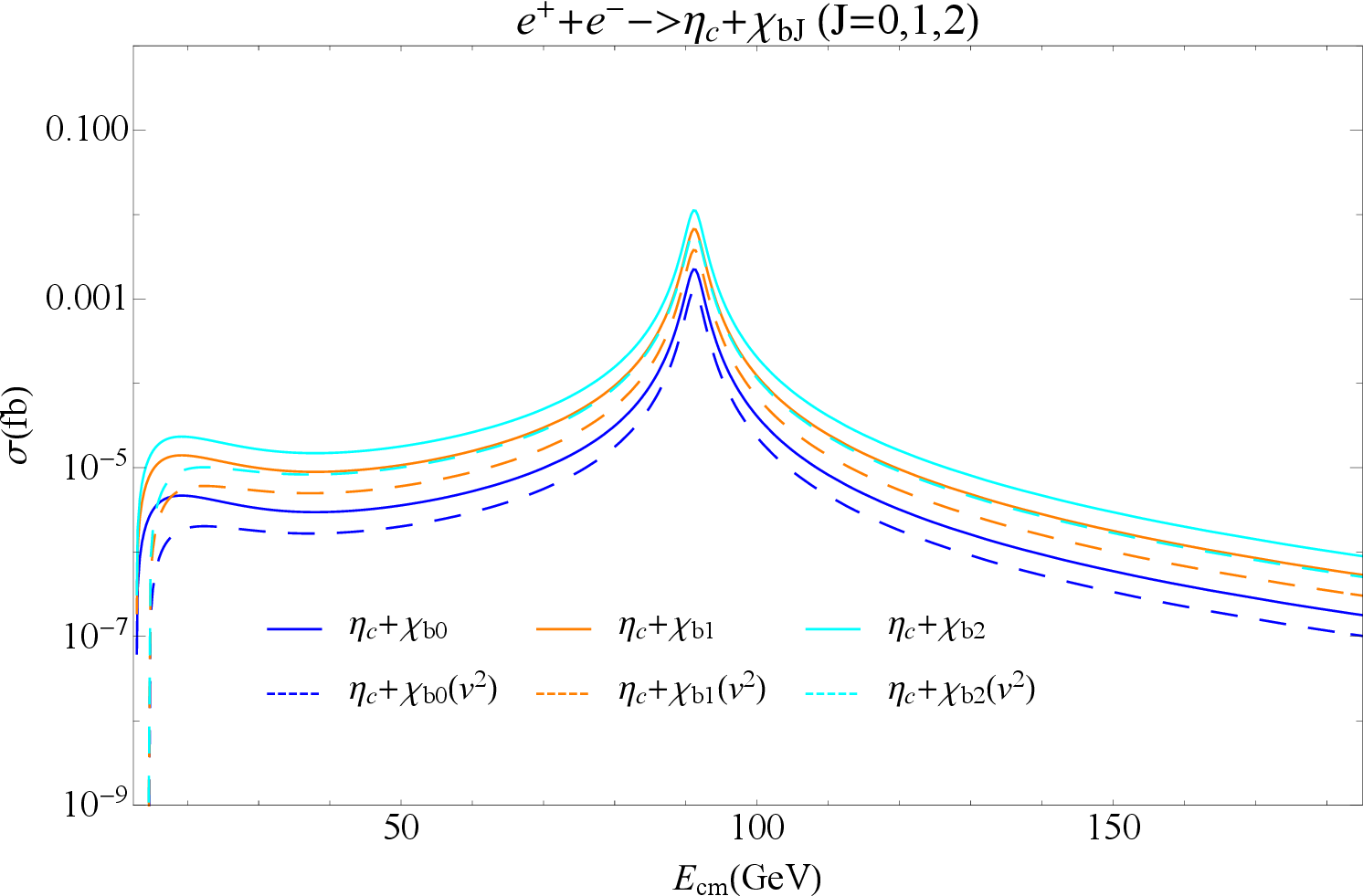}
			\end{tabular}
			\begin{tabular}{c c c }		
				\includegraphics[width=0.333\textwidth]{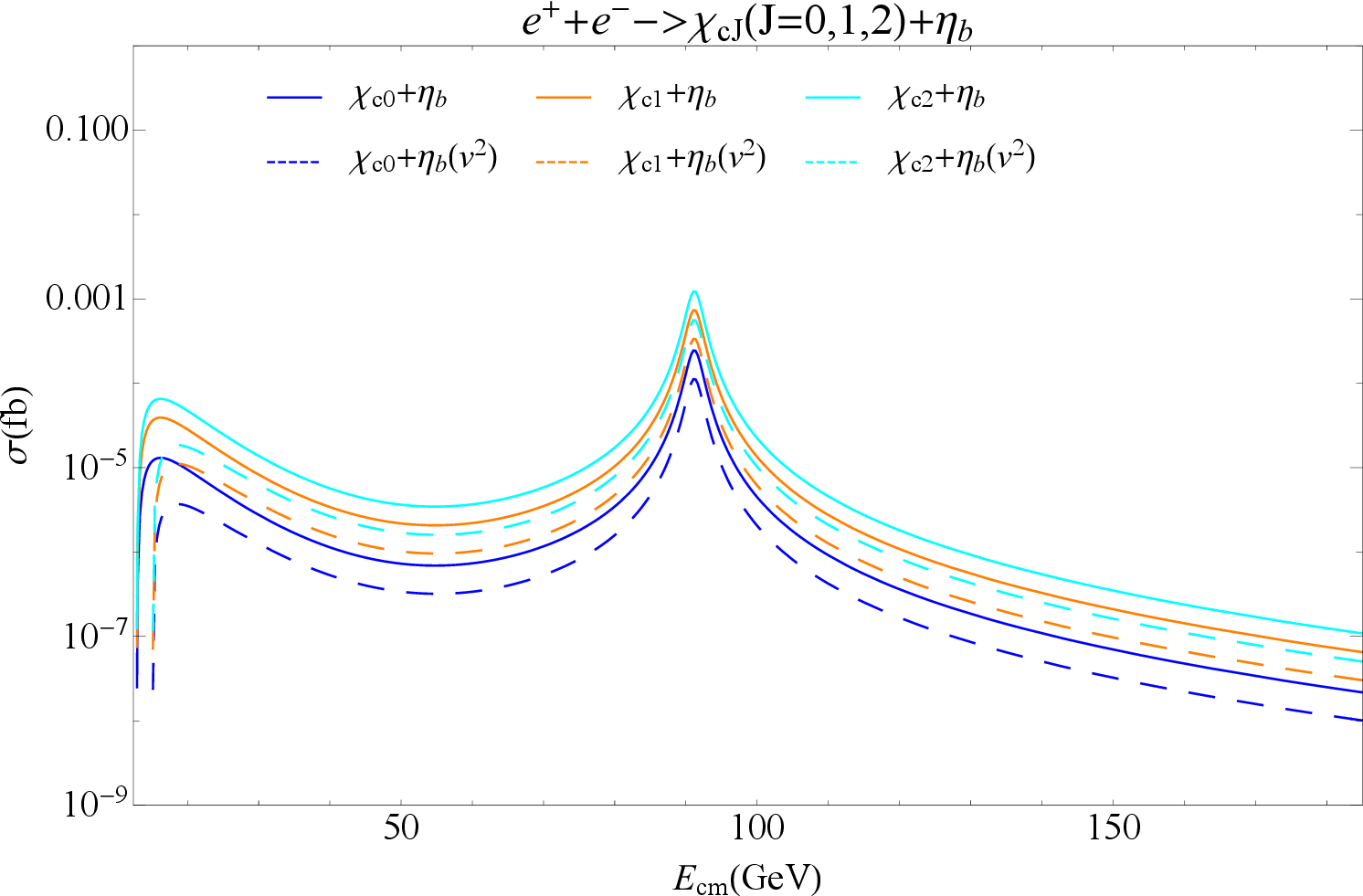}
				\includegraphics[width=0.333\textwidth]{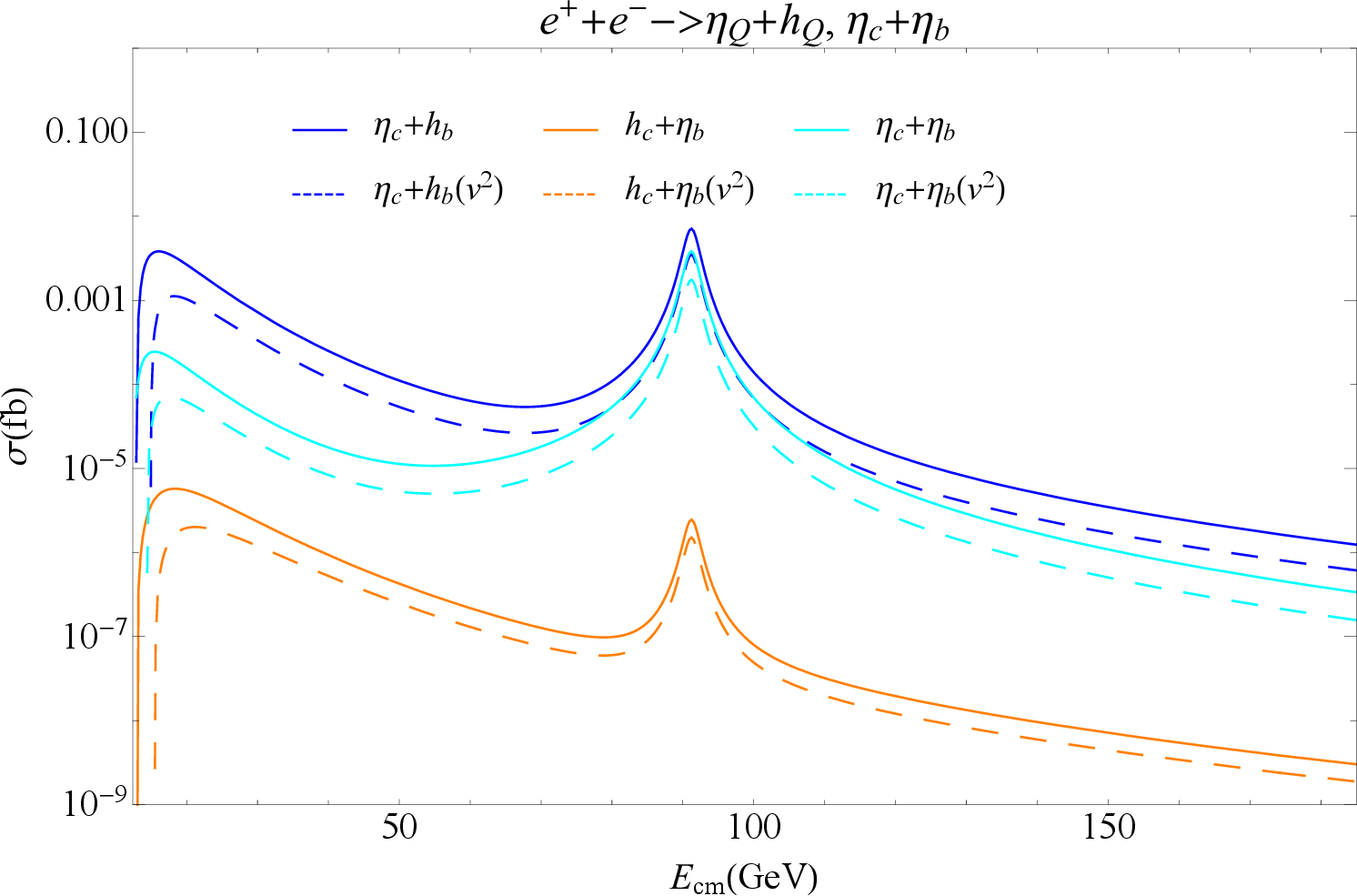}
			\end{tabular}
			
			\caption{(Color online) Cross section $~\sigma~$ versus c.m. energy~$E_{cm}$~$(E_{cm}=\sqrt{s})$. The solid line represents LO  and dashed line represents NLO($v^2$) result. The red line represent CS channel, green line represent total CO channels, black line represent the sum of  CS and CO. And there are only CO channels for the last $\eta_Q+\chi_{Q'J}, \eta_Q+h_Q', \eta_c+\eta_b$ processes.   }
			\label{z0cc}
		\end{figure*}
	\end{widetext}

\begin{widetext}
		\begin{figure*}[htbp]
			\begin{tabular}{c c c}
				\includegraphics[width=0.333\textwidth]{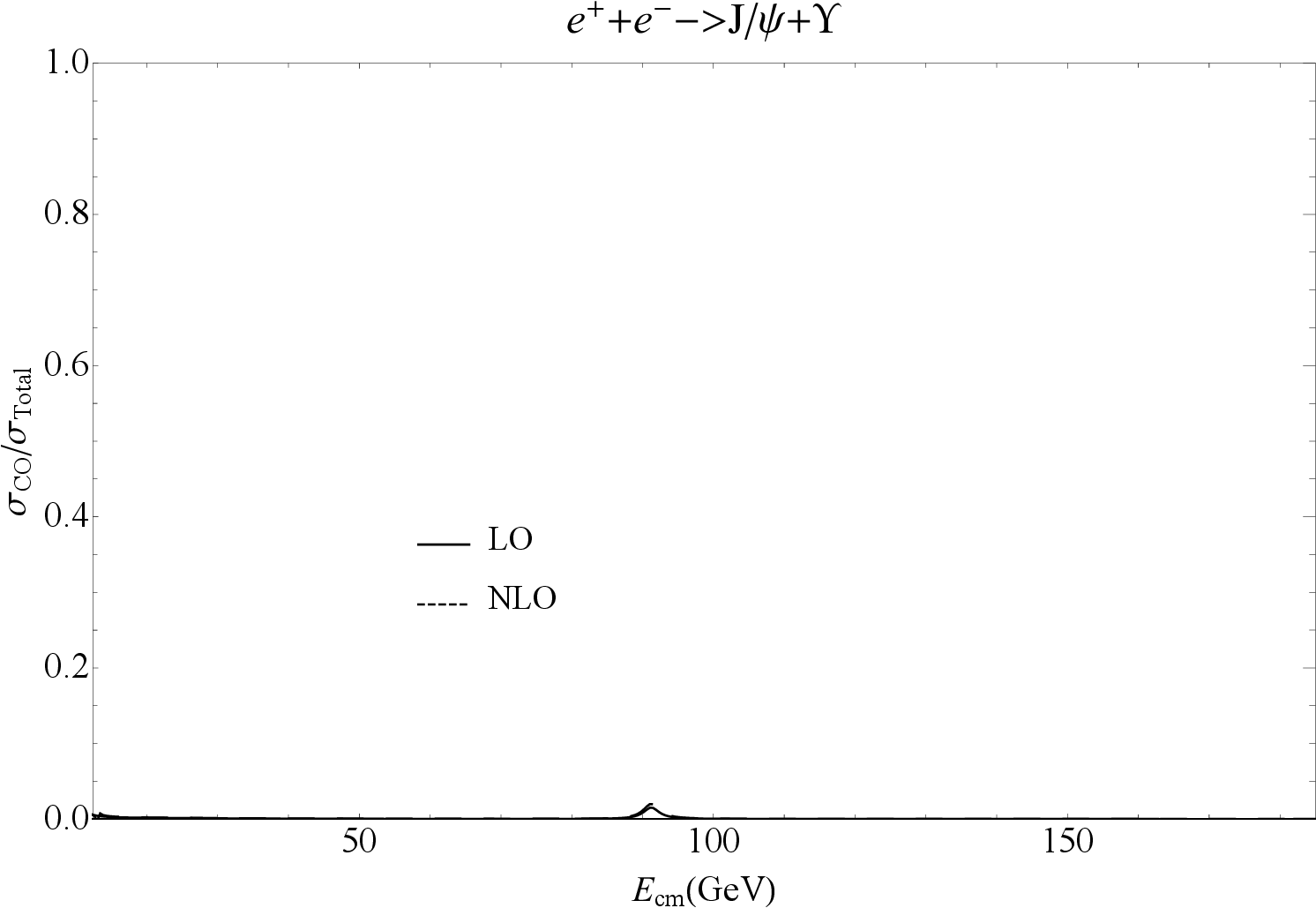}
				\includegraphics[width=0.333\textwidth]{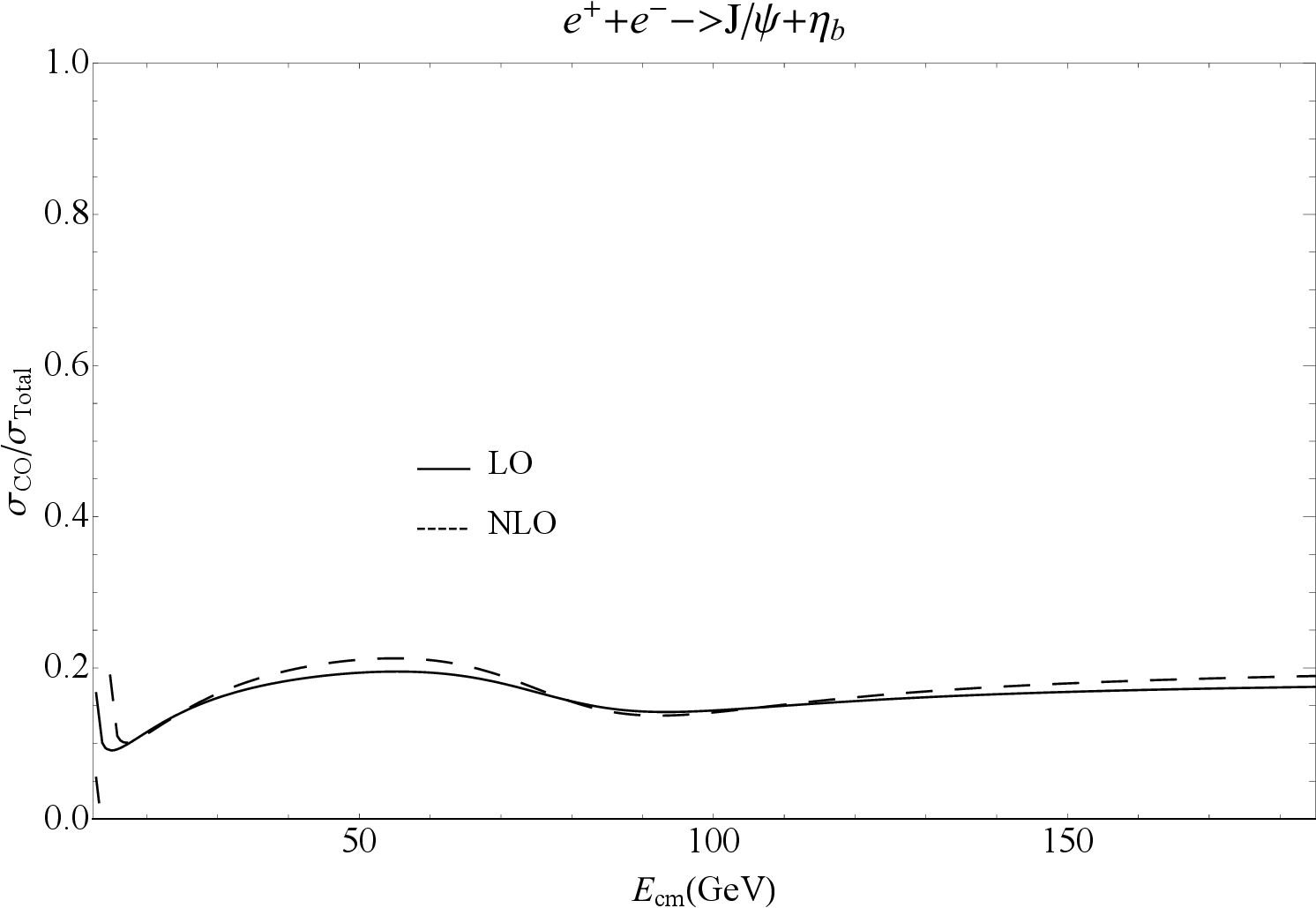}
				\includegraphics[width=0.333\textwidth]{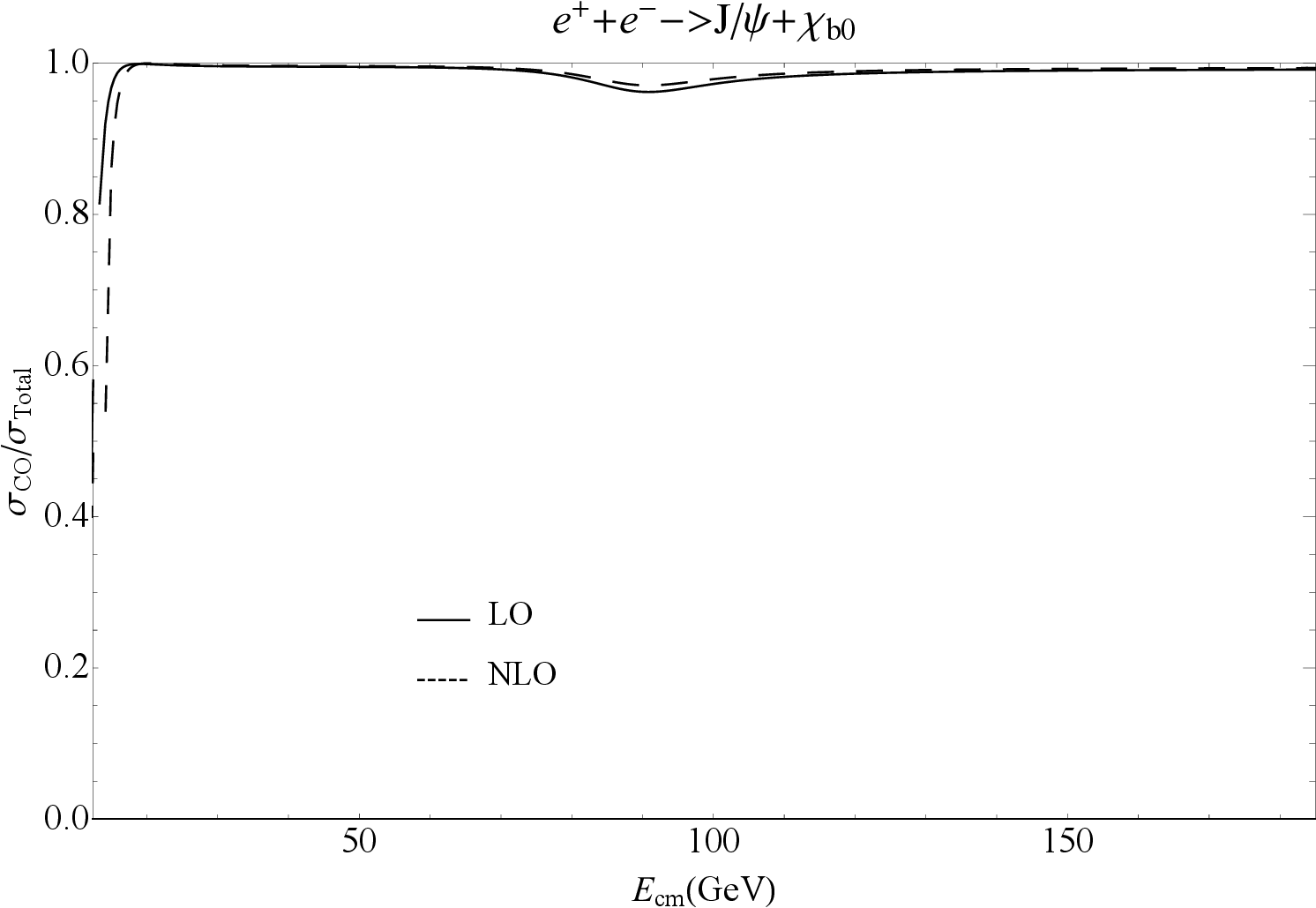}
			\end{tabular}
			\begin{tabular}{c c c}						
				\includegraphics[width=0.333\textwidth]{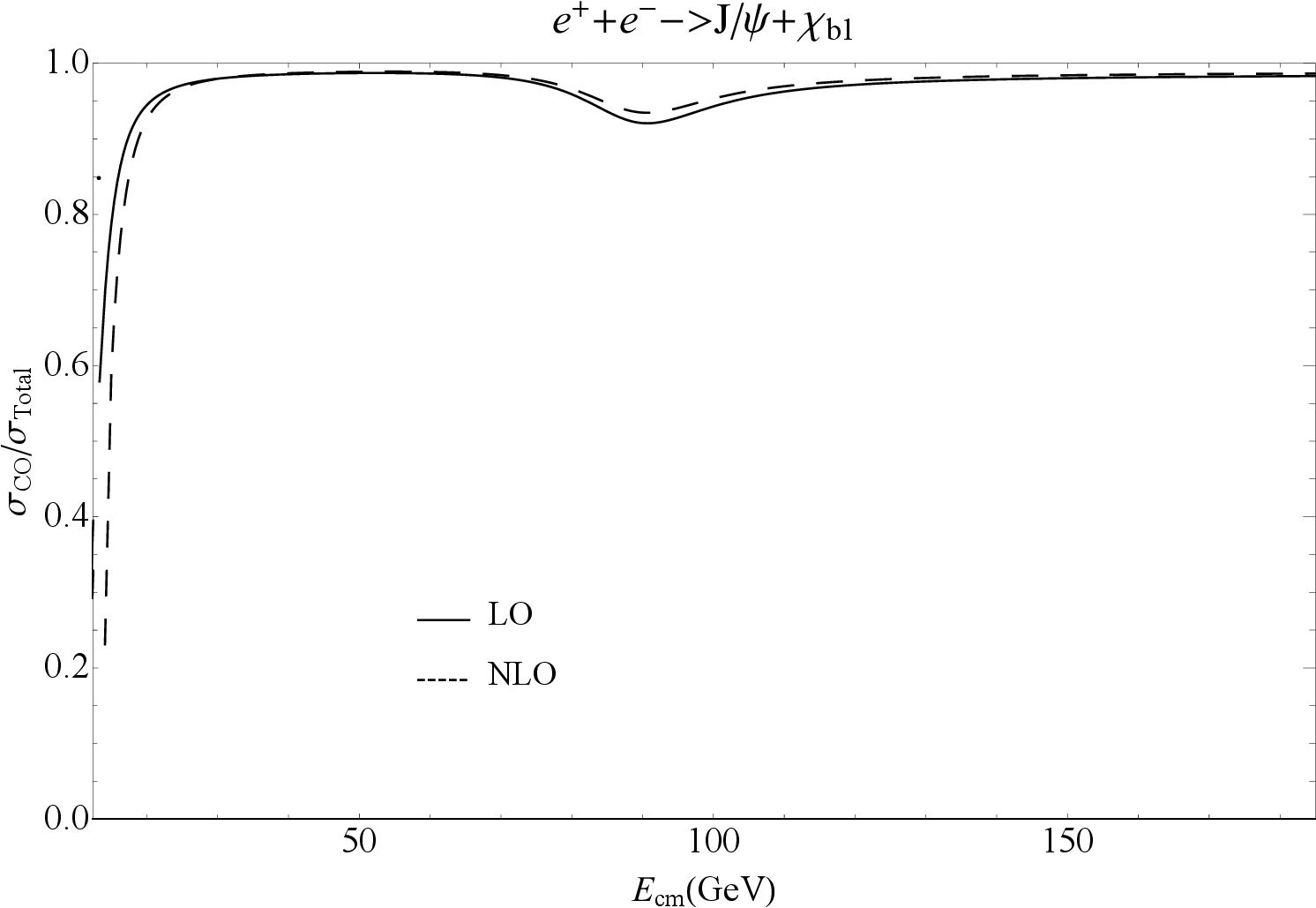}
				\includegraphics[width=0.333\textwidth]{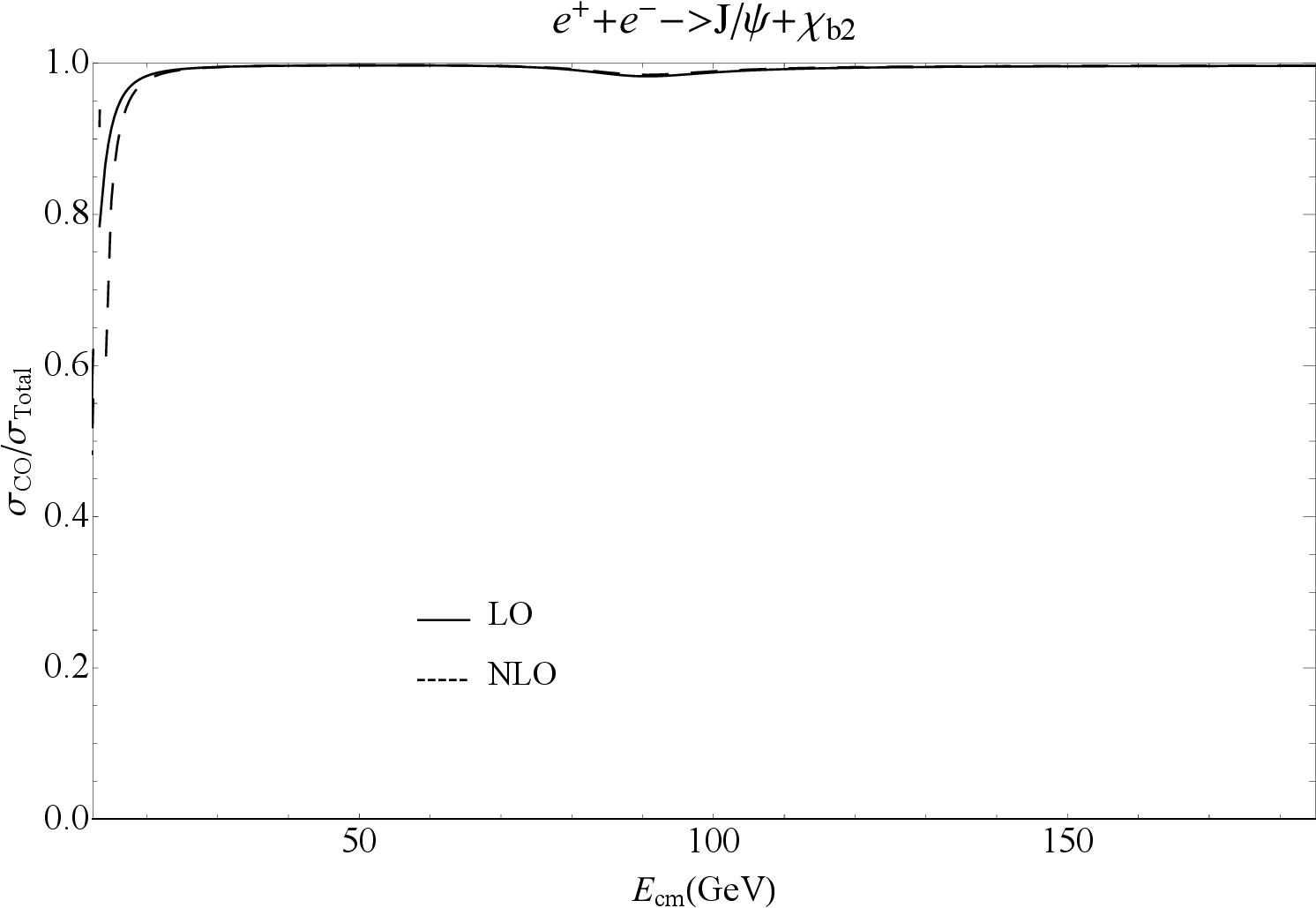}
				\includegraphics[width=0.333\textwidth]{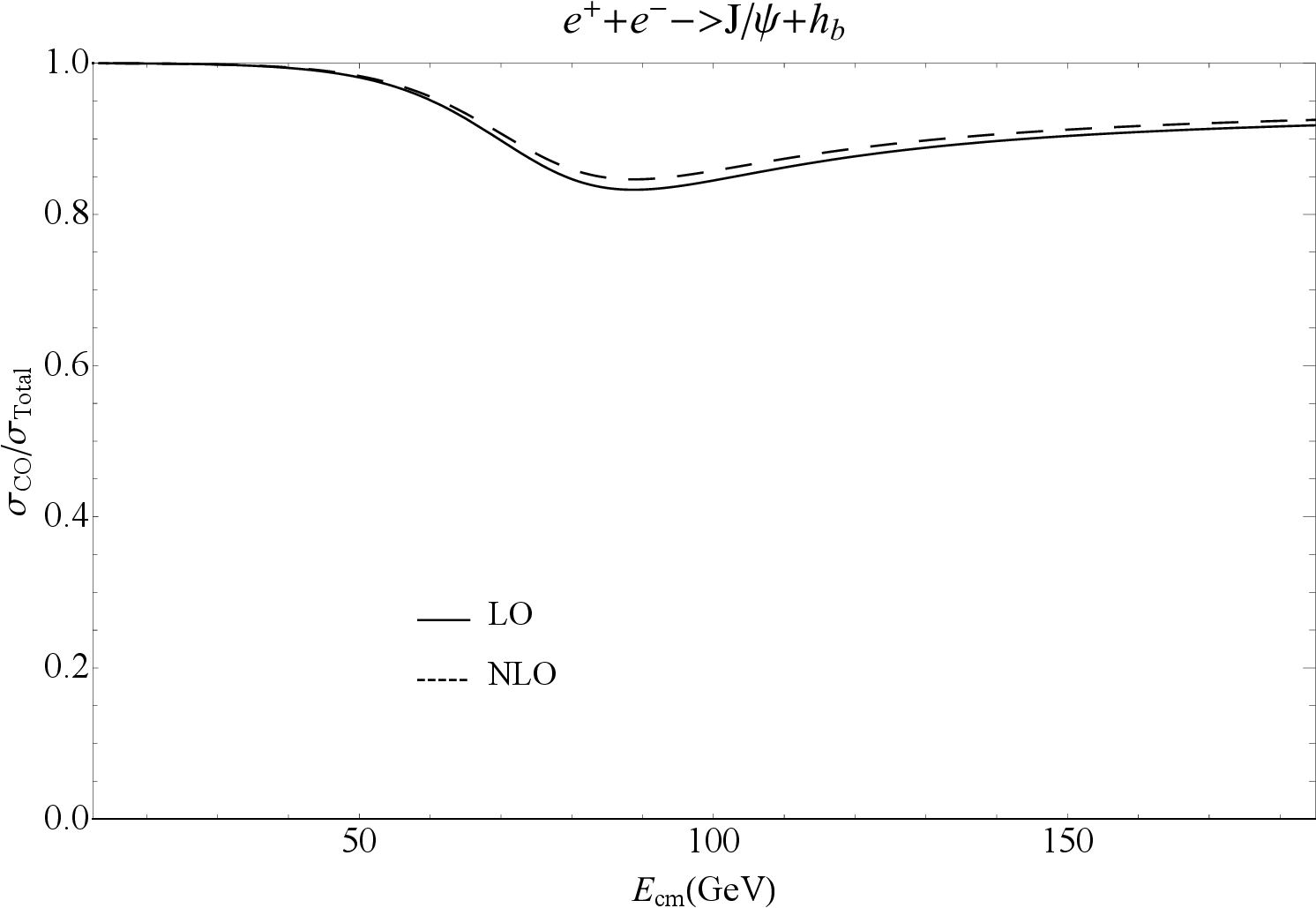}
			\end{tabular}
			
			\begin{tabular}{c c c }
				
				\includegraphics[width=0.333\textwidth]{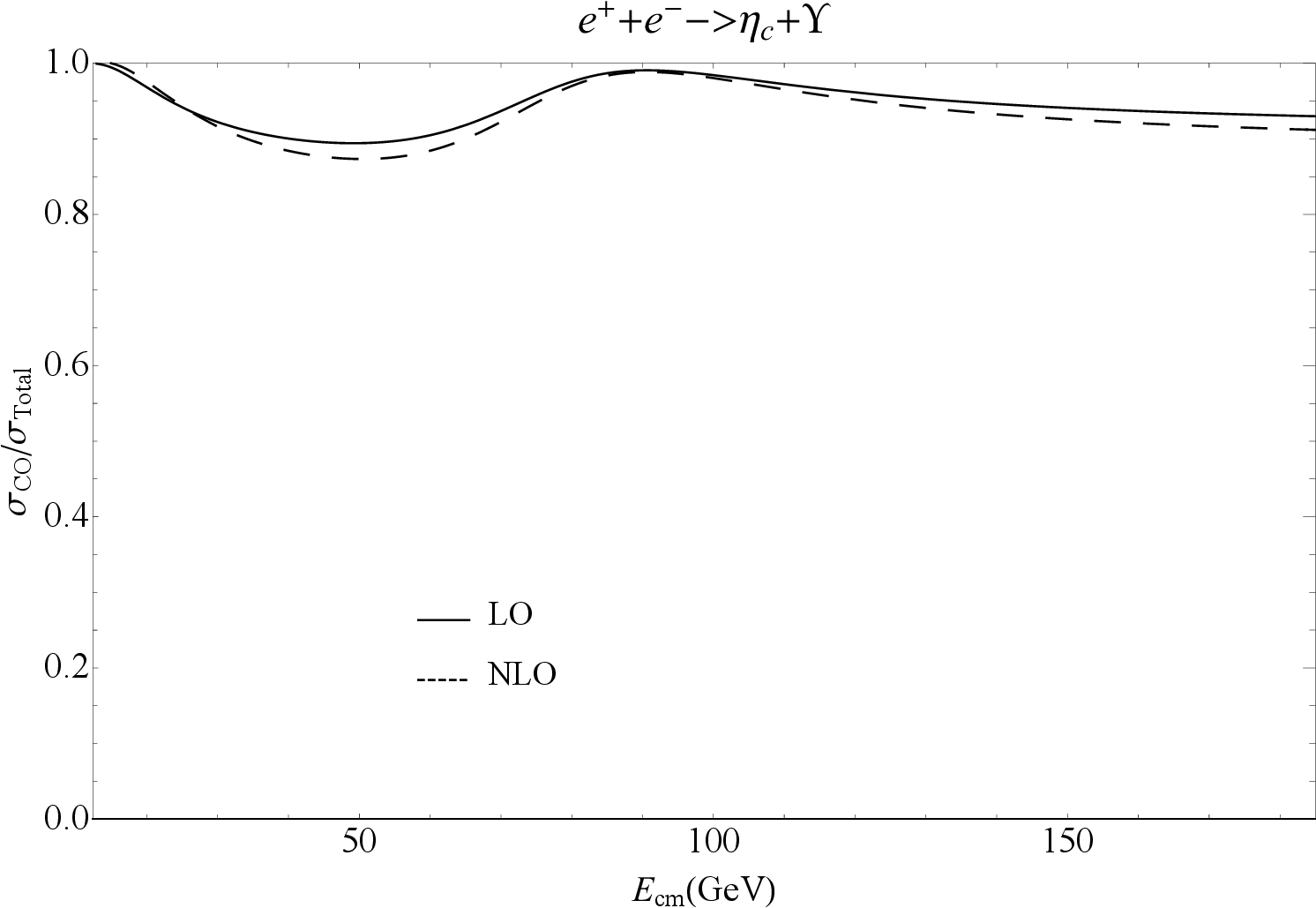}
				\includegraphics[width=0.333\textwidth]{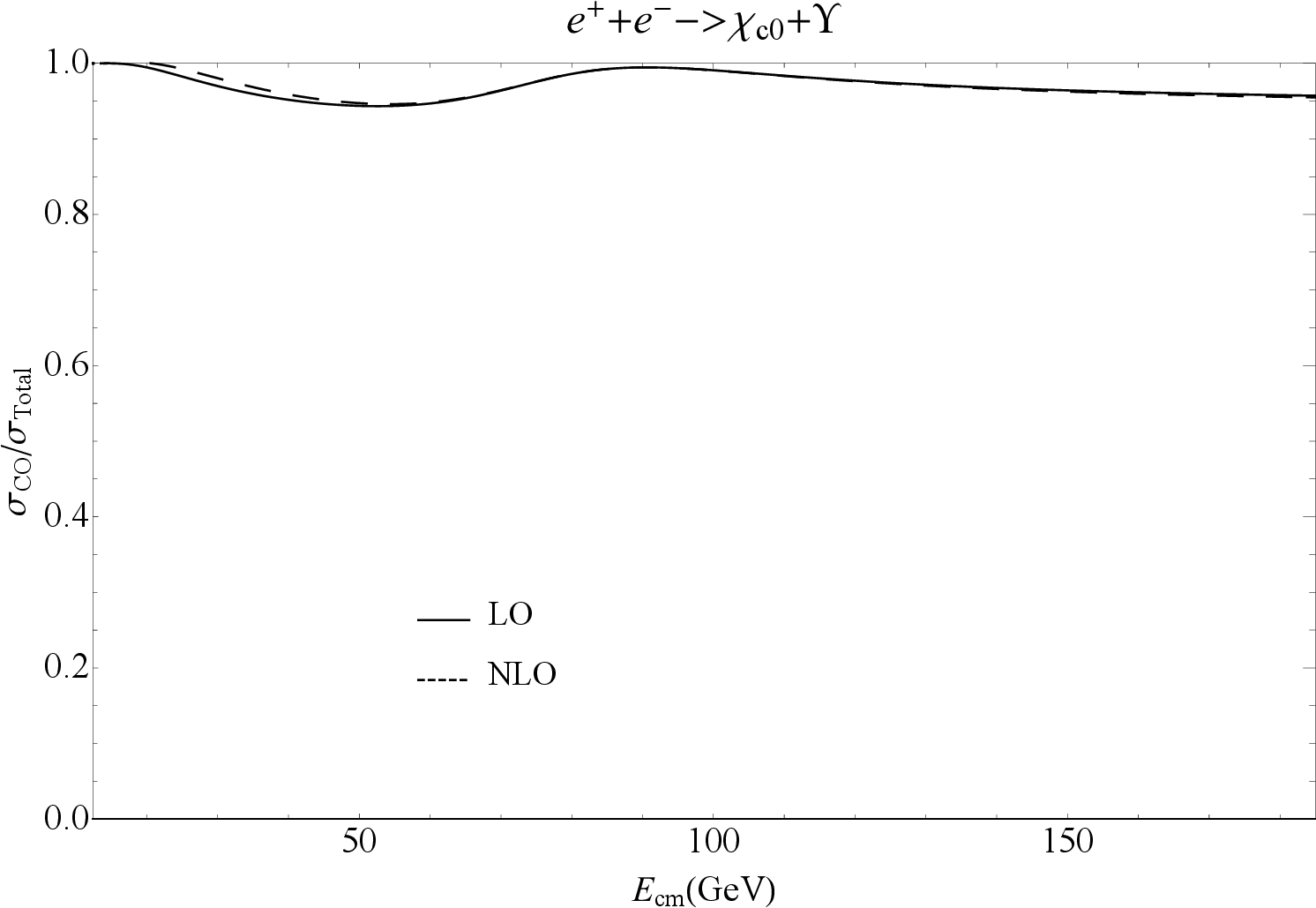}
				\includegraphics[width=0.333\textwidth]{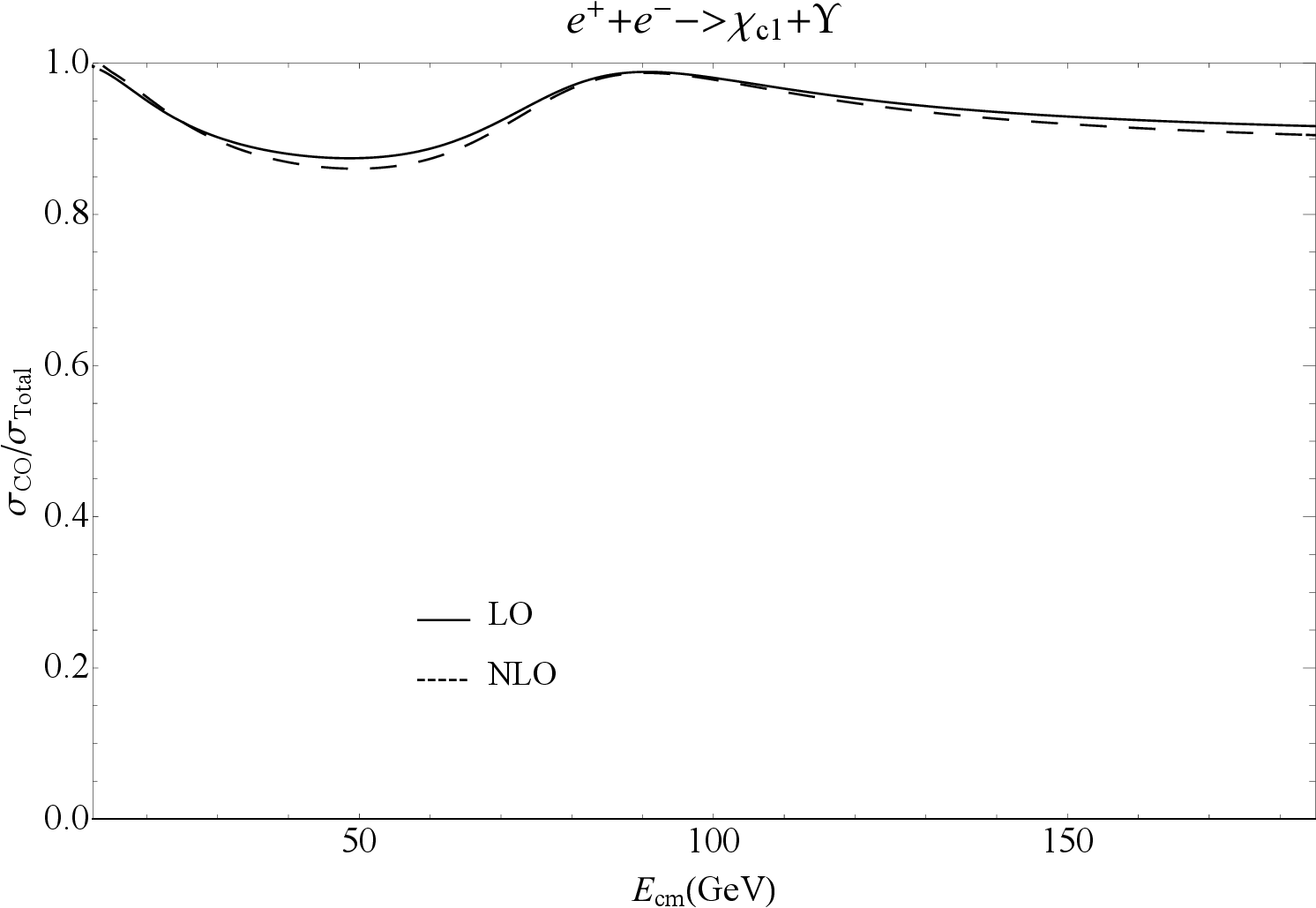}
			\end{tabular}
			\begin{tabular}{c c c }
				\includegraphics[width=0.333\textwidth]{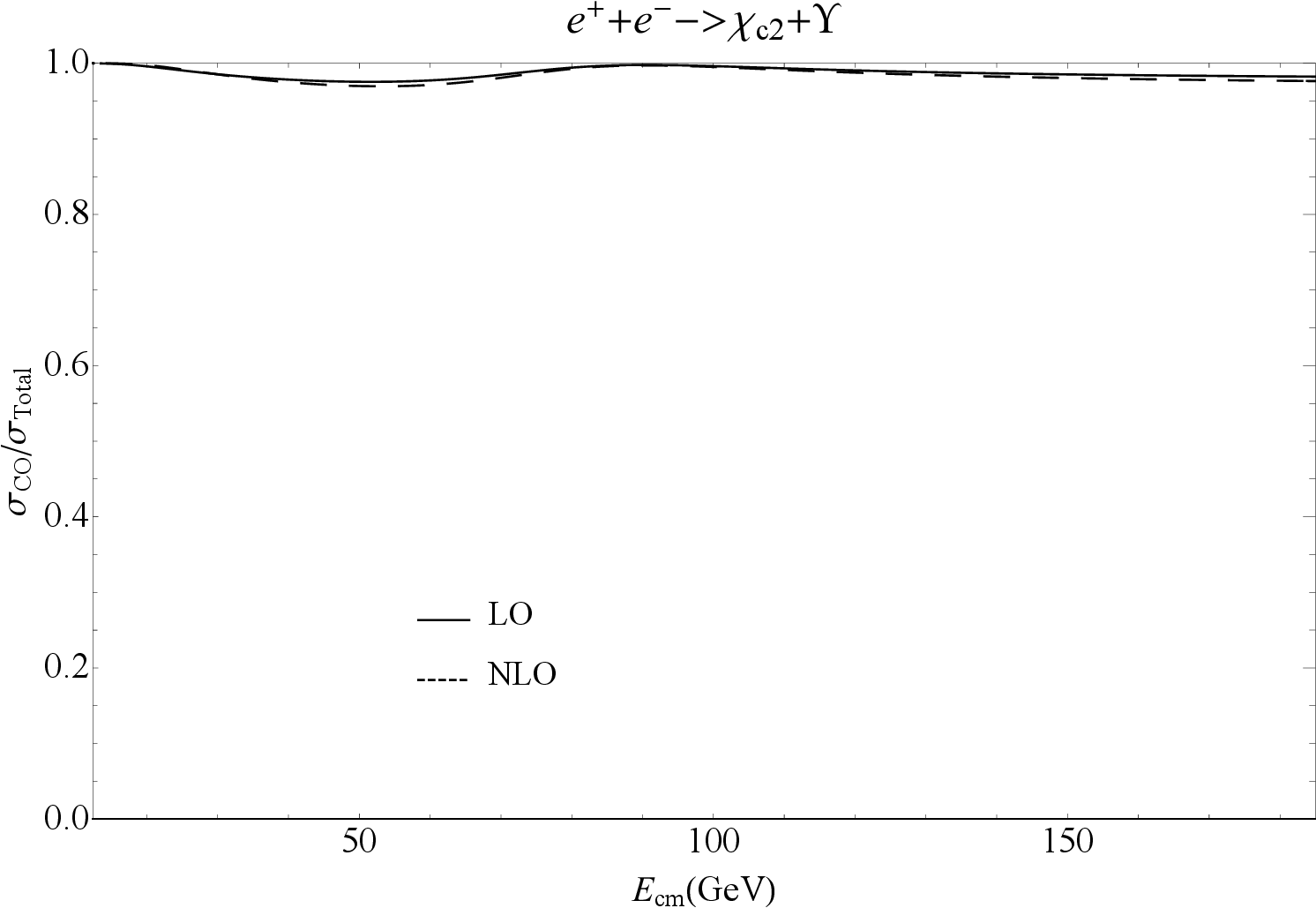}
				\includegraphics[width=0.333\textwidth]{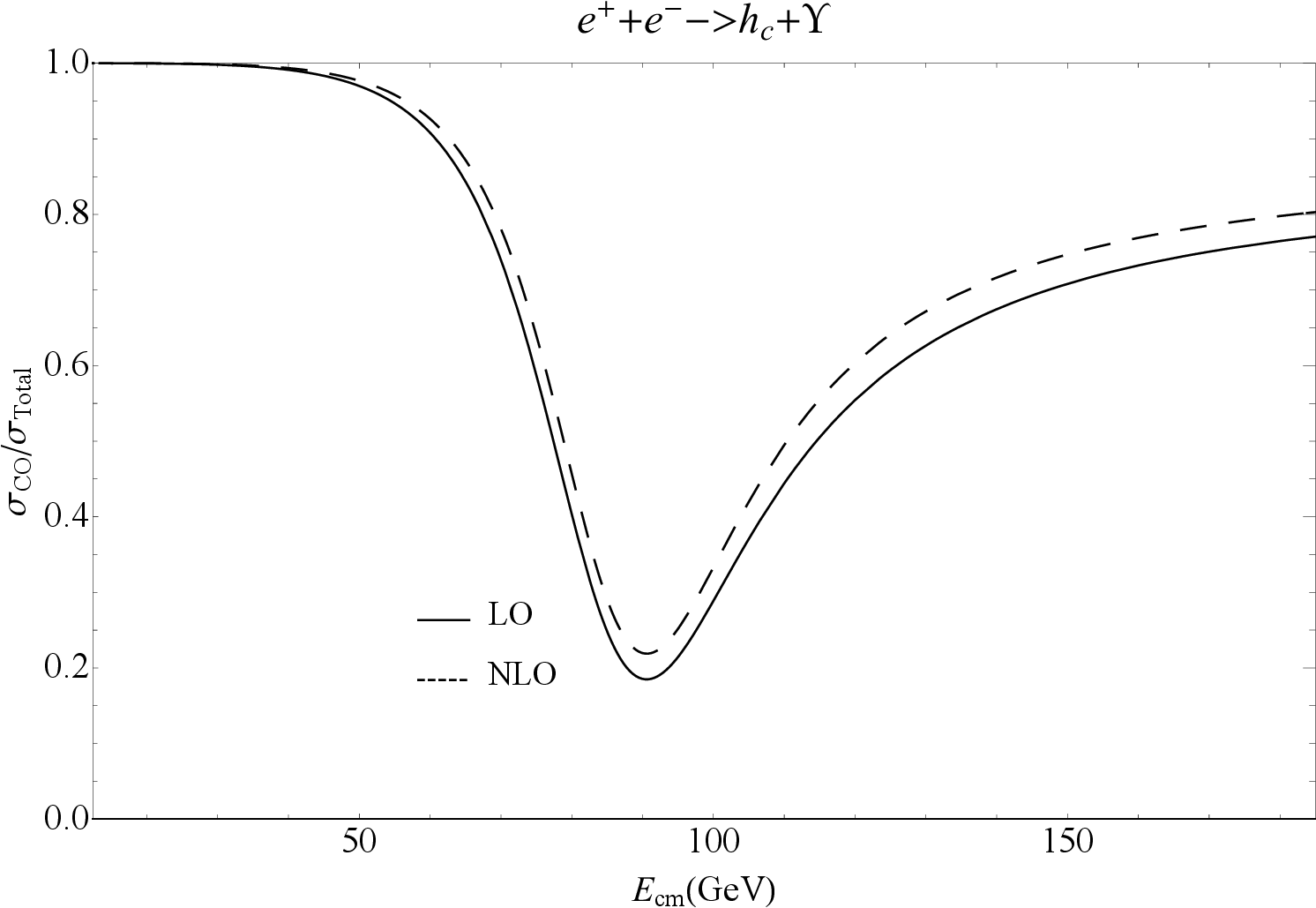}
				
			\end{tabular}	
			\caption{ $\sigma_{CO}/\sigma_{Total} $ as a function of c.m. energy $E_{cm}$. }
			\label{z0bbco}
		\end{figure*}
	\end{widetext}

	The cross sections versus the centre-of-mass (c.m.) energy are shown in Fig.~\ref{z0cc}. In the low energy region, the $\gamma$-propagated processes are dominant, whereas the $Z^0$-propagated processes dominate around the Z-factory energy region, due to the  $Z^0$ resonance effects. The ratios of the CO cross section to the total cross section at LO and at NLO$(v^2)$ as a function of the c.m. energy are shown in Fig.~\ref{z0bbco}. Except for the production of $J/\psi+\Upsilon, J/\psi+\eta_b$, and $ h_c+\Upsilon$, the CO proportion is surprisingly large (typically $0.8\sim1$) and nearly a constant from the threshold to $2M_Z$, which is different from the cases of double charmonium/bottomonium production\cite{Wang:2025sbx}, where the CO proportion gradually increases with $E_{cm}$. This is due to the lack of a non-fragmentation mechanism. If we disregard the  non-fragmentation diagrams, i.e., the  (a), (b), (e), and (f) diagrams in Fig.~1 of Ref.~\cite{Wang:2025sbx}, the CO proportion versus the c.m. energy is consistent with the present study. 
	It can be explained from the perspectives of amplitude and cross section. For double charmonium/bottomonium production, e.g., the $J/\psi+\eta_c$ process, asymptotically the total cross section $\sigma_{tot}=\sigma_{CS}+\sigma_{CO}$ falls off with increase of the energy as $\sigma_{tot}\sim\frac{1}{s^2}(A+B+\mathcal{O}(\frac{1}{s^2}))$, where $A\sim B \sim\mathcal{O}(1)$. The main contributions A and B originate from the  gluon fragmentation CO process (c, d diagrams in Fig.~1 of Ref.~\cite{Wang:2025sbx}) and the photon fragmentation CS process (g, h diagrams  in Fig.~1 of Ref.~\cite{Wang:2025sbx}), respectively. The last $\mathcal{O}(\frac{1}{s^2})$ term\footnote{The amplitudes of the QCD non-fragmentation CS process and QED fragmentation CS process have the same Lorentz structure $\mathcal{A}\sim\epsilon_{\nu}^{\Upsilon}p_{3\rho}p_{4\sigma}\varepsilon^{\mu\nu\rho\sigma}$ and the amplitude ratio between them is $\frac{\mathcal{A}_{QCD}}{\mathcal{A}_{QED}}|_{s\rightarrow\infty}\sim\frac{1}{s}$. Therefore, the CS cross section $\sigma_{CS}\simeq\sigma_{QCD}+\sigma_{QED}$ asymptotically falls off with the increase of the energy as $\sigma_{CS}\sim\frac{1}{s^2}(1+\mathcal{O}(\frac{1}{s^2}))$, where the main contribution originates from the QED fragmentation process.} originates from the QCD non-fragmentation CS process (a,b diagrams in Fig.~1 of Ref.~\cite{Wang:2025sbx}). As a result, the QCD non-fragmentation process dominates in the low energy region, while  with increasing c.m. energy, it becomes negligible, then the comparable gluon fragmentation and photon fragmentation processes are dominant and the CO ratio $\frac{\sigma_{CO}}{\sigma_{total}}\sim\frac{A}{A+B}$ exhibits a constant value (see Fig.~4 of Ref.~\cite{Wang:2025sbx}). However, for charmonium+bottomonium production, e.g., the $\eta_c+\Upsilon$ production, the amplitude ratio between the CO and CS processes, which comes from the coupling vertices and color factors, is $\frac{\mathcal{A}_{CO}}{\mathcal{A}_{CS}}\sim\frac{9\alpha_s\sqrt{C_AC_F}}{2\sqrt{2}N_c\alpha}$. Therefore, the total cross section $\sigma_{tot}|_{s\rightarrow\infty}\sim \frac{1}{s^2}(A+B)$, where A and B originate from the CO and CS channels, respectively. Further accounting for numerous CO channels and the suppressed CO LDMEs, the cross section ratio $\frac{\sigma_{CO}}{\sigma_{tot}}$  behaves approximately as a constant $\frac{A}{A+B}$. 
The CO ratios $\frac{\sigma_{CO}}{\sigma_{total}}$ at the high energy limit ($\sqrt{s}\gg m_Z$) are presented in Table~\ref{tab:co_ratio_he}.

\begin{table}[htbp]
  \centering
  \caption{The ratios between the color octet and the total cross sections $\frac{\sigma_{CO}}{\sigma_{total}}$ for various charmonium+bottomonium production processes in the high energy limit ($\sqrt{s}\gg m_Z$). The values outside and inside the brackets are the ratios for leading order and next-to-leading order in $v^2$.}
  \label{tab:co_ratio_he}
  \begin{tabular}{lc} 
    \hline
    Process & $\frac{\sigma_{CO}}{\sigma_{total}}$ \\
    \hline
    $J/\psi+\Upsilon$   & 0.00(0.00) \\
    $J/\psi+\eta_b$     & 0.19(0.20) \\
    $J/\psi+\chi_{b0}$  & 0.99(0.99) \\
    $J/\psi+\chi_{b1}$  & 0.99(0.99) \\
    $J/\psi+\chi_{b2}$  & 1.00(1.00) \\
    $J/\psi+h_b$        & 0.94(0.94) \\
    $\eta_c+\Upsilon$   & 0.91(0.89) \\
    $\chi_{c0}+\Upsilon$& 0.95(0.94) \\
    $\chi_{c1}+\Upsilon$& 0.90(0.88) \\
    $\chi_{c2}+\Upsilon$& 0.98(0.97) \\
    $h_c+\Upsilon$      & 0.85(0.87) \\
    \hline
  \end{tabular}
\end{table}

\begin{widetext}
		\begin{figure*}[htbp]
			\begin{tabular}{c c c}
				\includegraphics[width=0.333\textwidth]{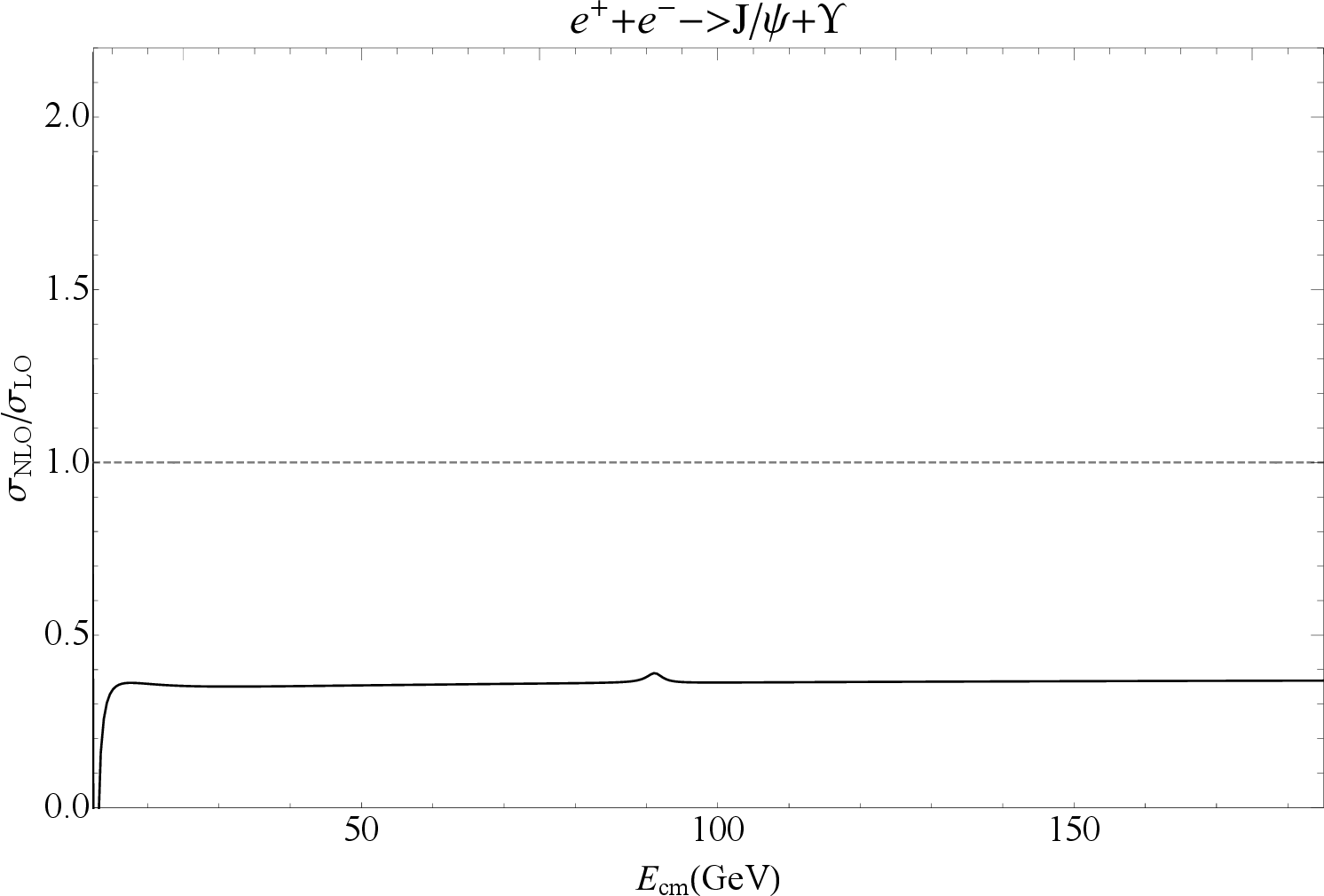}
				\includegraphics[width=0.333\textwidth]{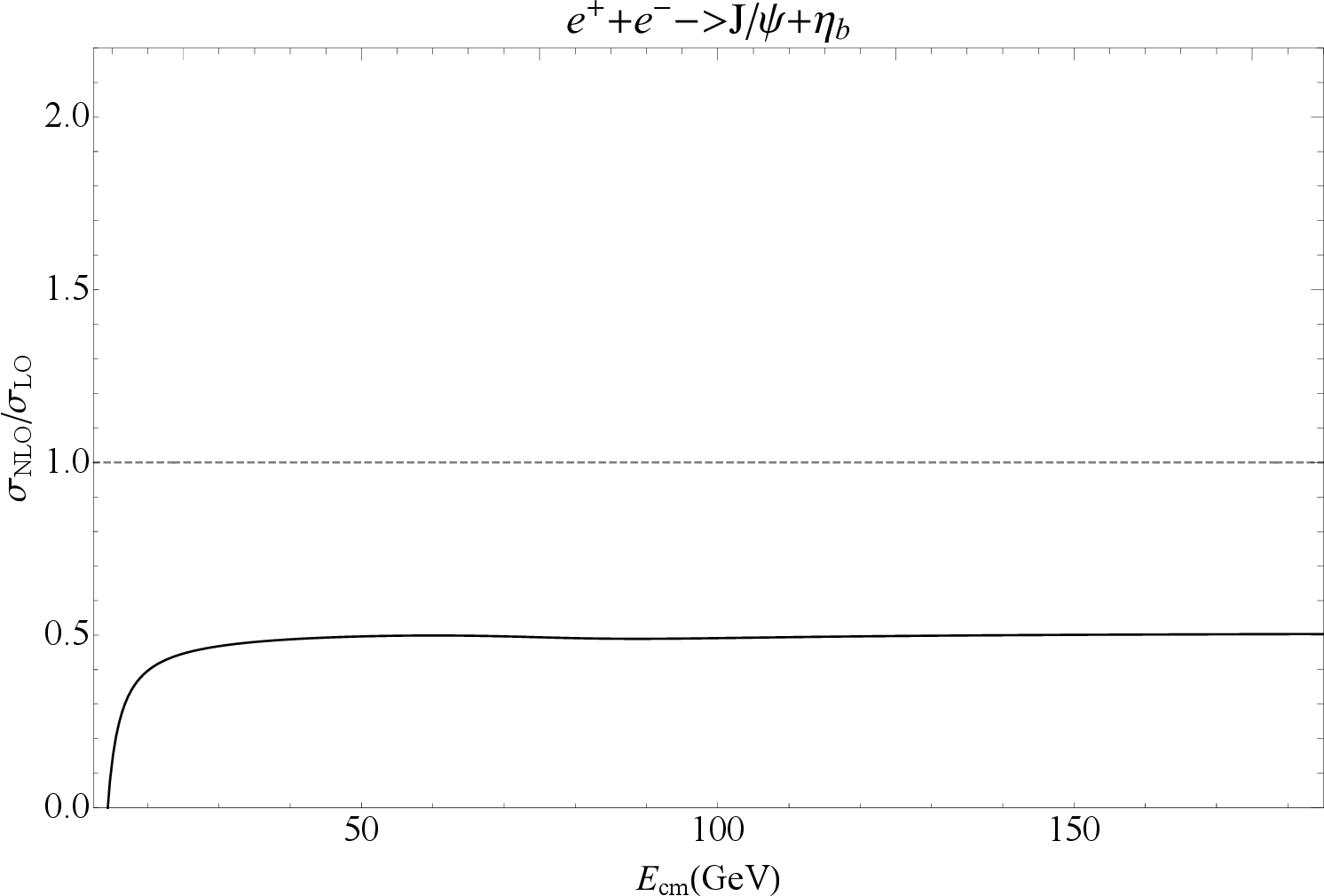}
				\includegraphics[width=0.333\textwidth]{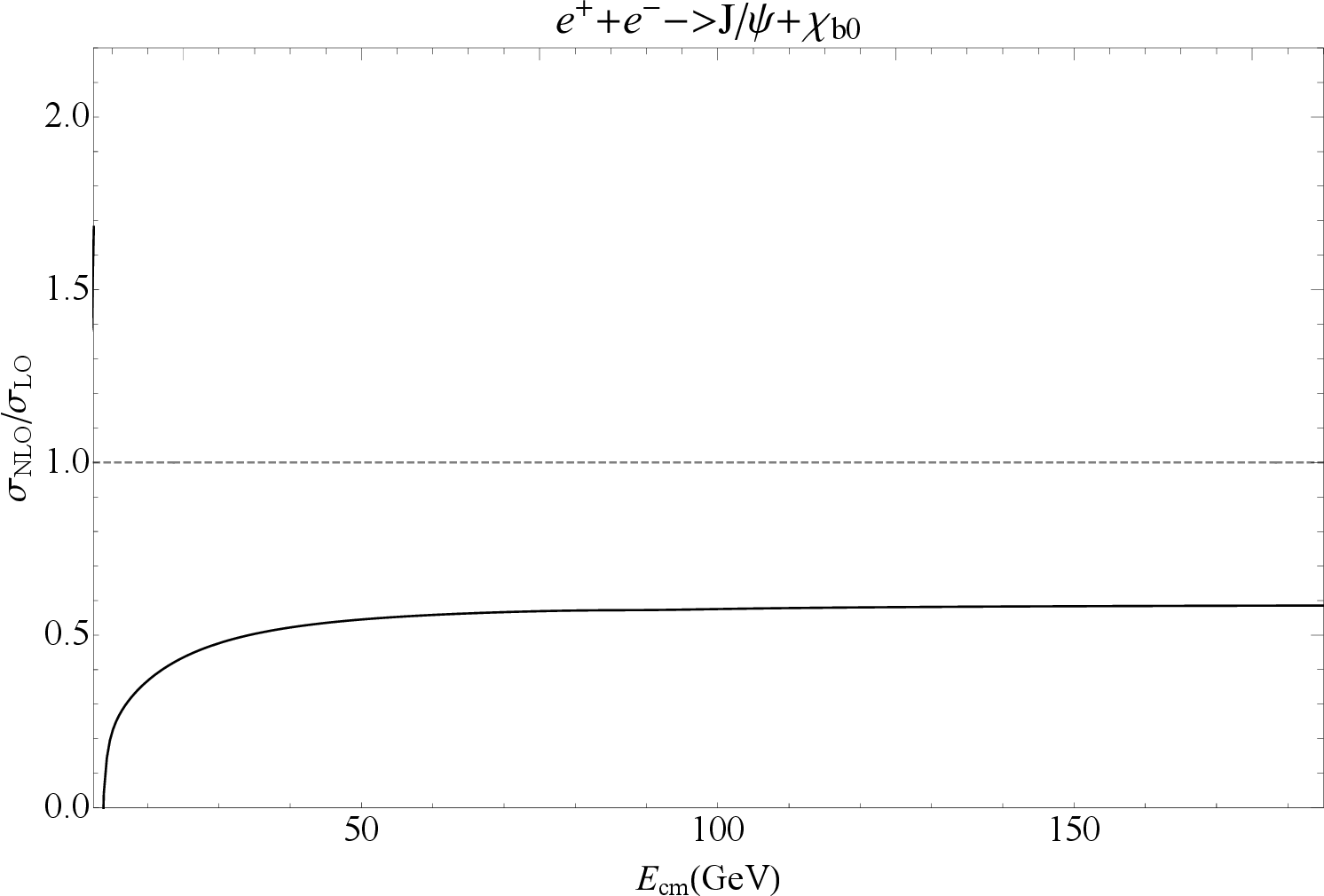}
			\end{tabular}
			\begin{tabular}{c c c}						
				\includegraphics[width=0.333\textwidth]{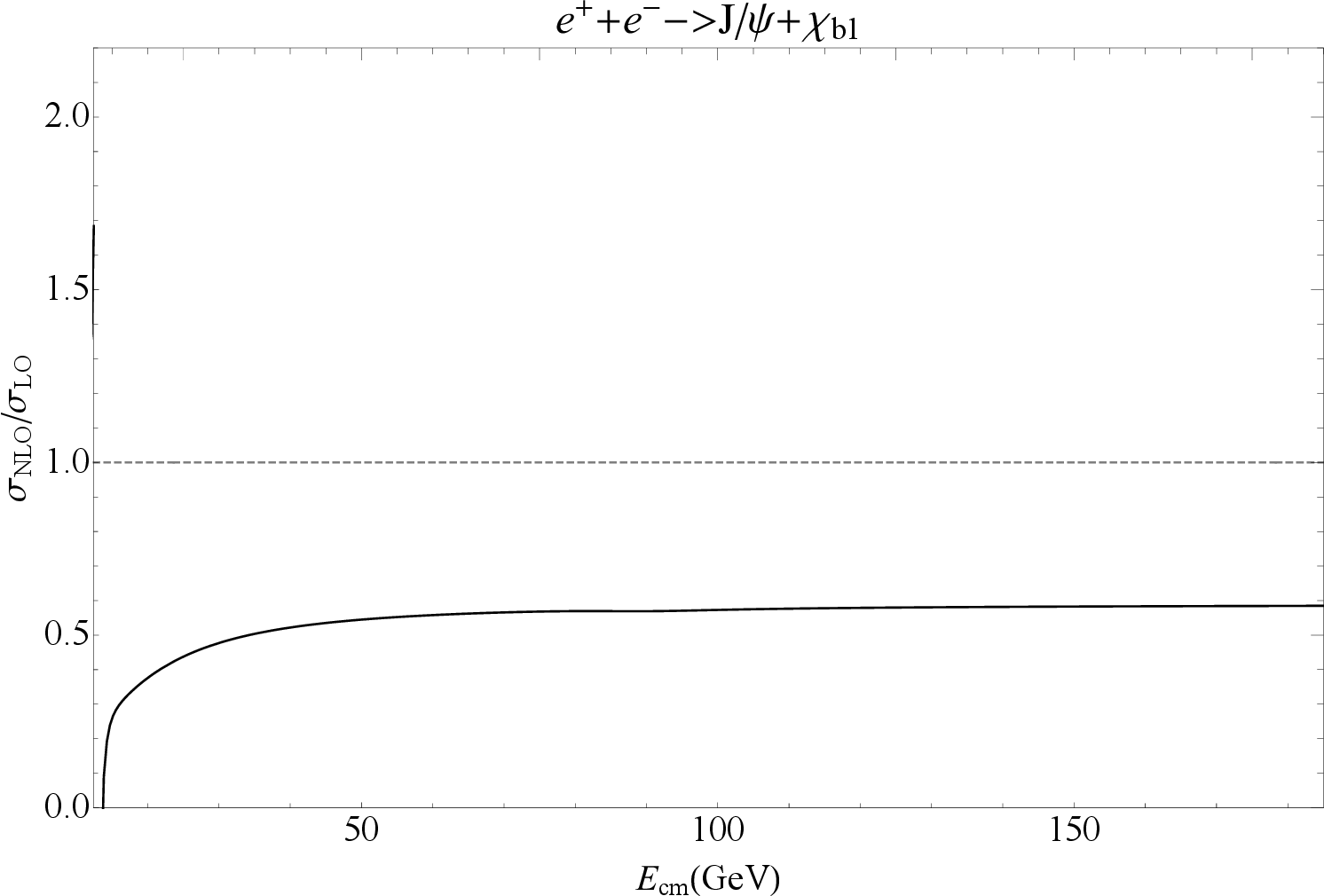}
				\includegraphics[width=0.333\textwidth]{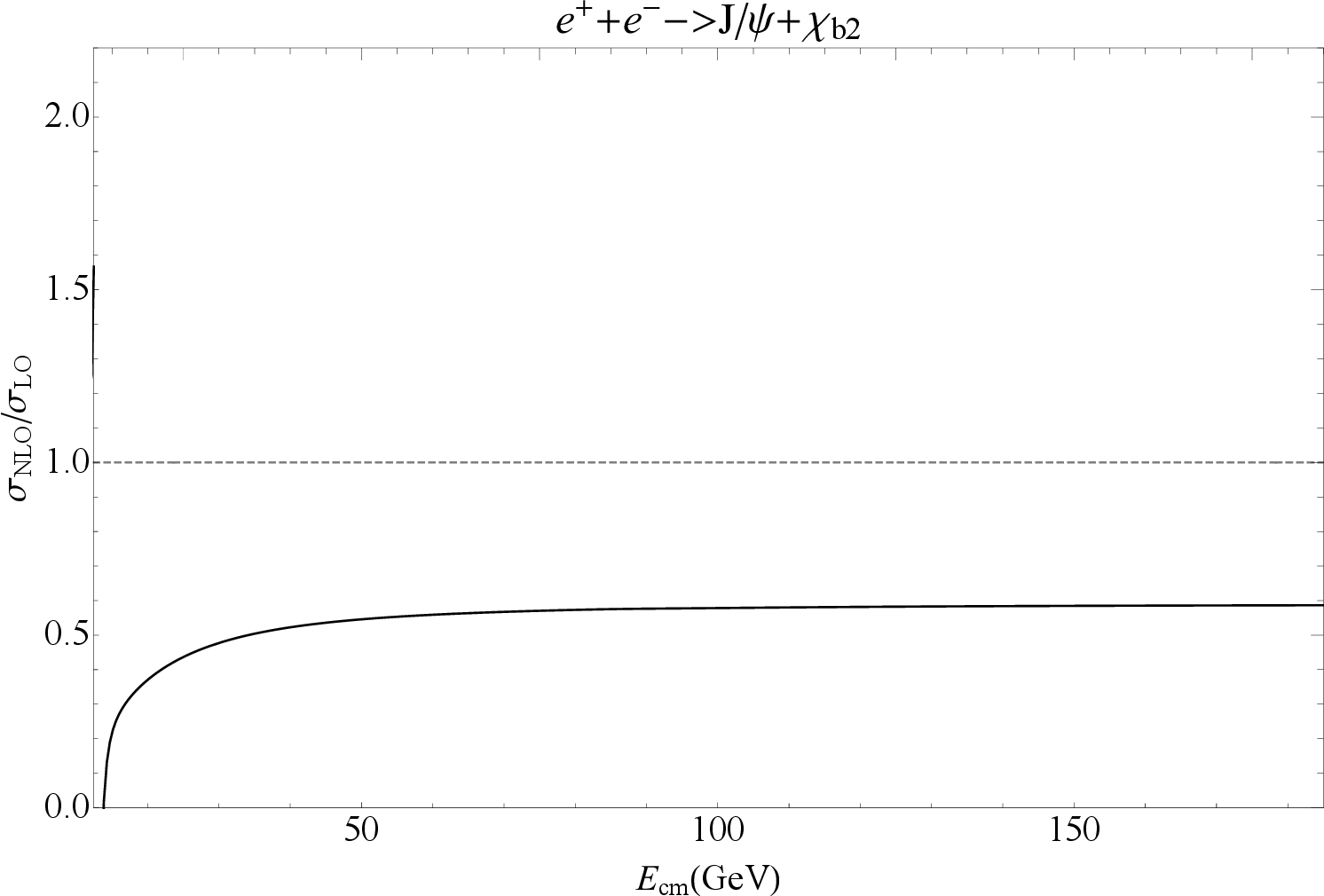}
				\includegraphics[width=0.333\textwidth]{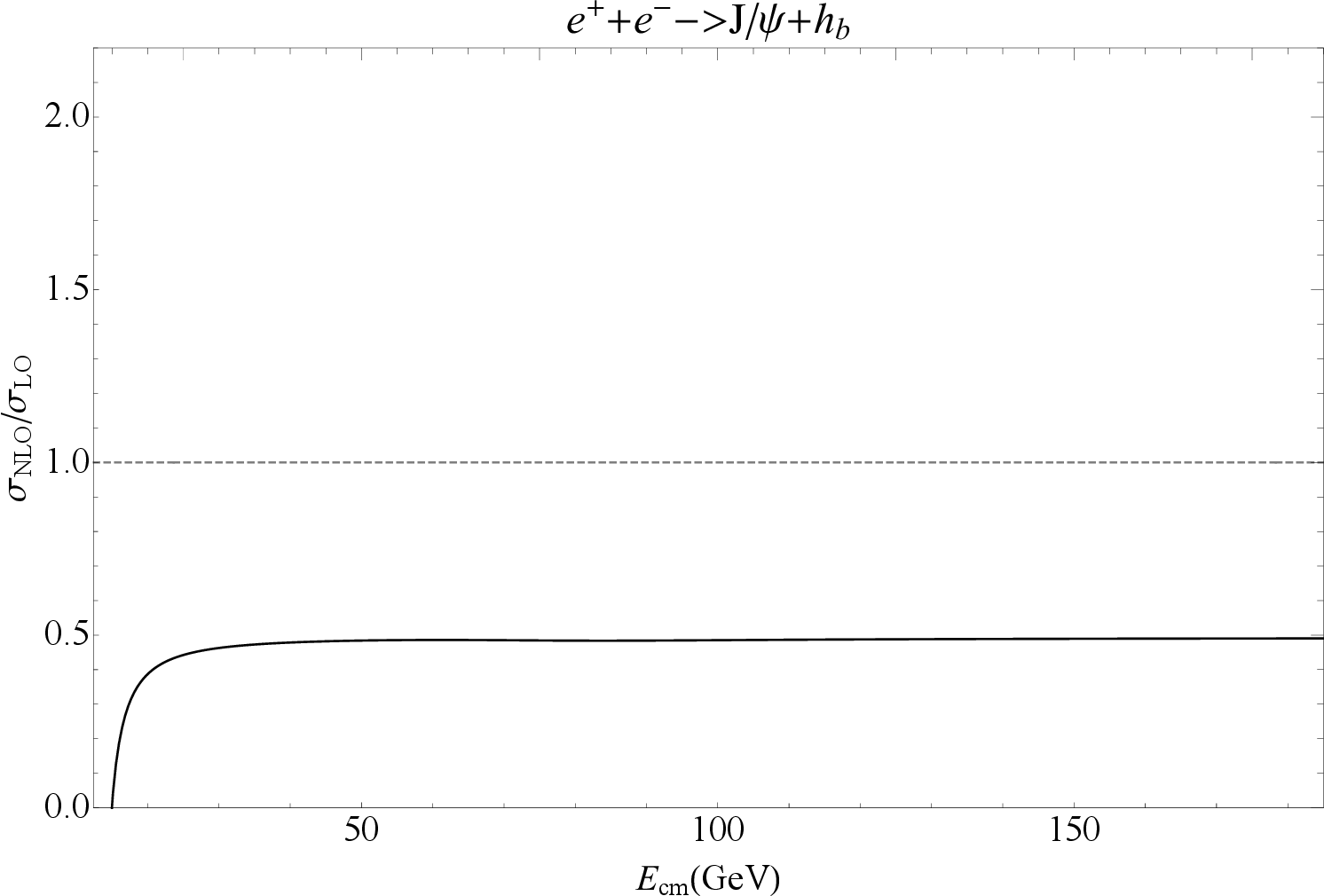}
			\end{tabular}
			
			\begin{tabular}{c c c }
				
				\includegraphics[width=0.333\textwidth]{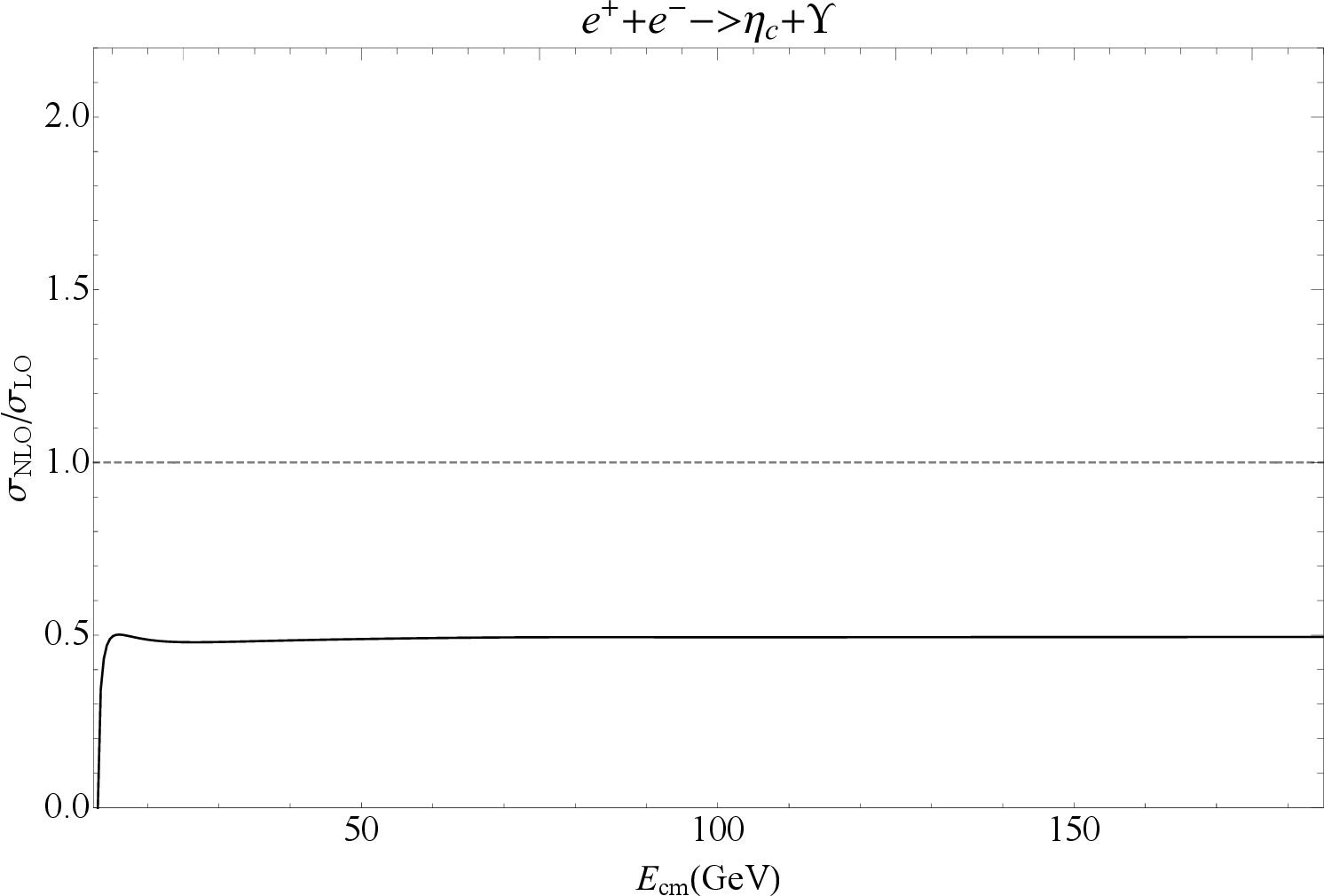}
				\includegraphics[width=0.333\textwidth]{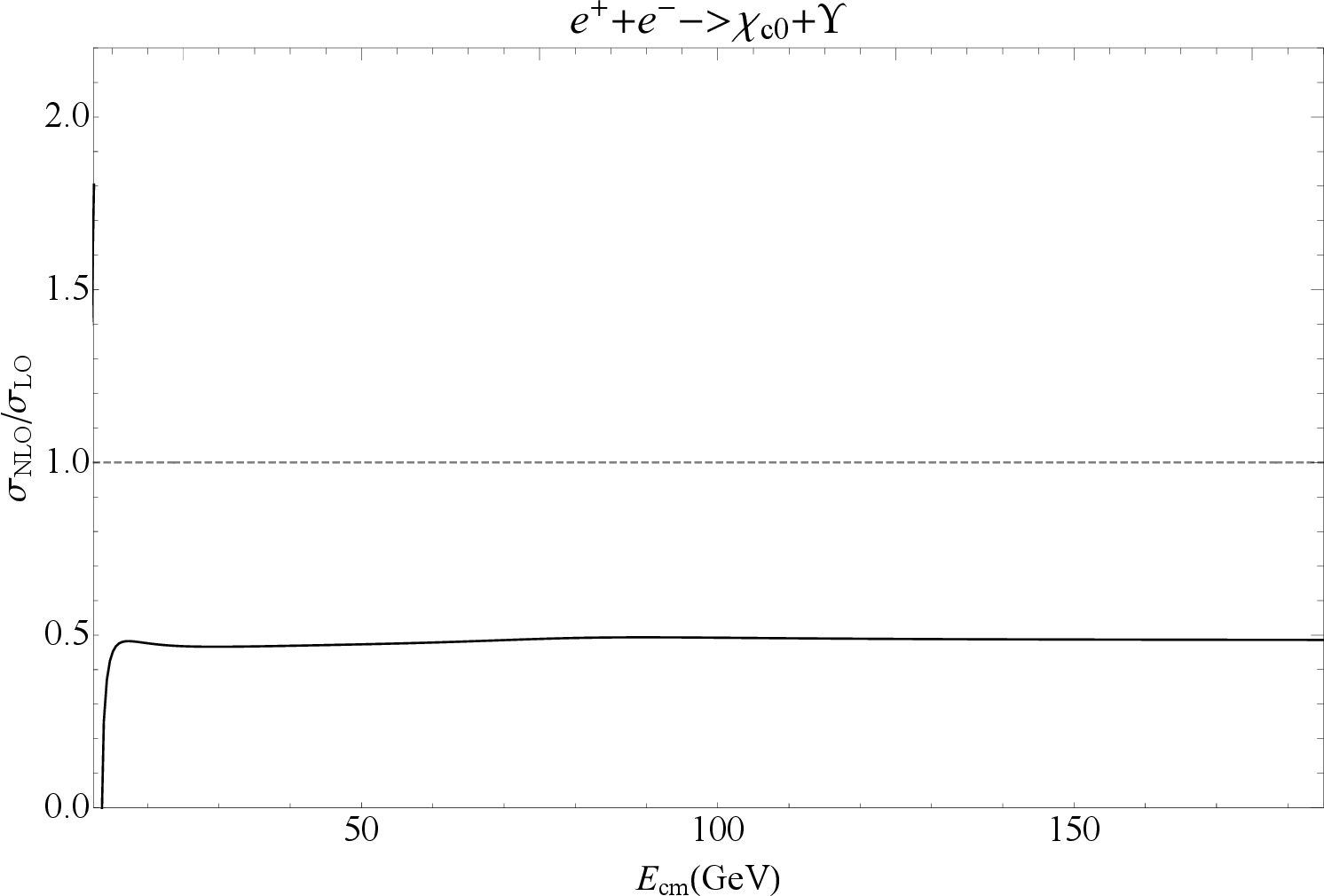}
				\includegraphics[width=0.333\textwidth]{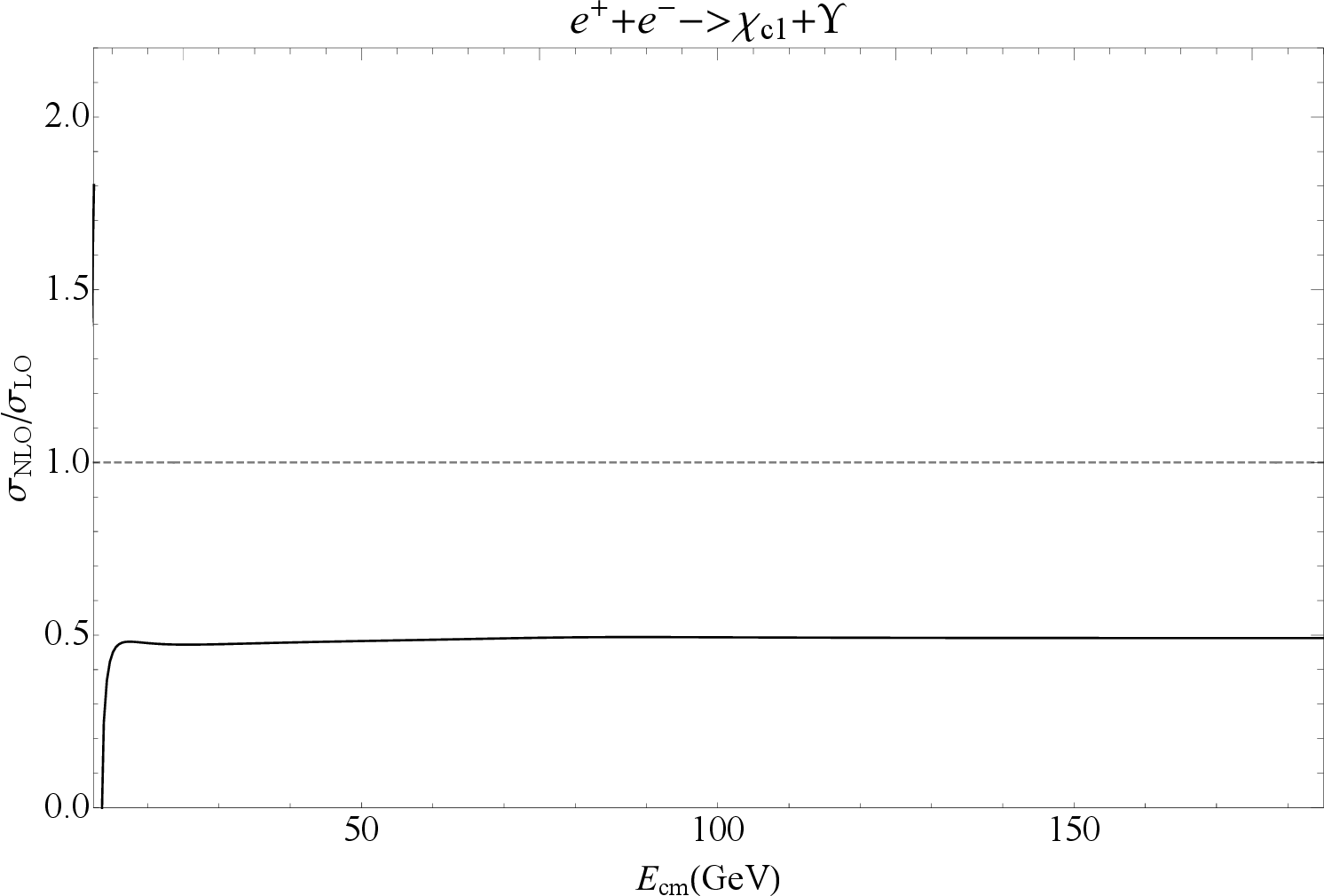}
			\end{tabular}
			\begin{tabular}{c c c }
				\includegraphics[width=0.333\textwidth]{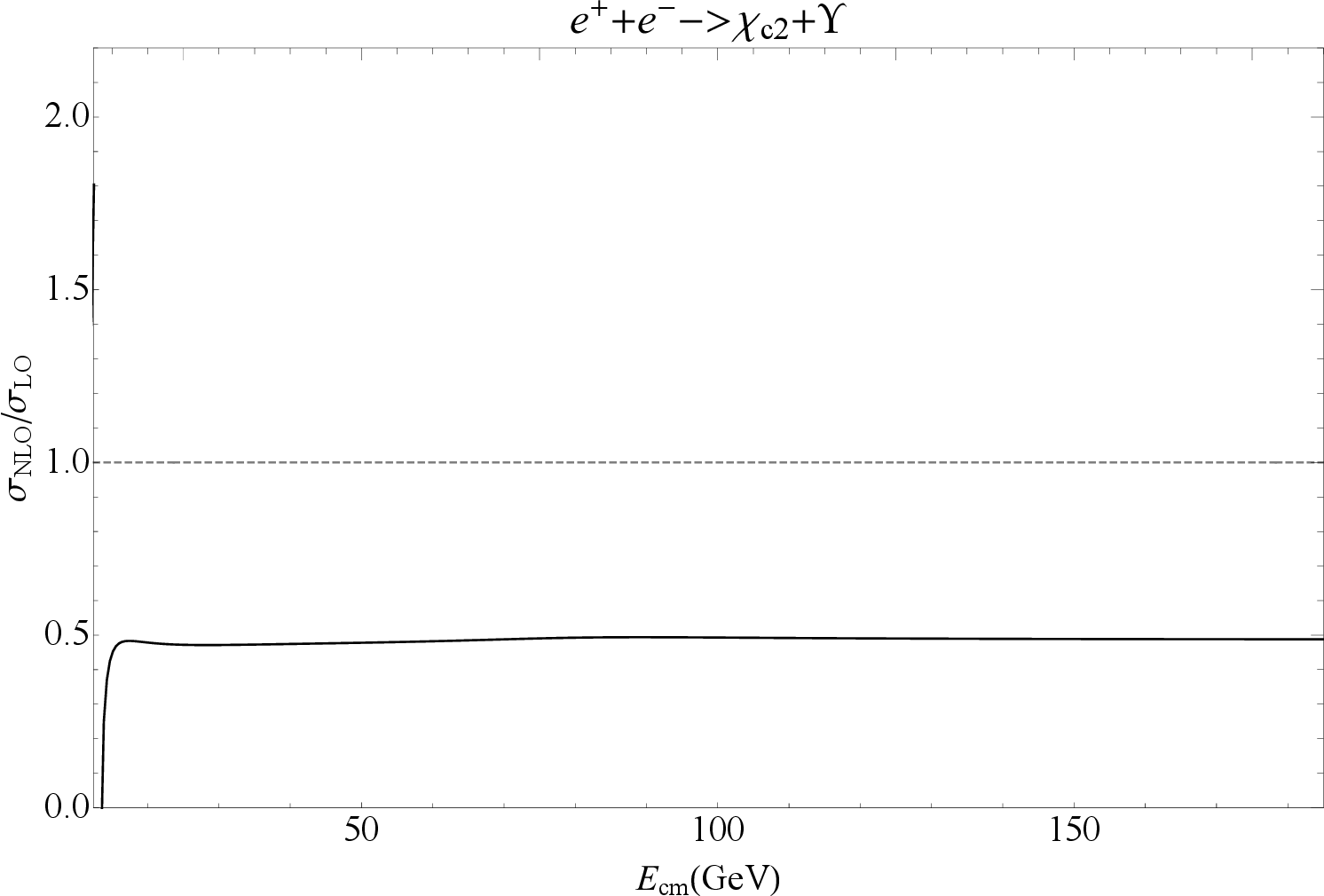}
				\includegraphics[width=0.333\textwidth]{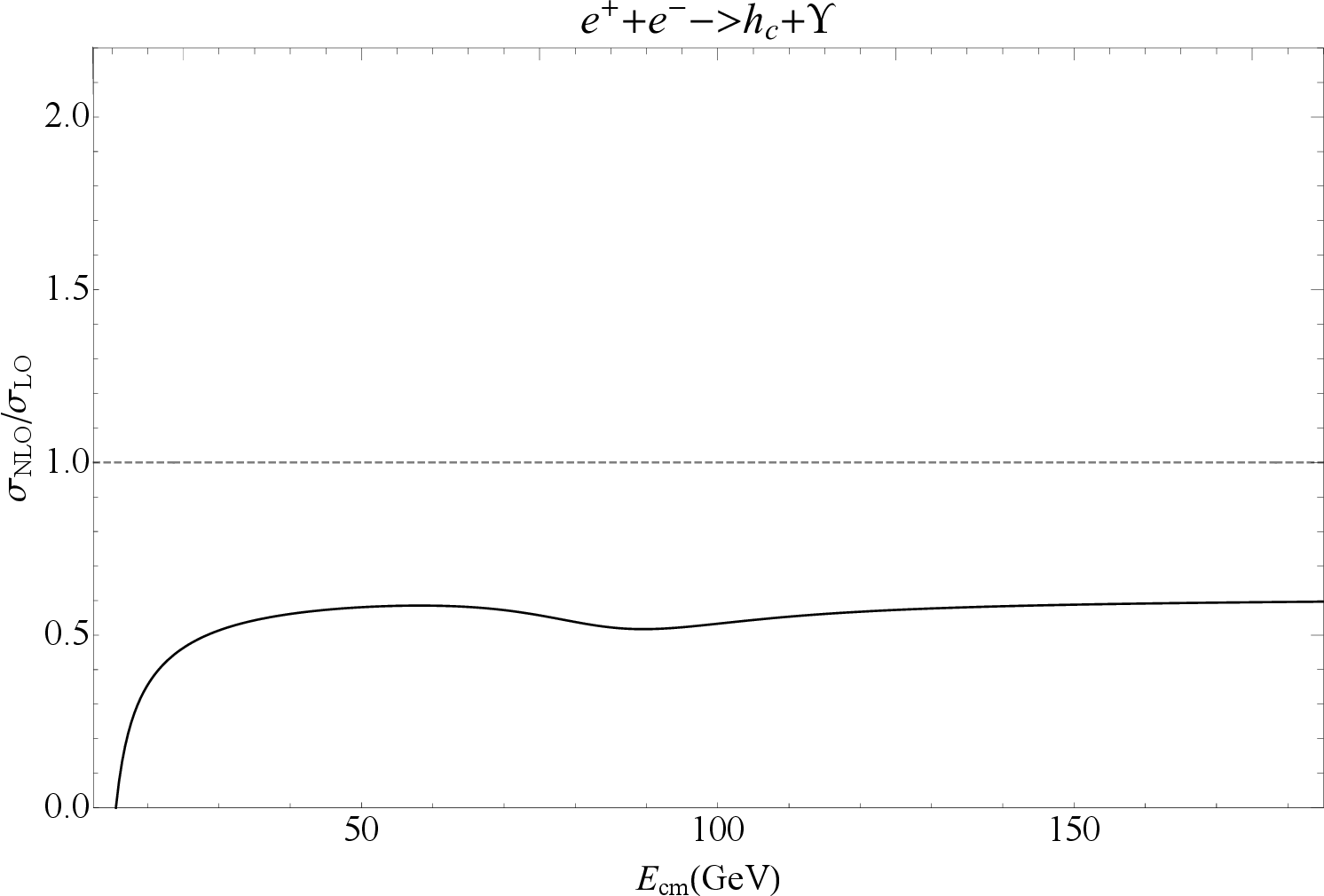}
				\includegraphics[width=0.333\textwidth]{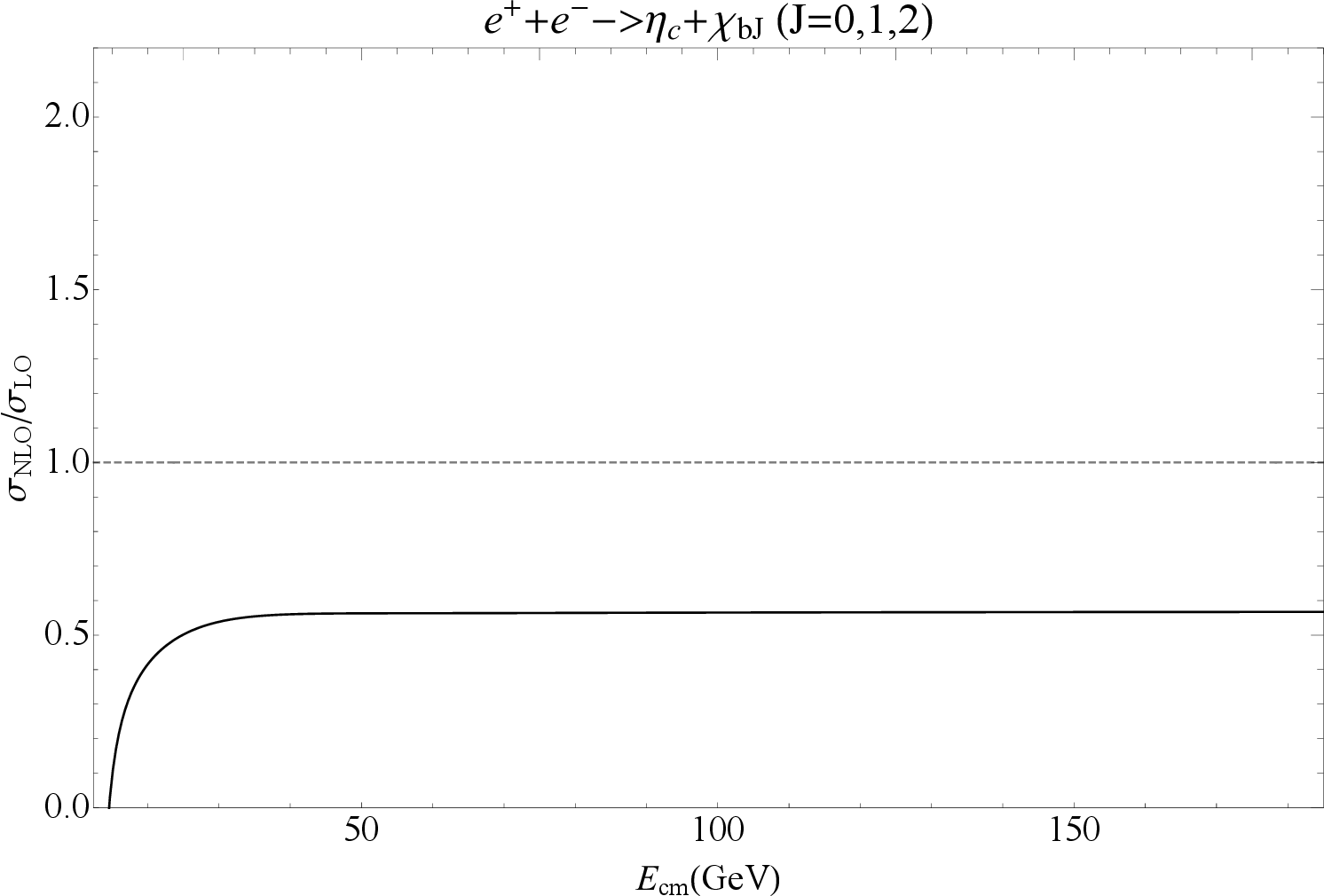}
			\end{tabular}
			\begin{tabular}{c c c }		
				\includegraphics[width=0.333\textwidth]{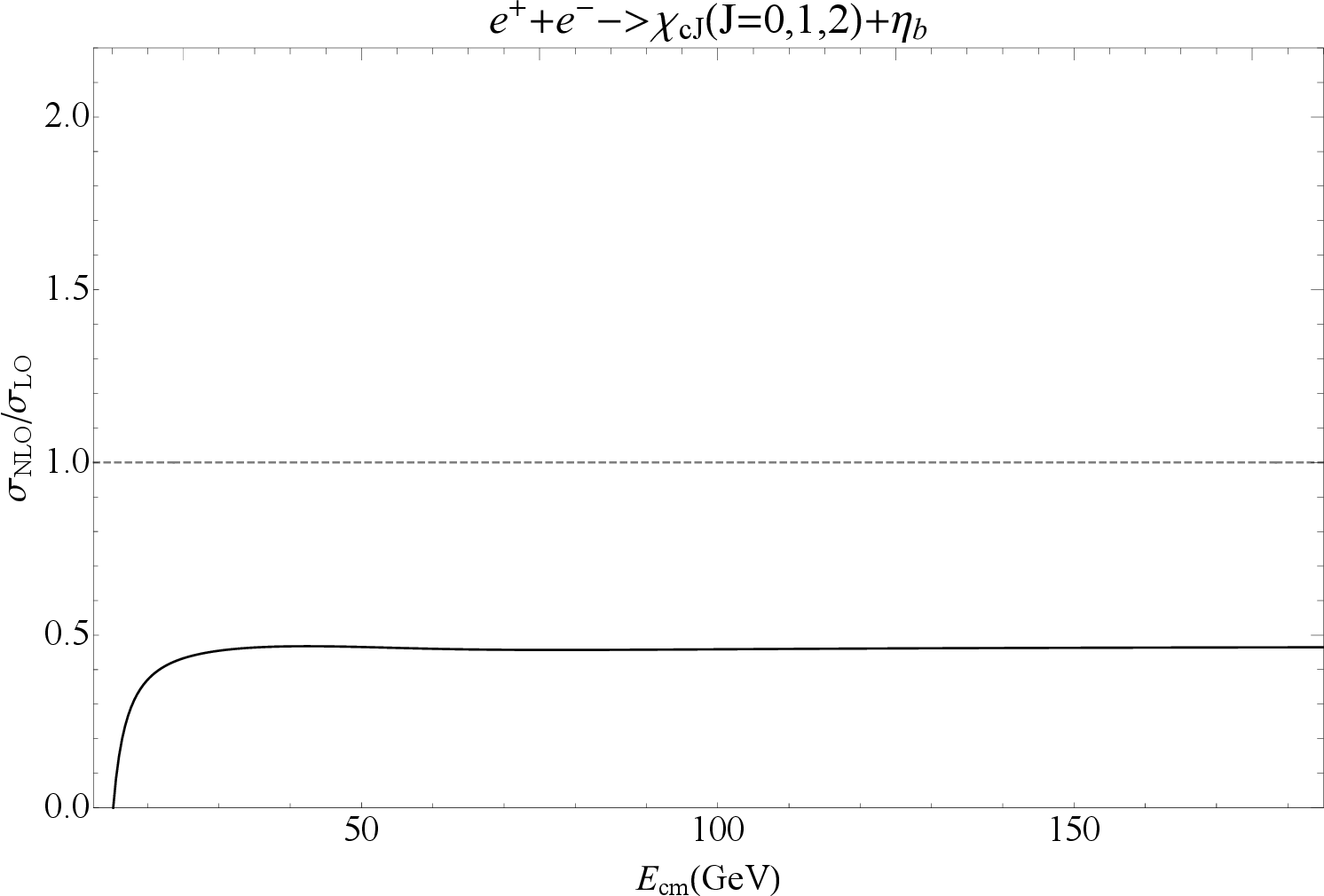}
				\includegraphics[width=0.333\textwidth]{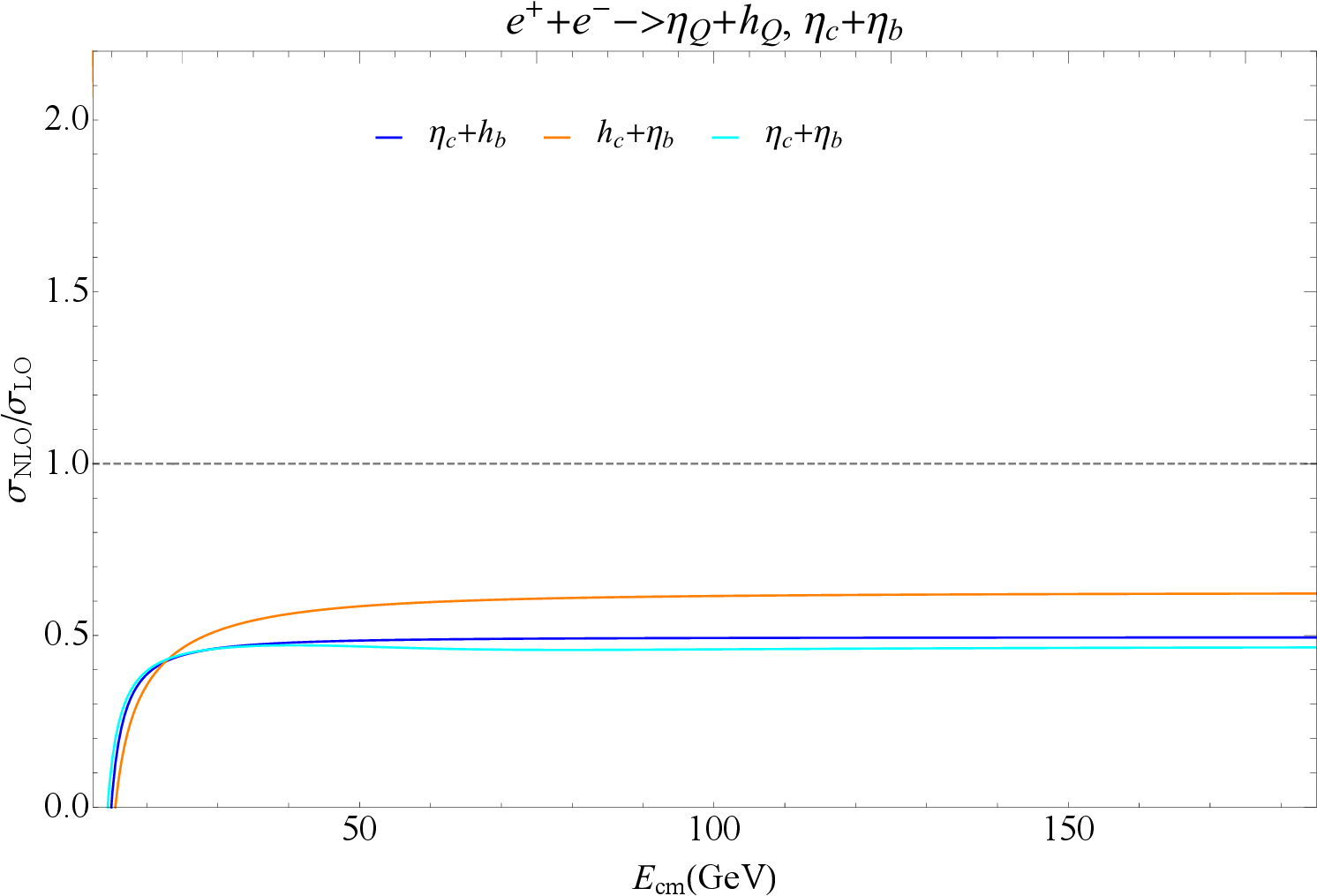}
			\end{tabular}
		 
			\caption{(Color online) The K factors ($\sigma_{\text{NLO}(v^2)}/\sigma_{\text{LO}}$) as functions of c.m. energy $E_{cm}$. }
			\label{z0cck}
		\end{figure*}
	\end{widetext}

\begin{widetext}
		\begin{figure*}[htbp]
		\begin{tabular}{c c c}
			\includegraphics[width=0.333\textwidth]{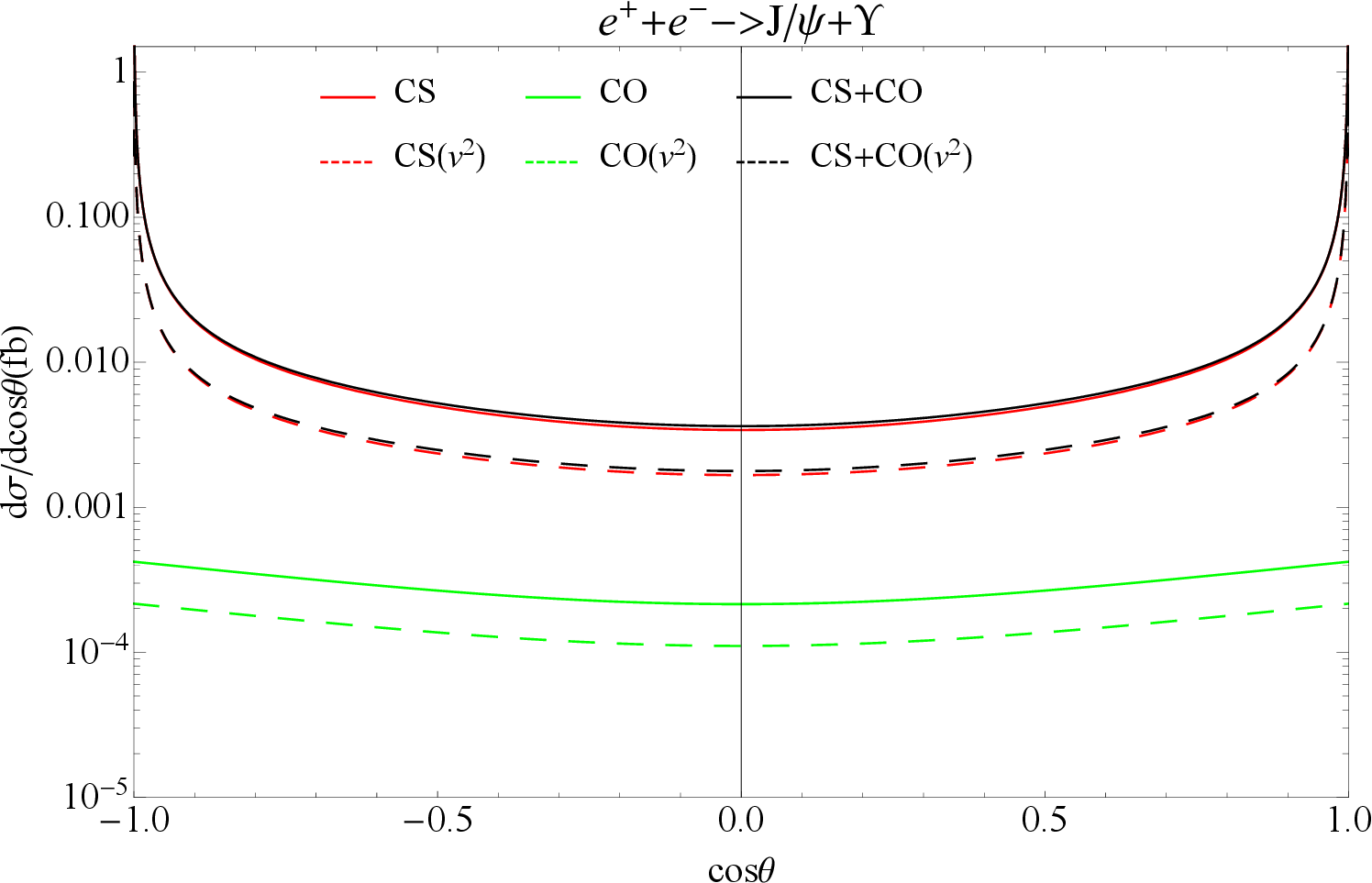}
			\includegraphics[width=0.333\textwidth]{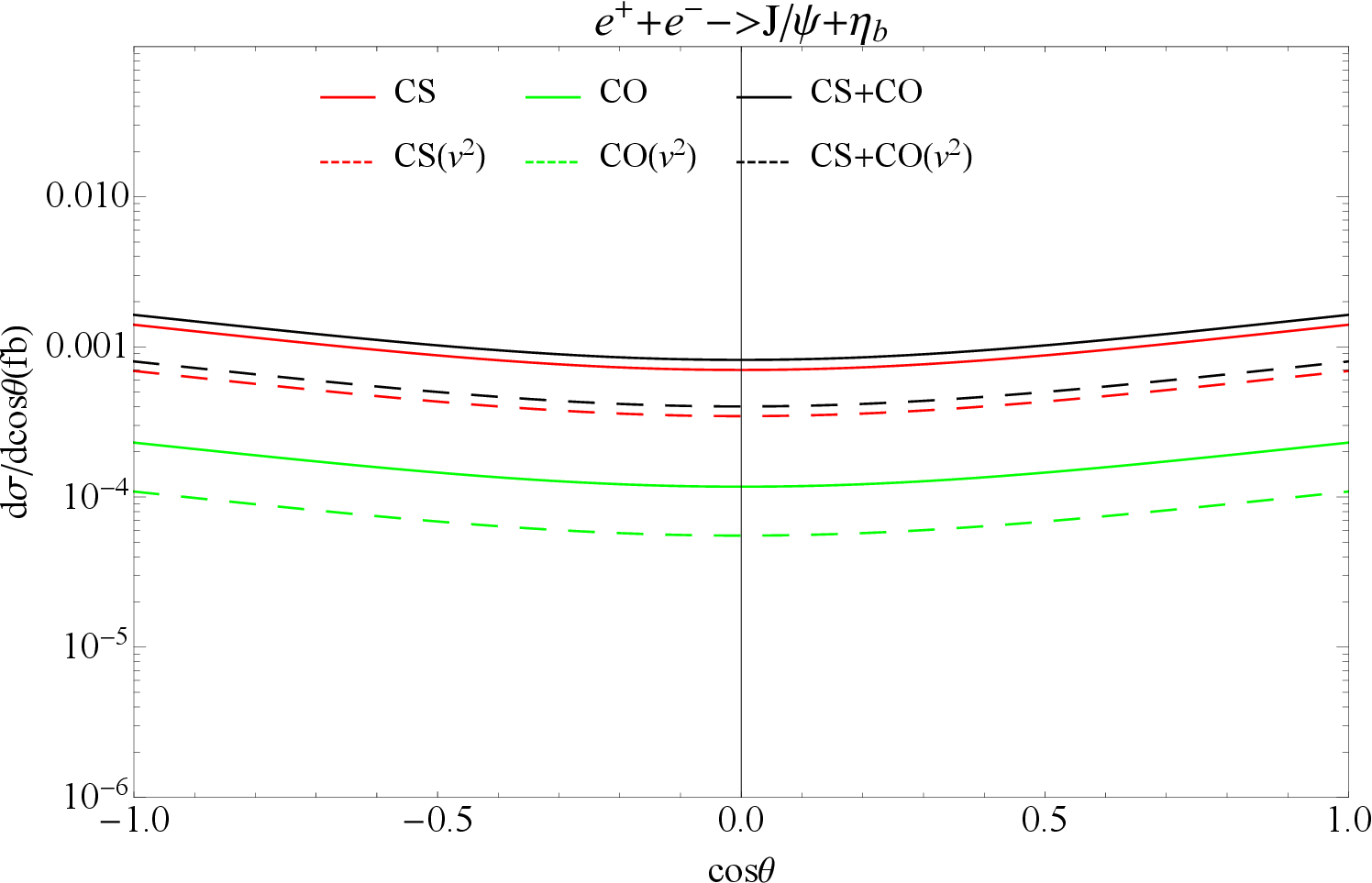}
			\includegraphics[width=0.333\textwidth]{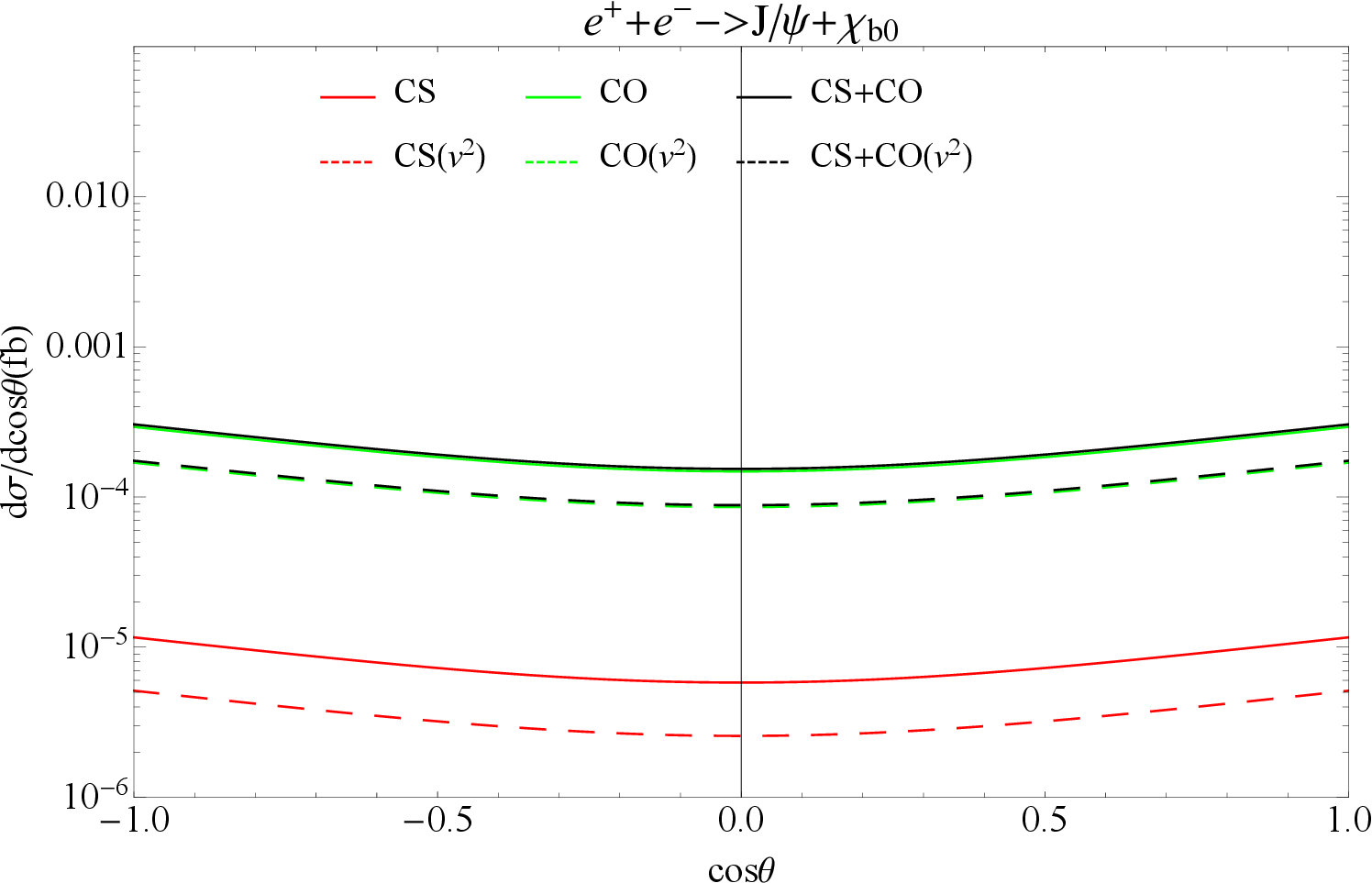}
		\end{tabular}
		\begin{tabular}{c c c}						
			\includegraphics[width=0.333\textwidth]{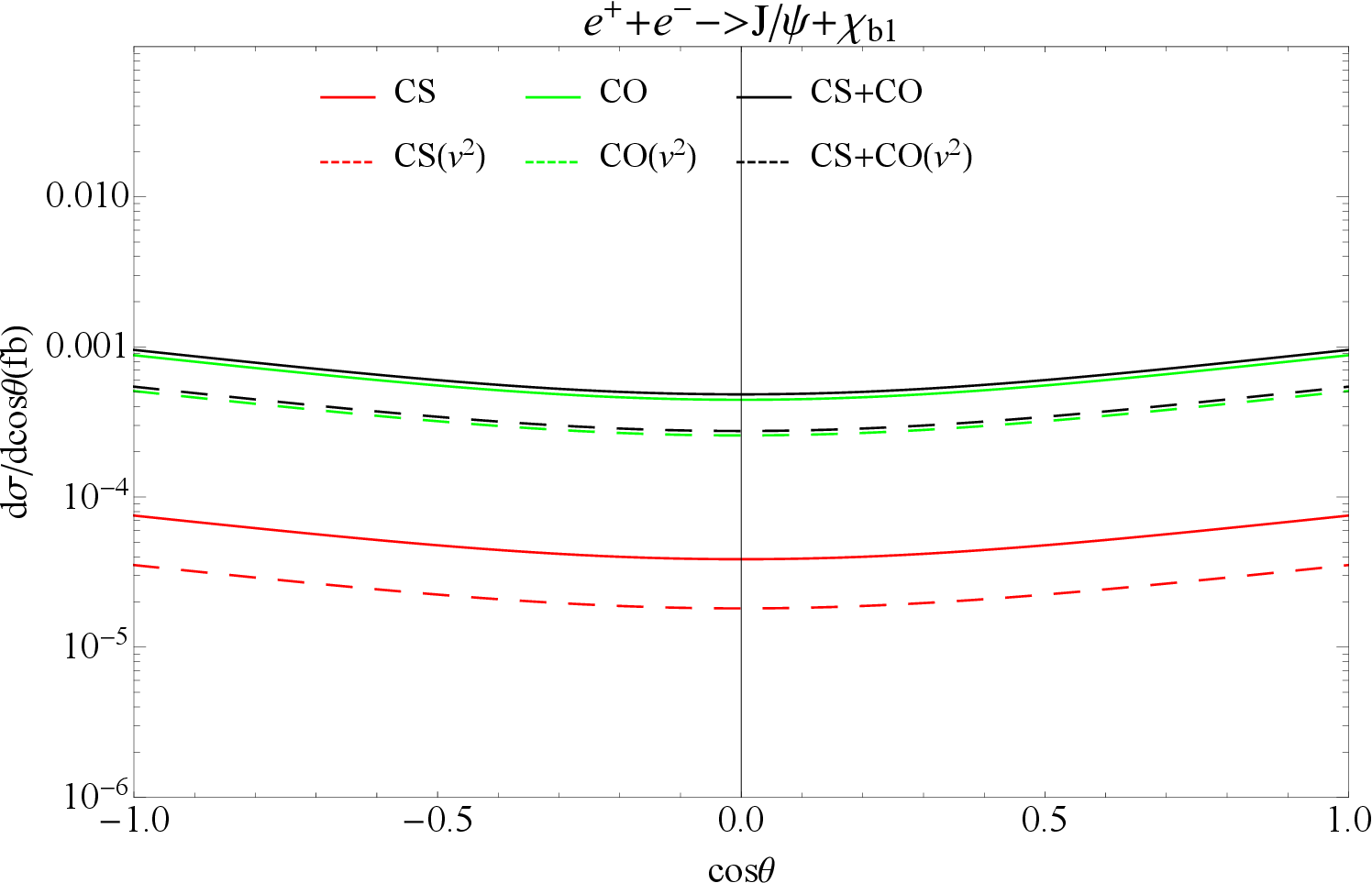}
			\includegraphics[width=0.333\textwidth]{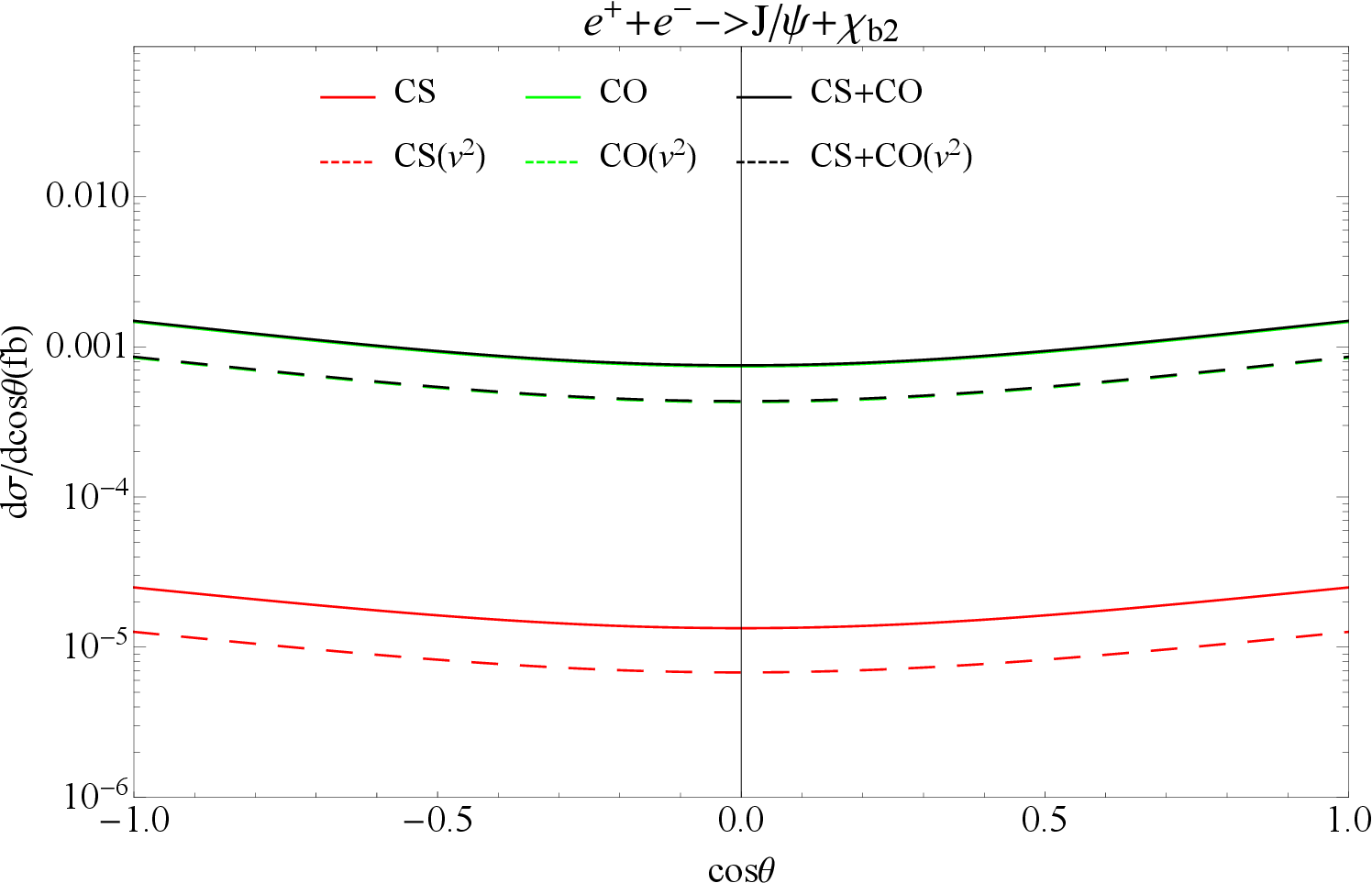}
			\includegraphics[width=0.333\textwidth]{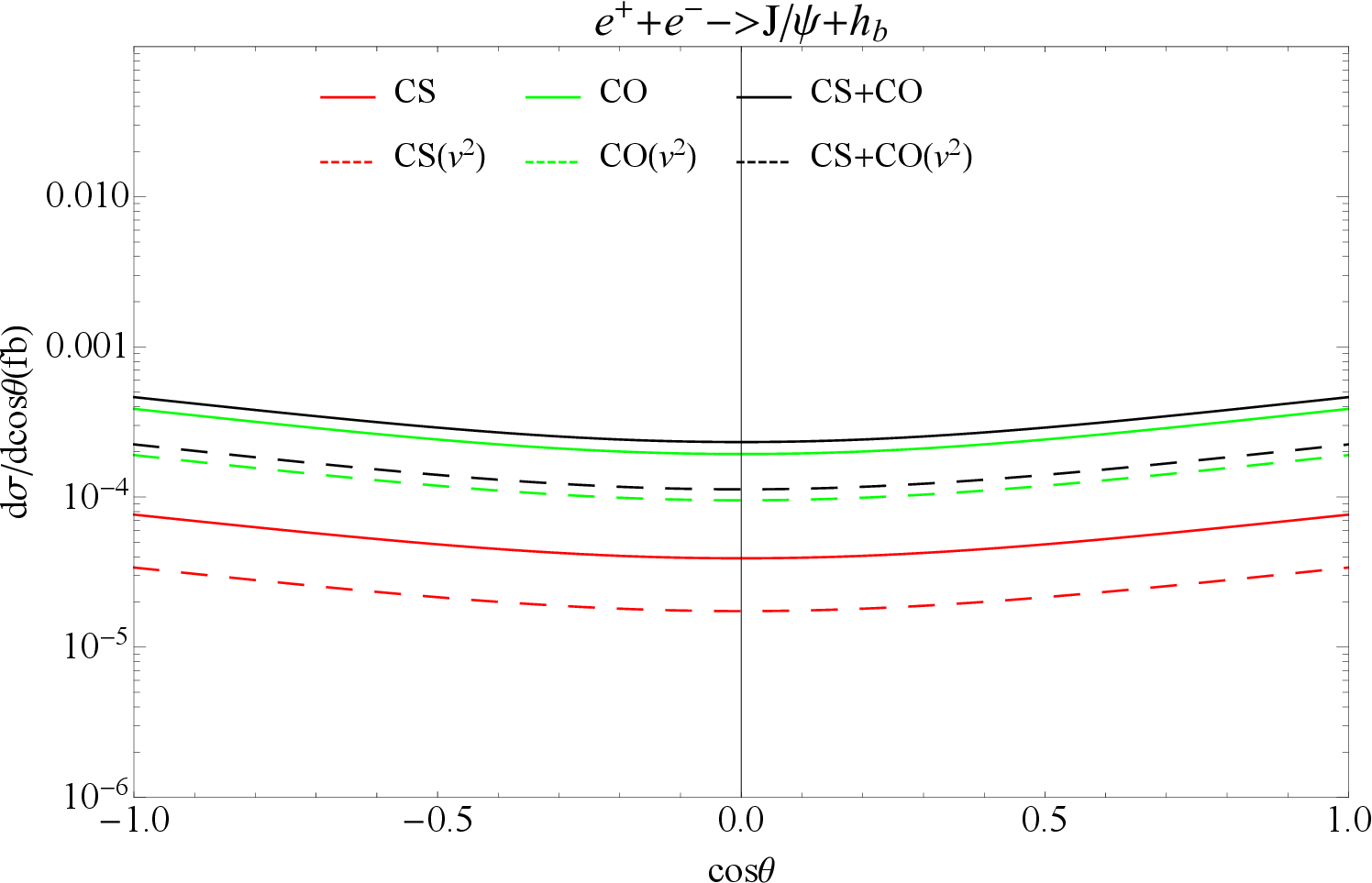}
		\end{tabular}
		
		\begin{tabular}{c c c }
			
			\includegraphics[width=0.333\textwidth]{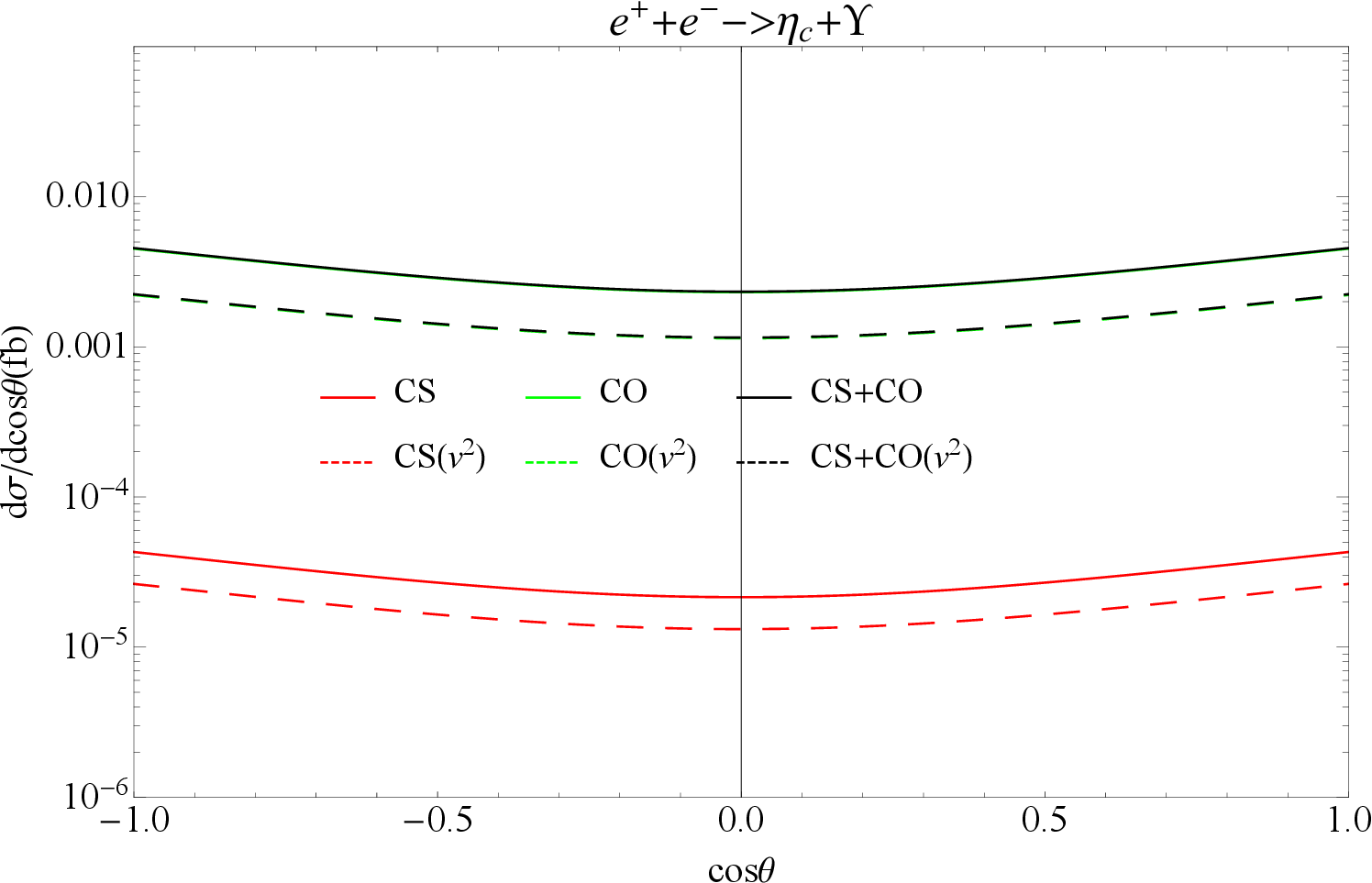}
			\includegraphics[width=0.333\textwidth]{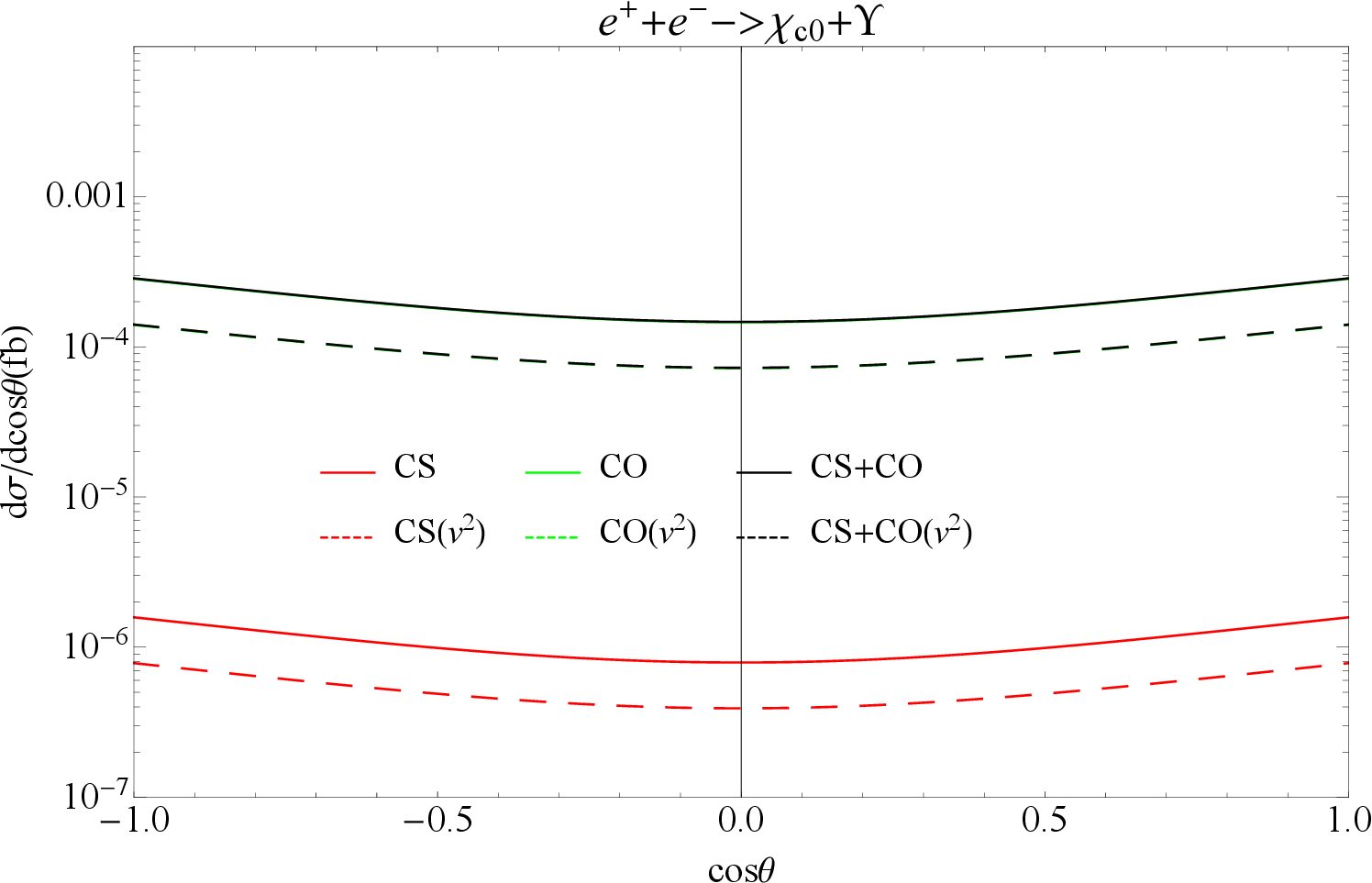}
			\includegraphics[width=0.333\textwidth]{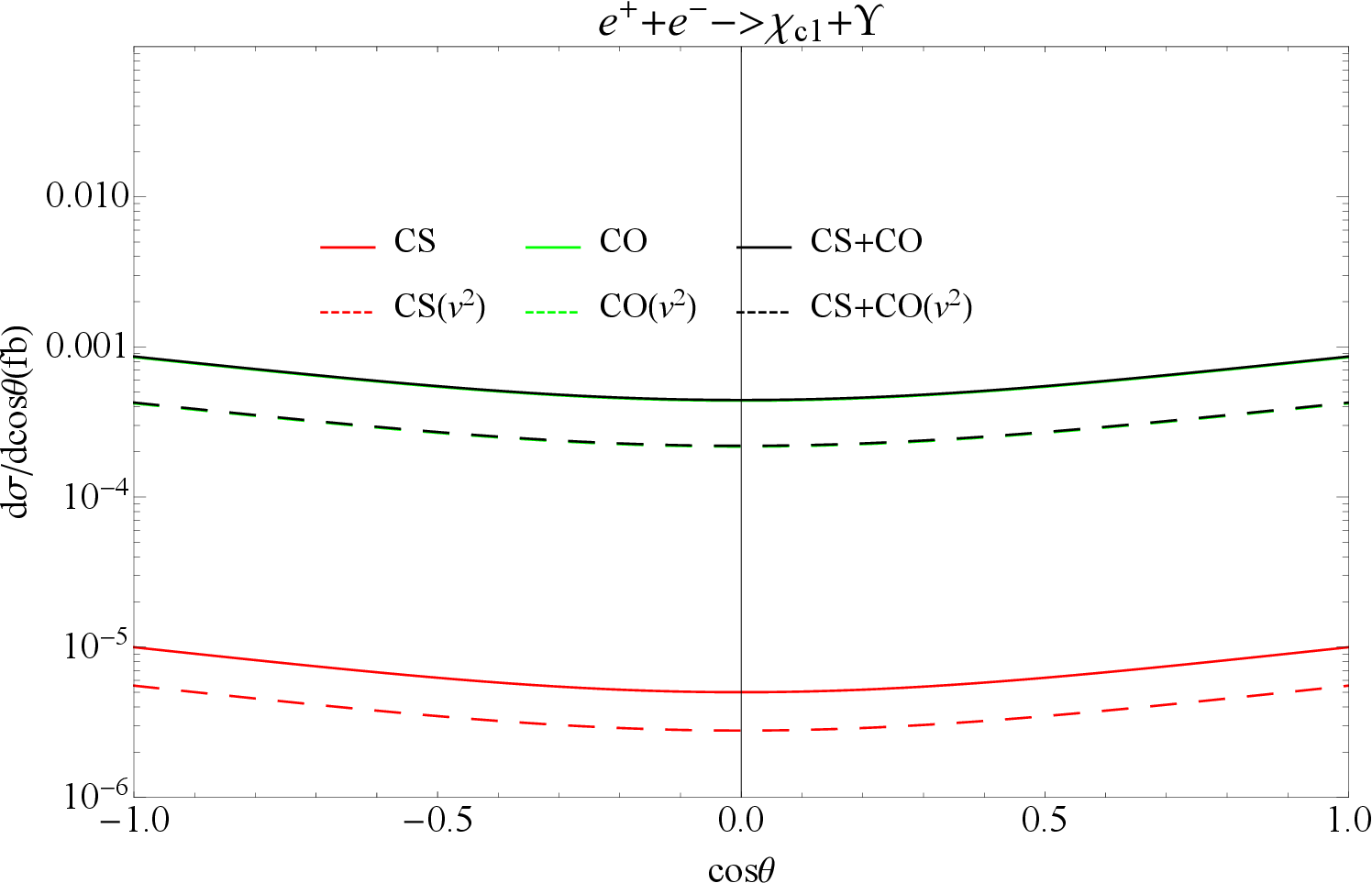}
		\end{tabular}
		\begin{tabular}{c c c }
			\includegraphics[width=0.333\textwidth]{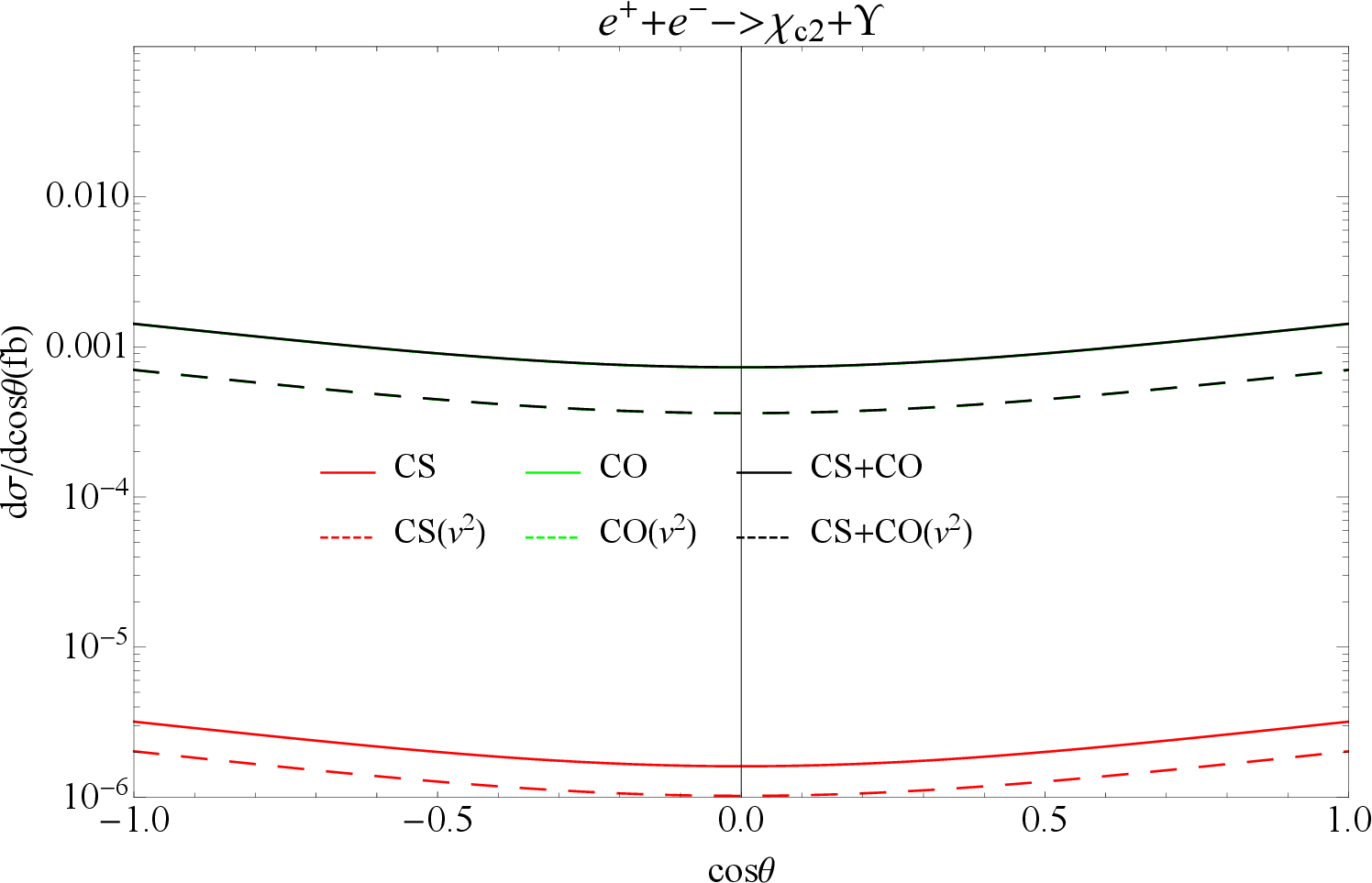}
			\includegraphics[width=0.333\textwidth]{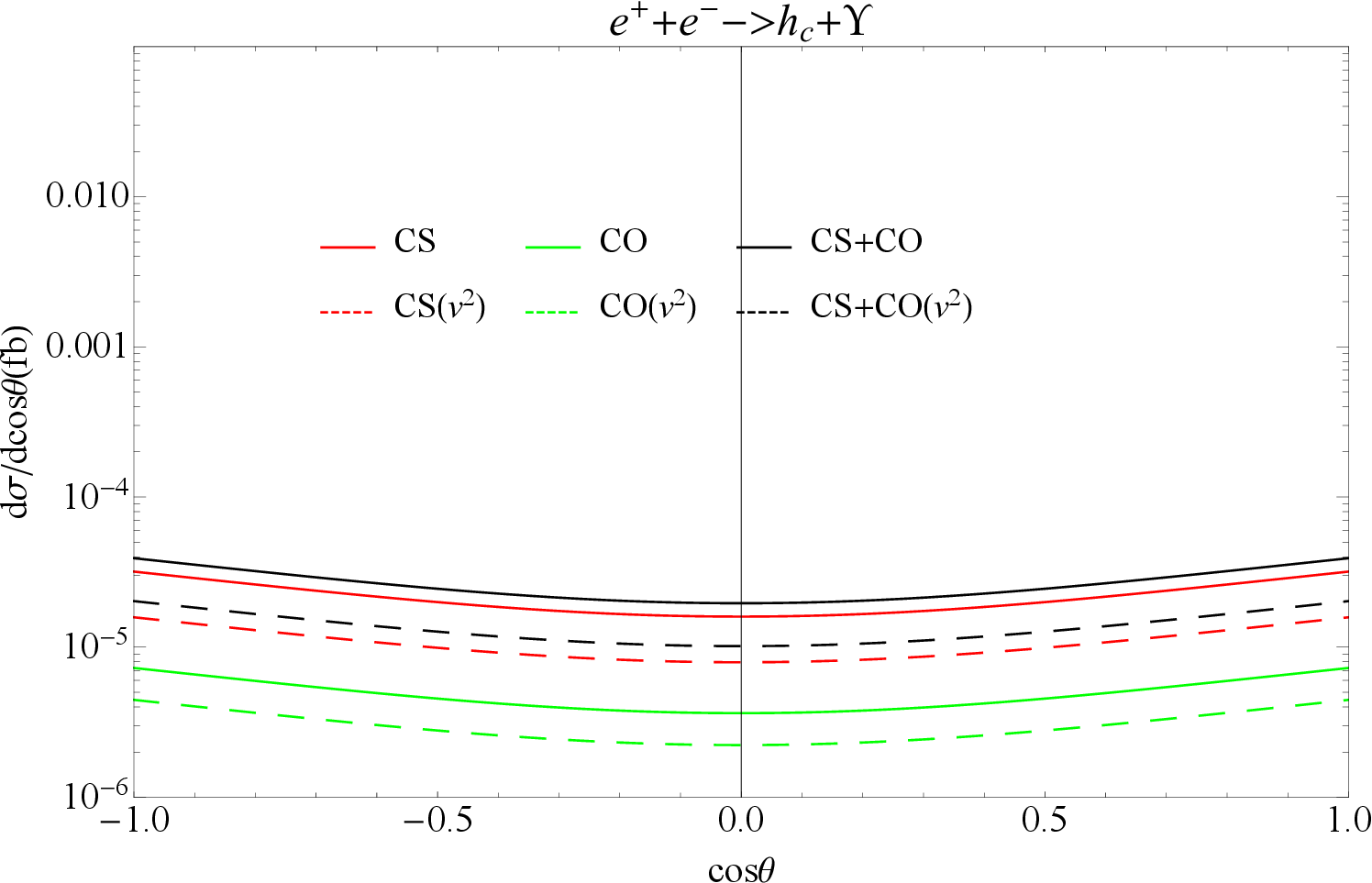}
			\includegraphics[width=0.333\textwidth]{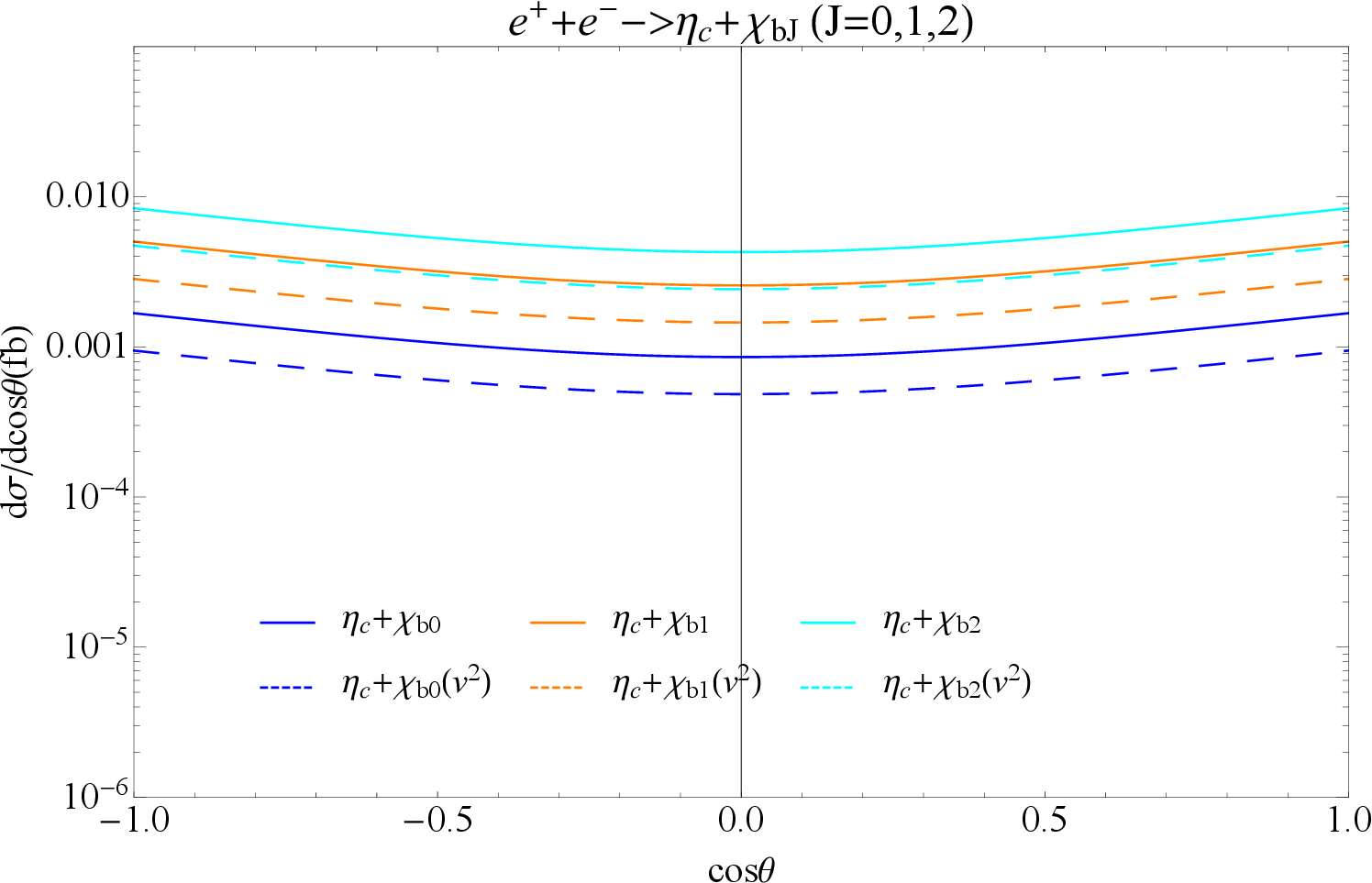}
		\end{tabular}
		\begin{tabular}{c c c }		
			\includegraphics[width=0.333\textwidth]{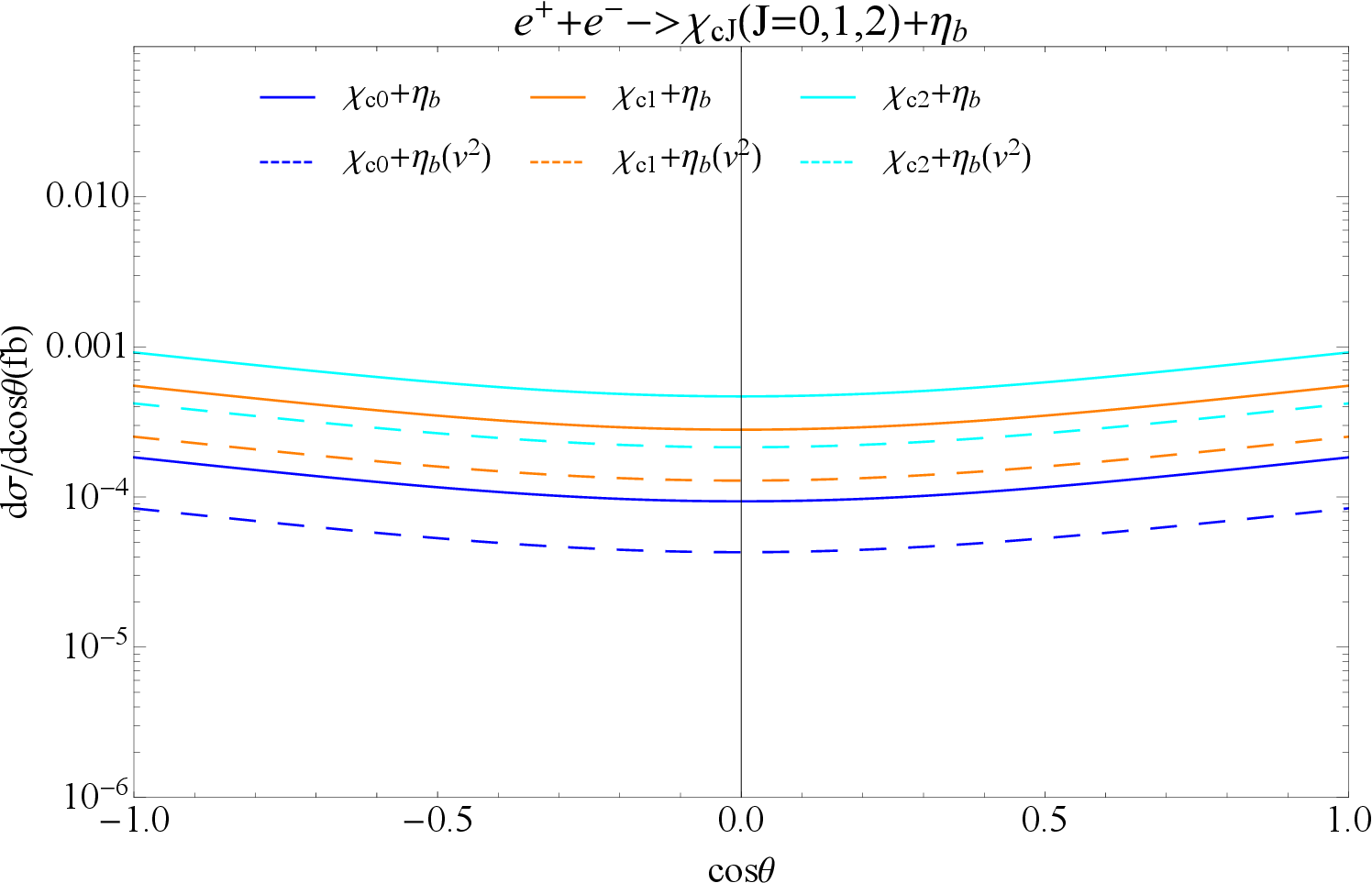}
			\includegraphics[width=0.333\textwidth]{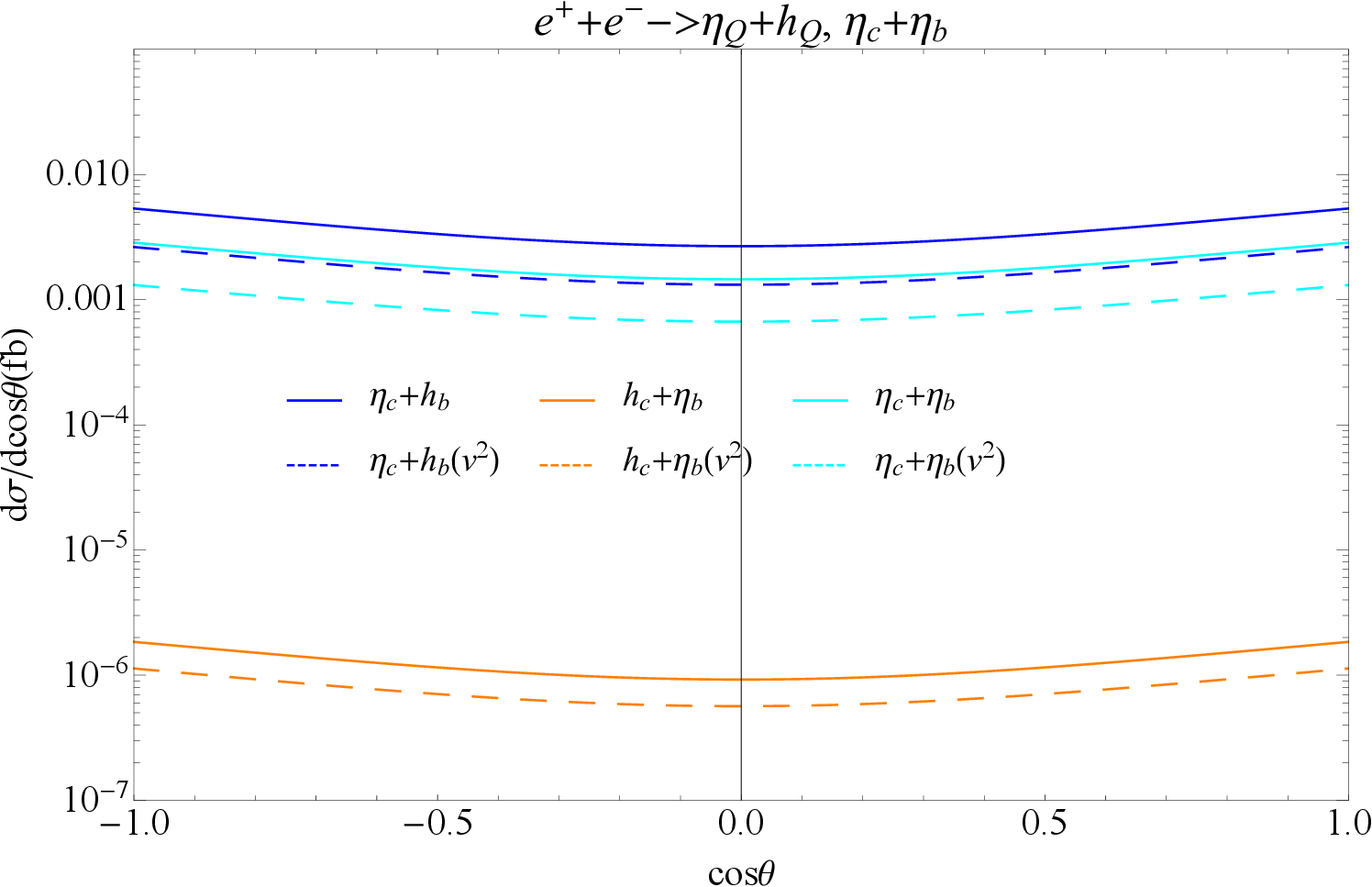}
		\end{tabular}
			\caption{(Color online) The differential cross sections  $d\sigma/dcos\theta$  at $\sqrt{s}$=$m_Z$. The solid line represents LO  and dashed line represents NLO($v^2$) result.  }
		\label{z0cccos}
			
		\end{figure*}
	\end{widetext}

	The K factors ($\sigma_{\text{NLO}(v^2)}/\sigma_{\text{LO}}$) as a function of $\sqrt{s}$ are shown in Fig. \ref{z0cck}. 
The relativistic corrections are significant, which reduce the LO cross sections by approximately 50\%. 
In Appendix \ref{appdB}, we provide the ratios of the SDCs between the RC and LO in the high-energy limit ($\sqrt{s} \gg m_Q$), which also illustrate the effects of the relativistic corrections.
  The numerical results of the total cross section at the $Z^0$ peak are shown in Table~\ref{TCS1}, where we list the CS channel cross sections and those for the CO channels in both $\mathcal{O}(v^0)$\footnote{Our LO CS cross sections of processes $J/\psi+\eta_b, \Upsilon+\eta_c$, and $J/\psi+\Upsilon$ are in agreement with the findings presented in the latest arXiv version of Ref.~\cite{Belov:2021ftc}, when the same input parameters are adopted.} and  $\mathcal{O}(v^2)$. 
	 Fig.~\ref{z0cccos} depicts the differential cross sections $d\sigma/d\cos\theta$, where $\theta$ is the angle between the electron($p_1$) and the charmonium($p_3$). We find that the angular distributions exhibit a hollow structure for all processes in both the CS and CO channels.

\begin{widetext}
		\begin{figure*}[htbp]
		\begin{tabular}{c c c}
			\includegraphics[width=0.333\textwidth]{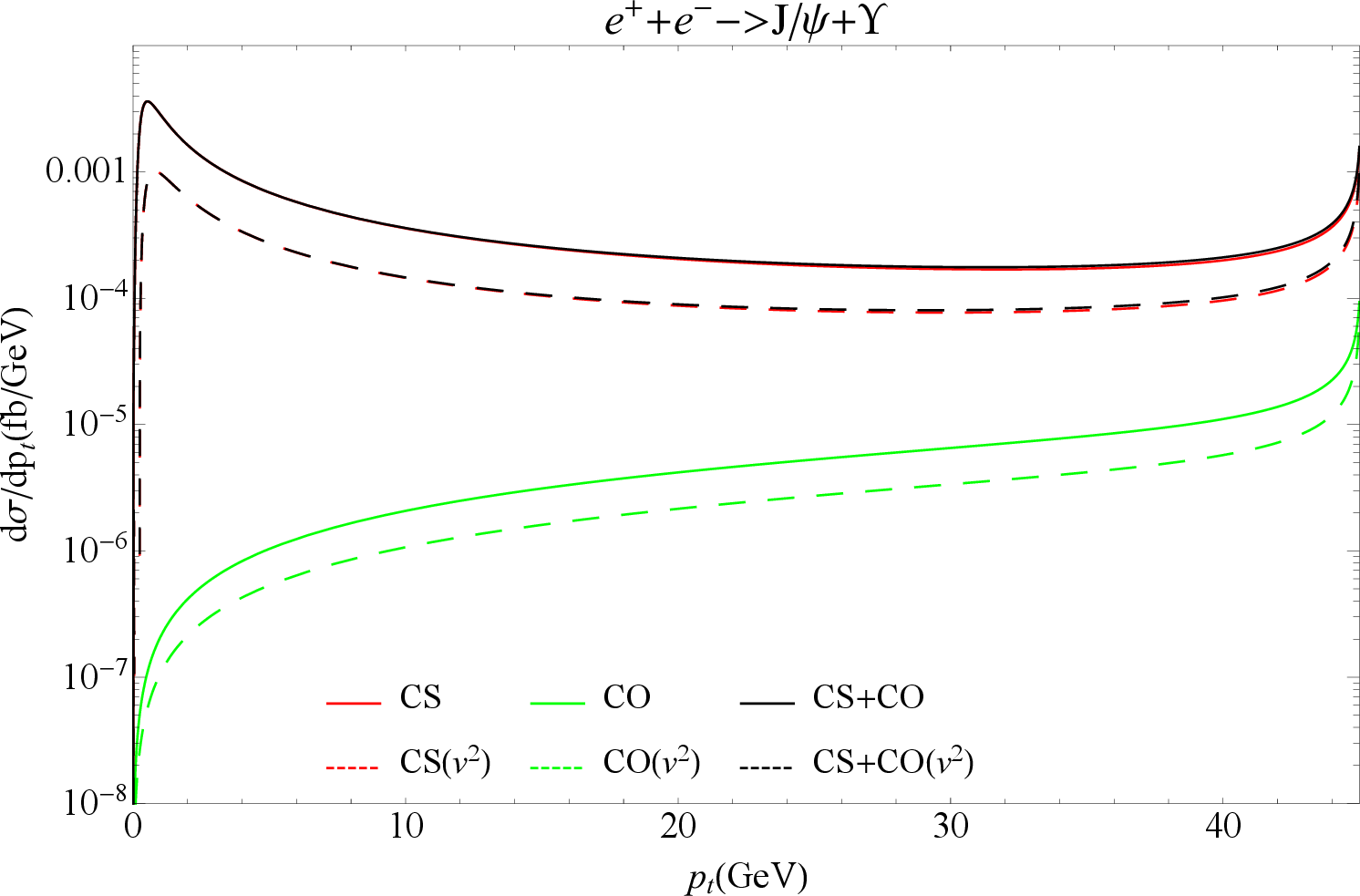}
			\includegraphics[width=0.333\textwidth]{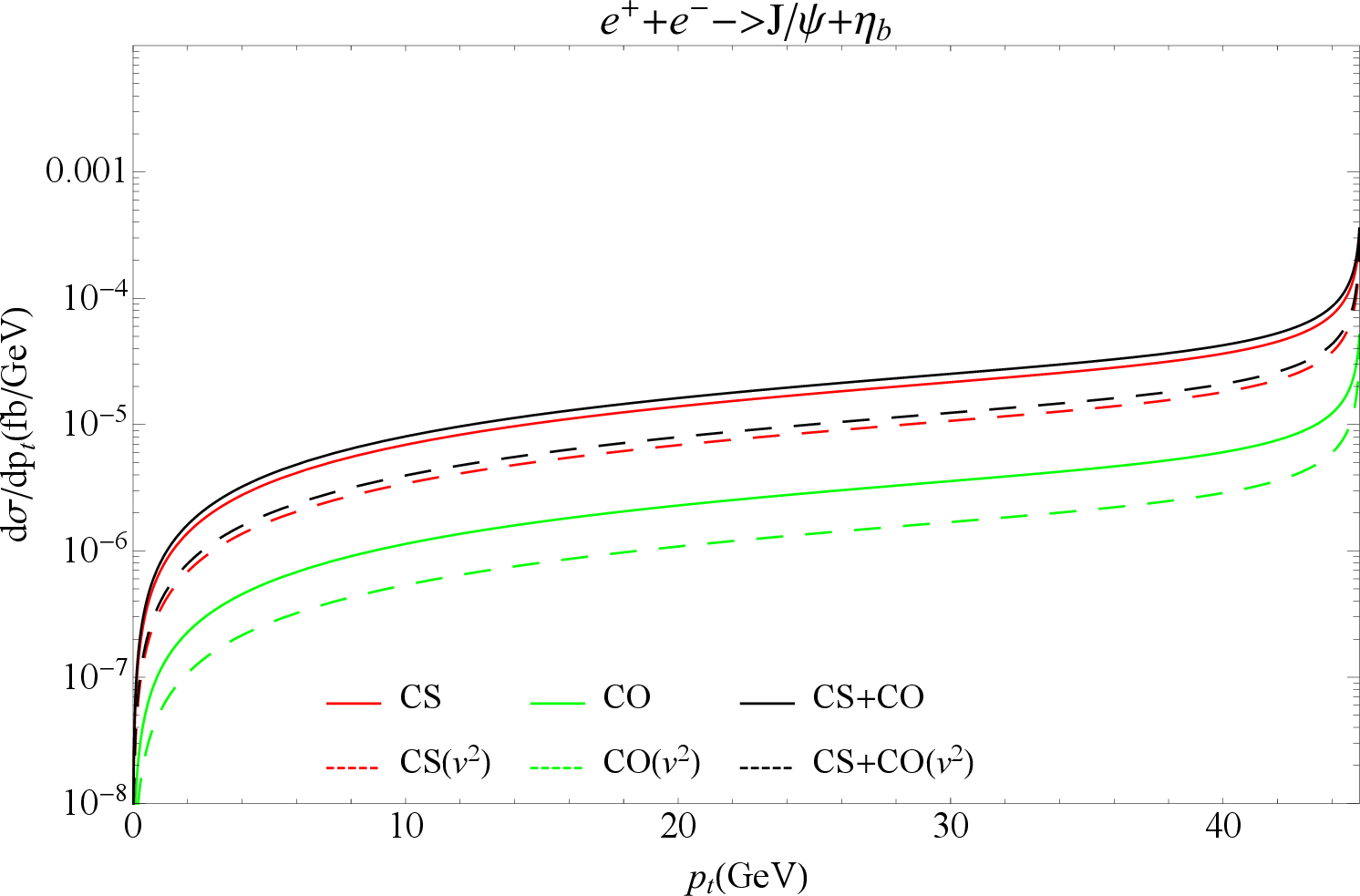}
			\includegraphics[width=0.333\textwidth]{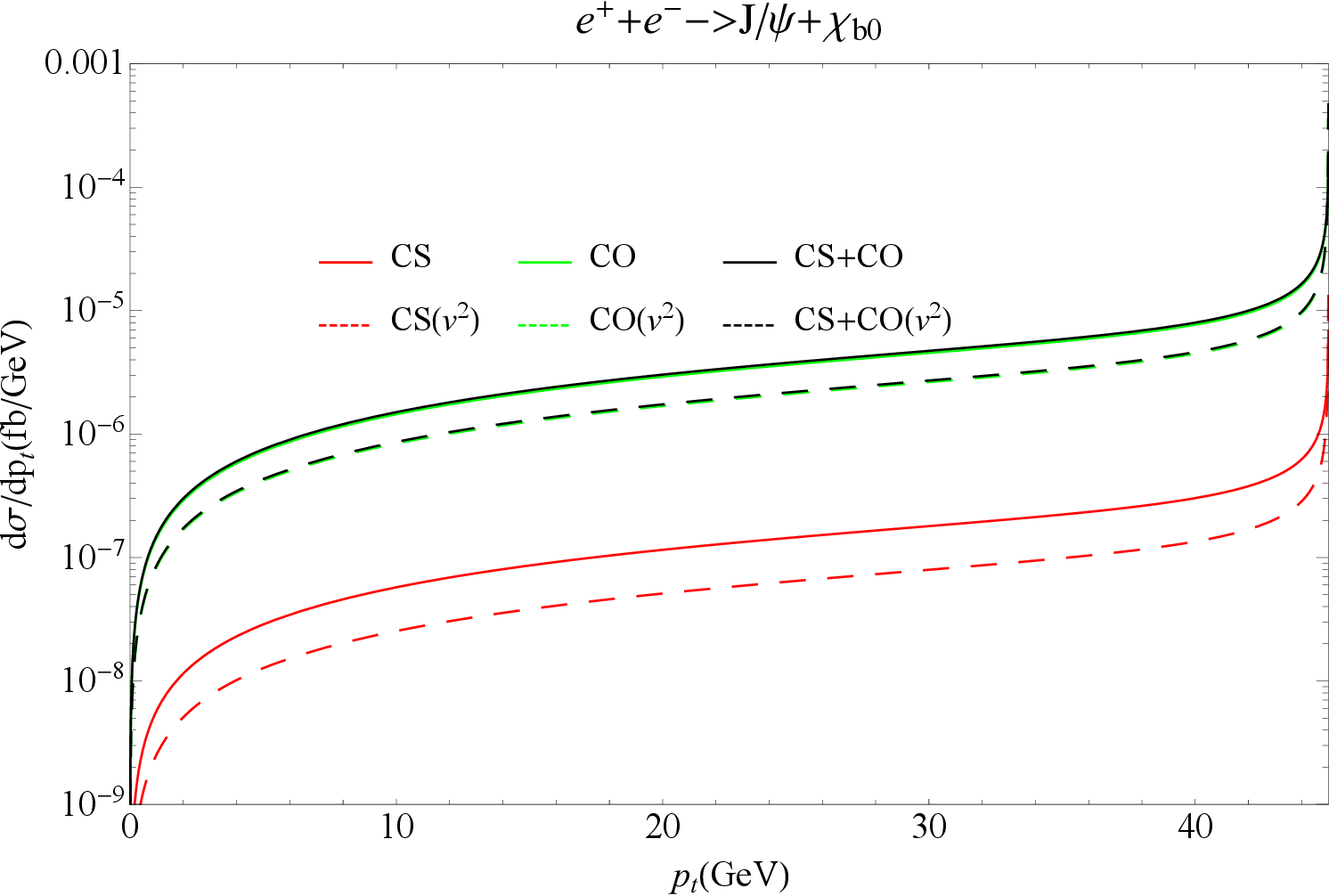}
		\end{tabular}
		\begin{tabular}{c c c}						
			\includegraphics[width=0.333\textwidth]{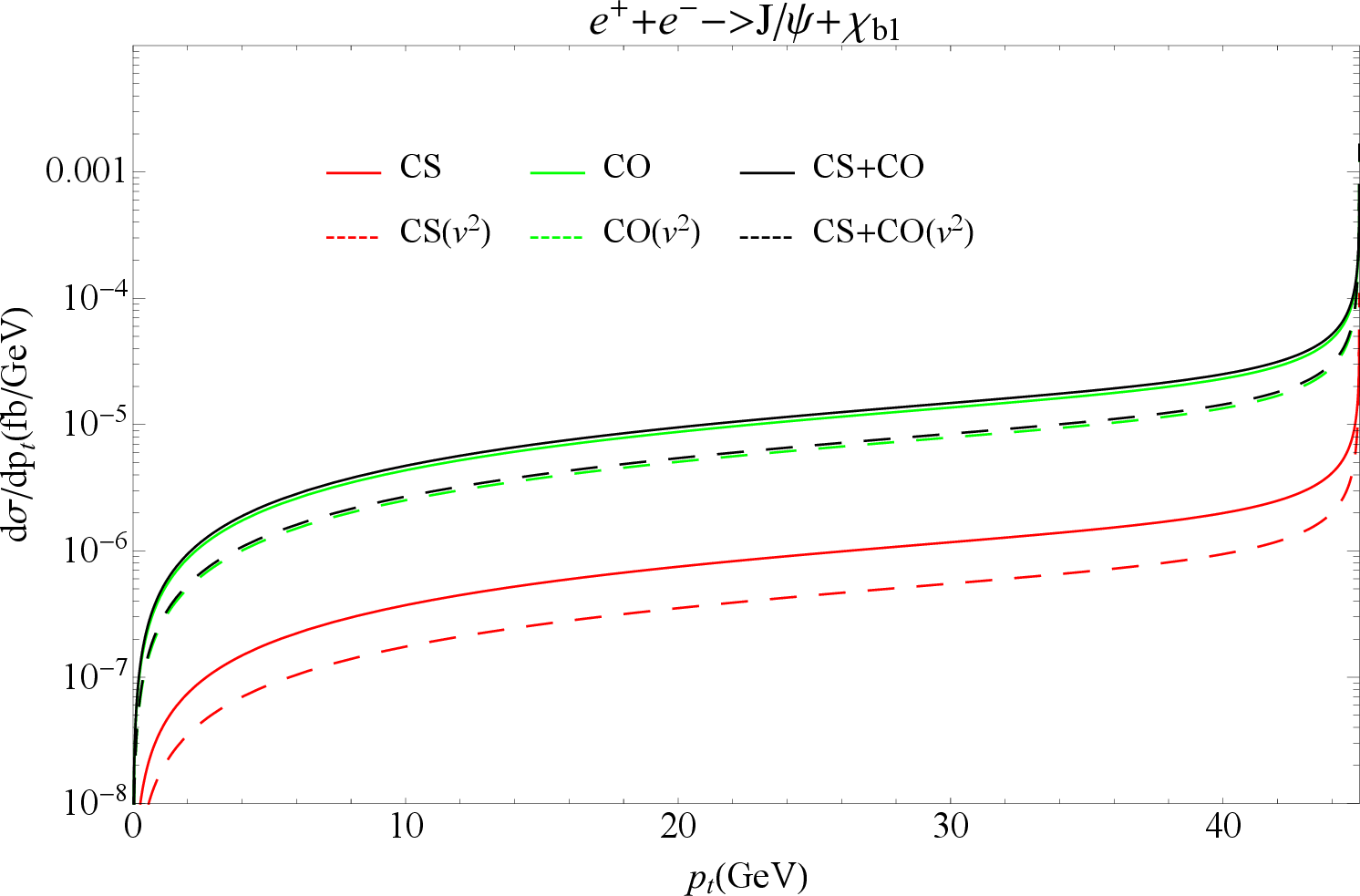}
			\includegraphics[width=0.333\textwidth]{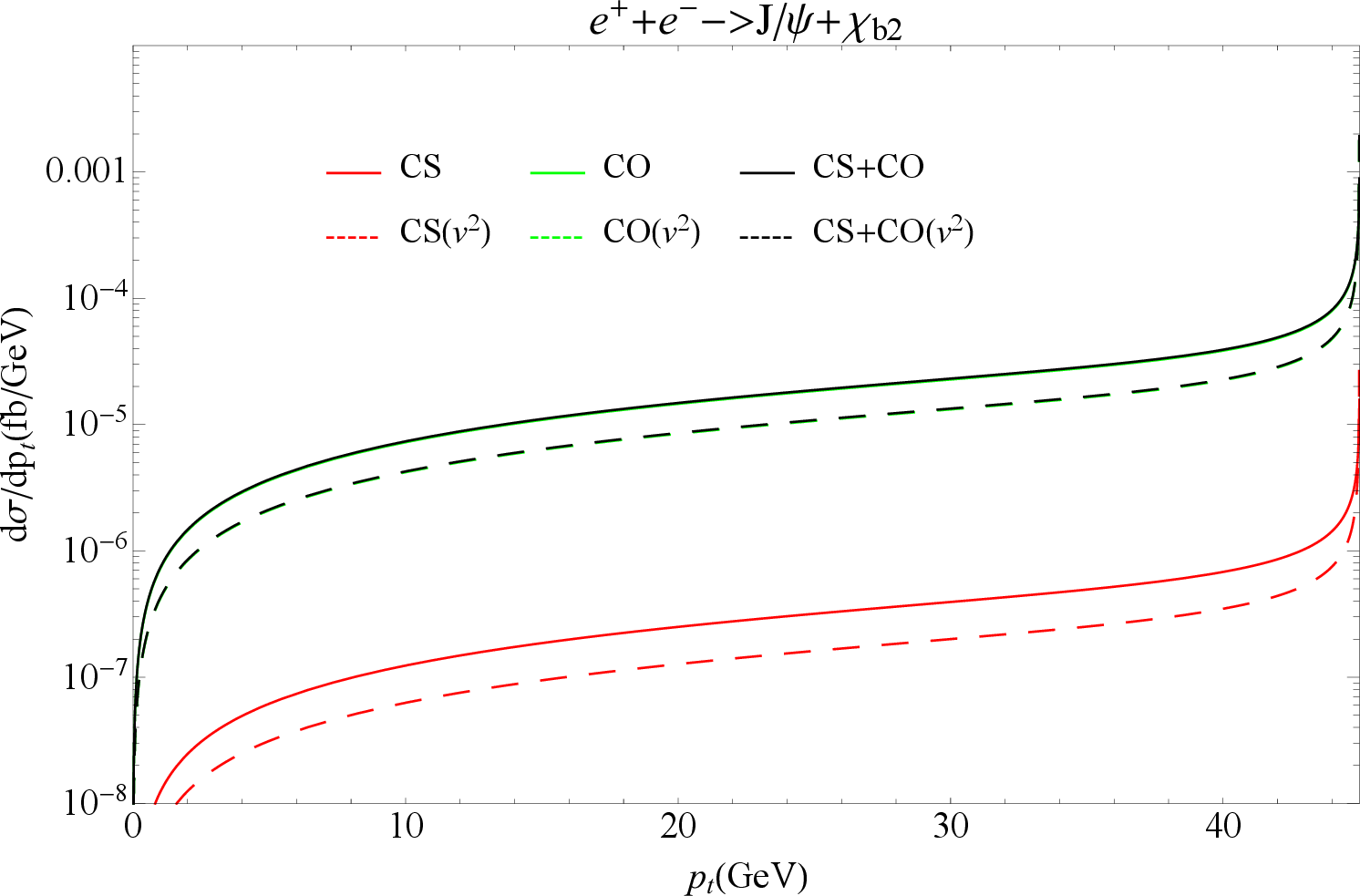}
			\includegraphics[width=0.333\textwidth]{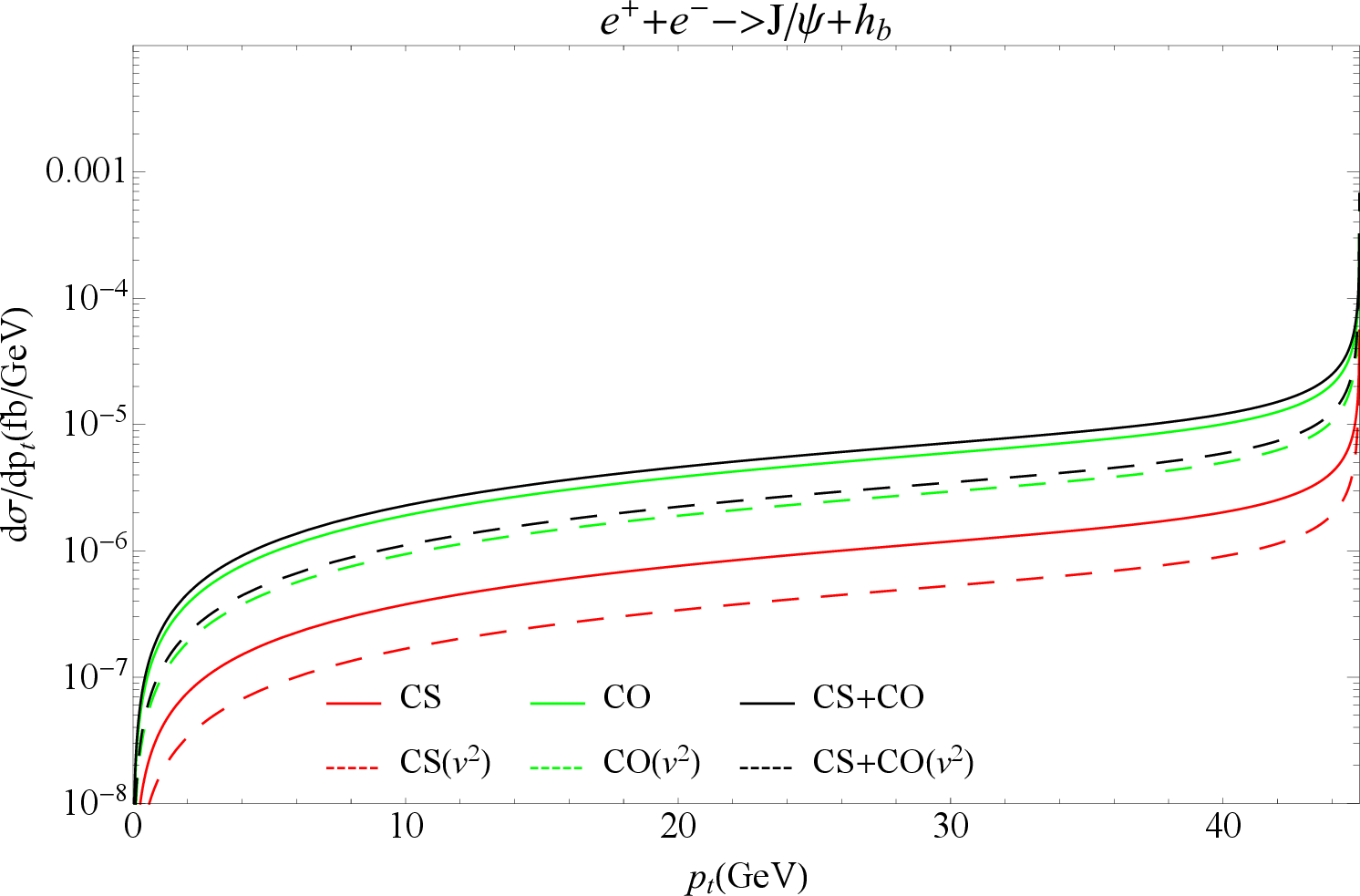}
		\end{tabular}
		
		\begin{tabular}{c c c }
			
			\includegraphics[width=0.333\textwidth]{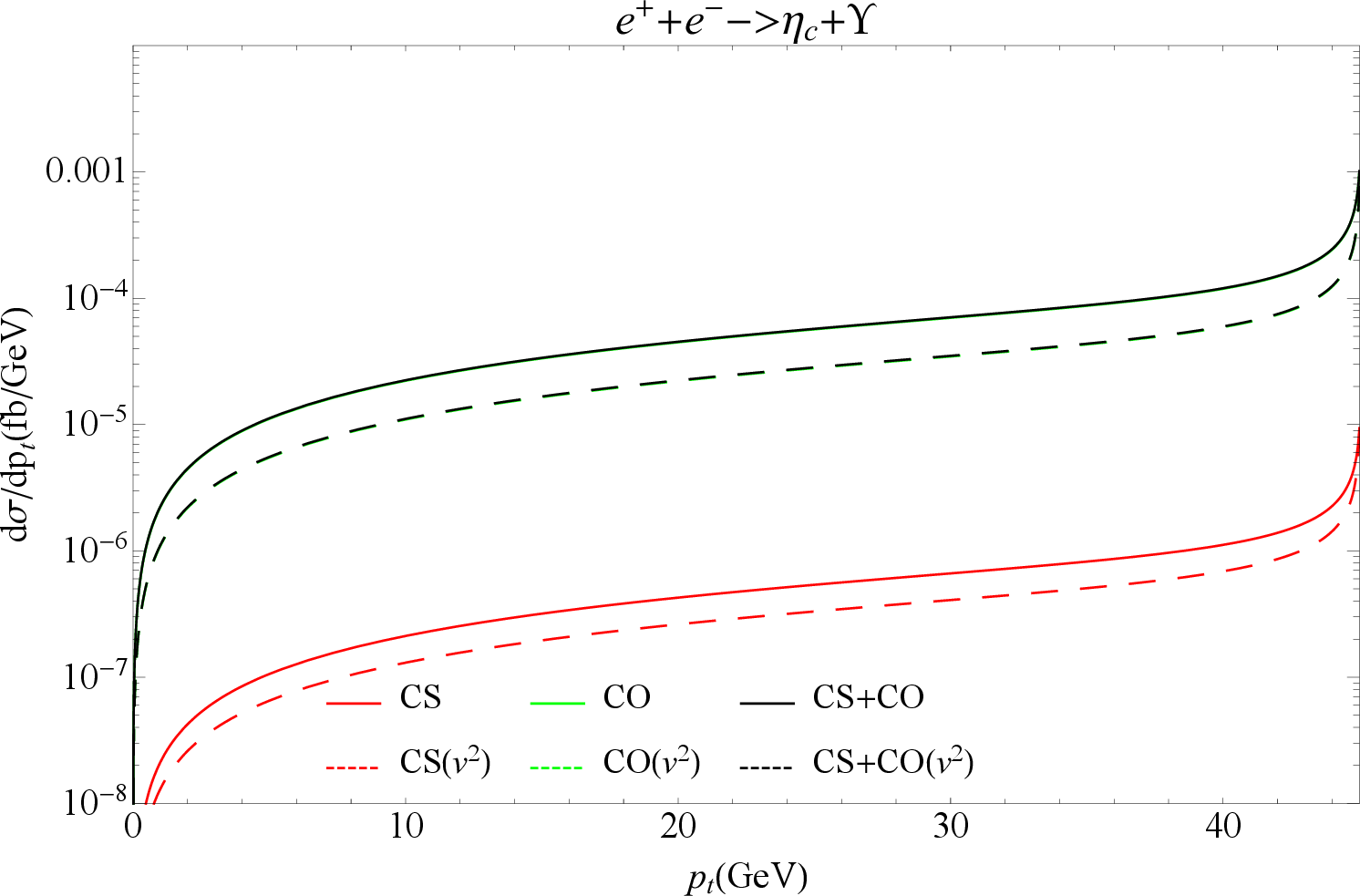}
			\includegraphics[width=0.333\textwidth]{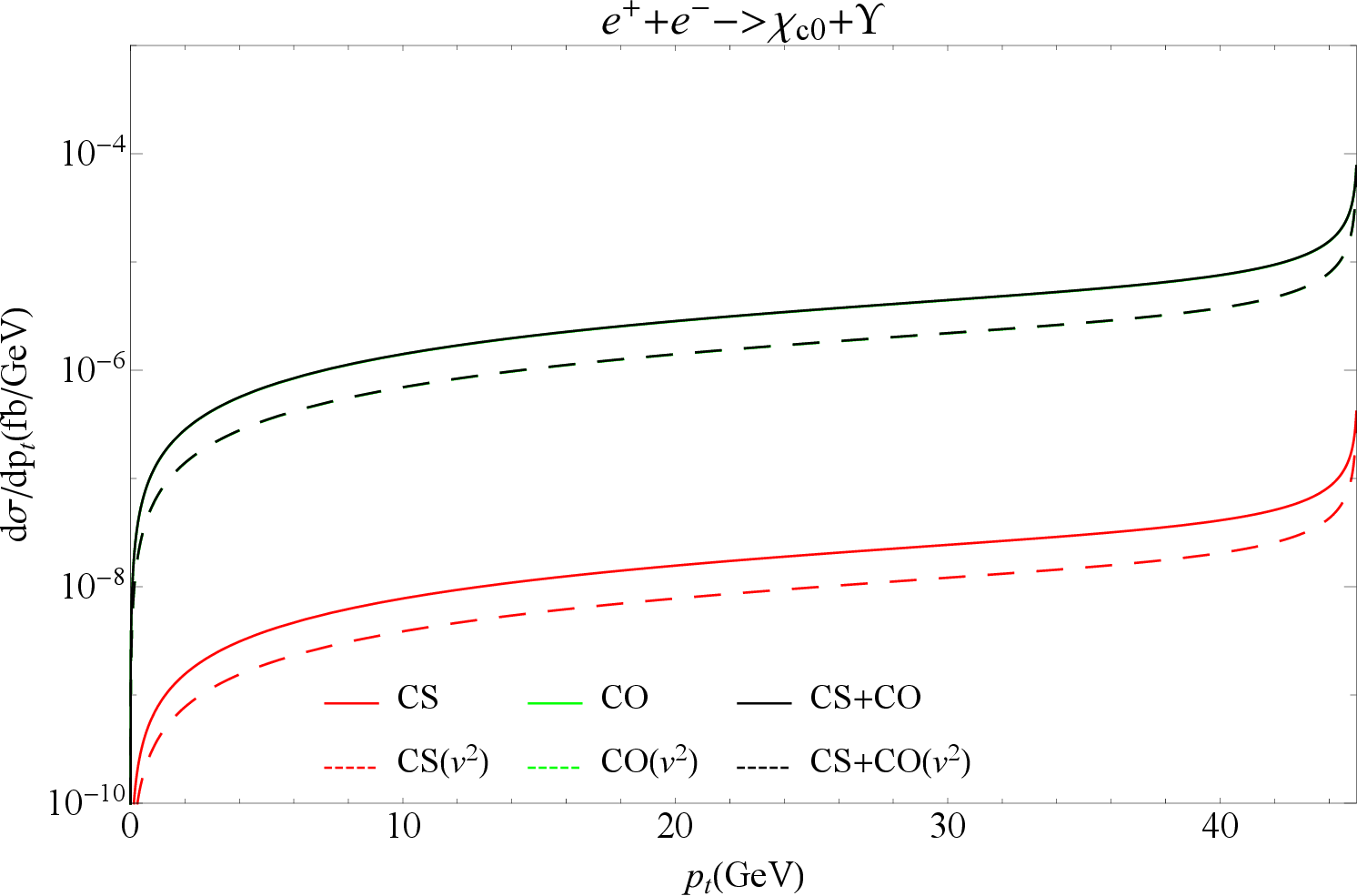}
			\includegraphics[width=0.333\textwidth]{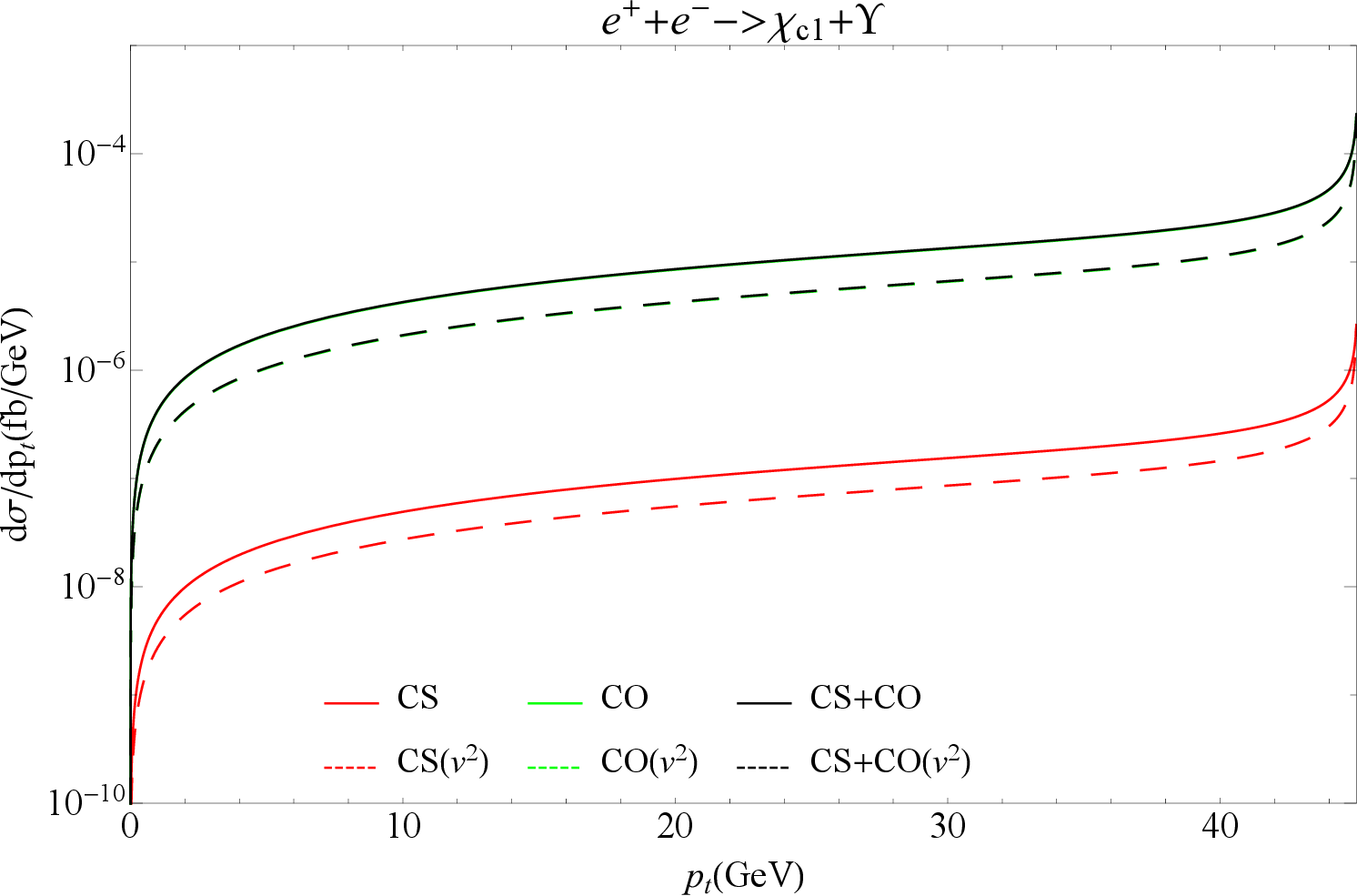}
		\end{tabular}
		\begin{tabular}{c c c }
			\includegraphics[width=0.333\textwidth]{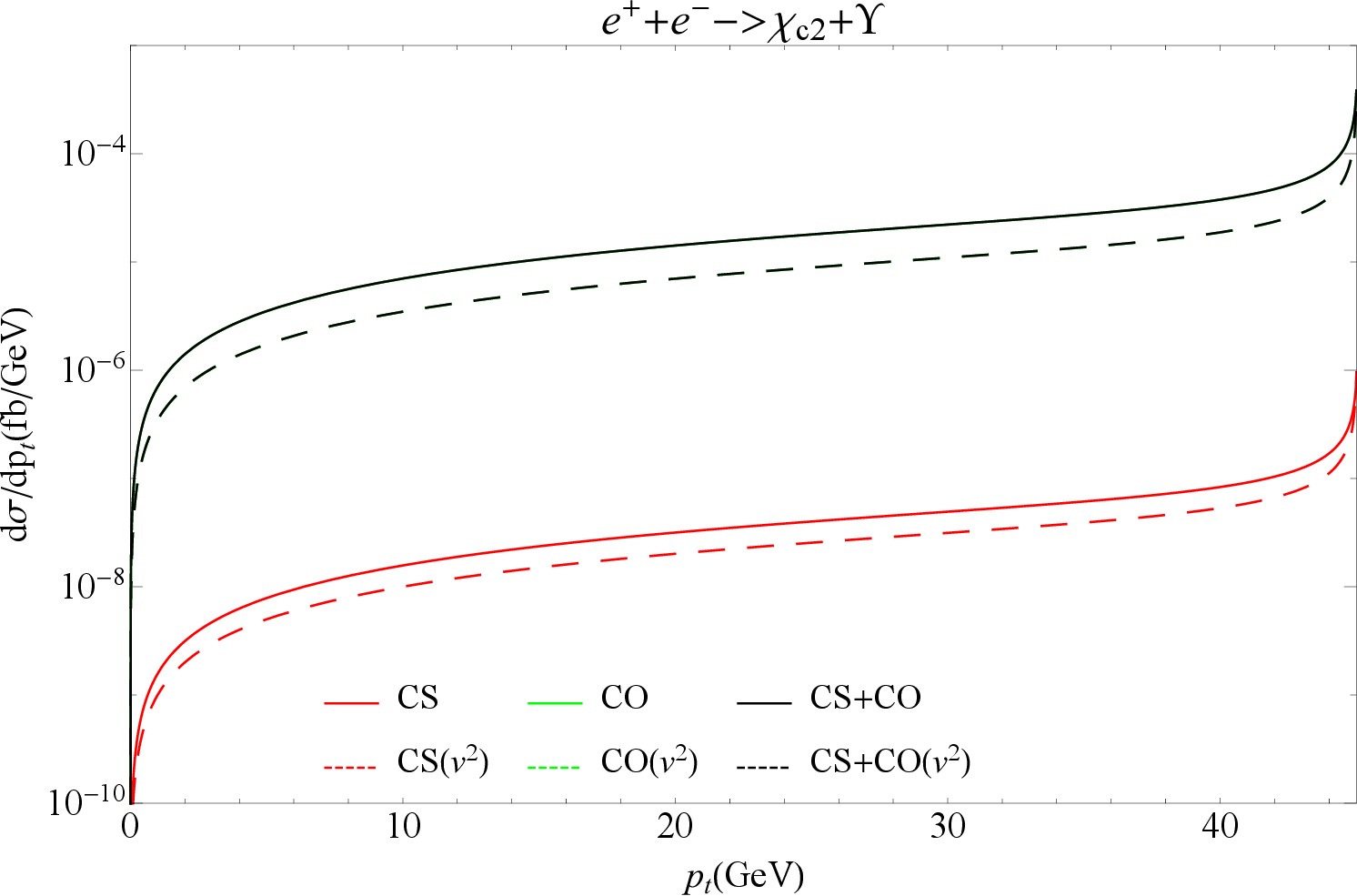}
			\includegraphics[width=0.333\textwidth]{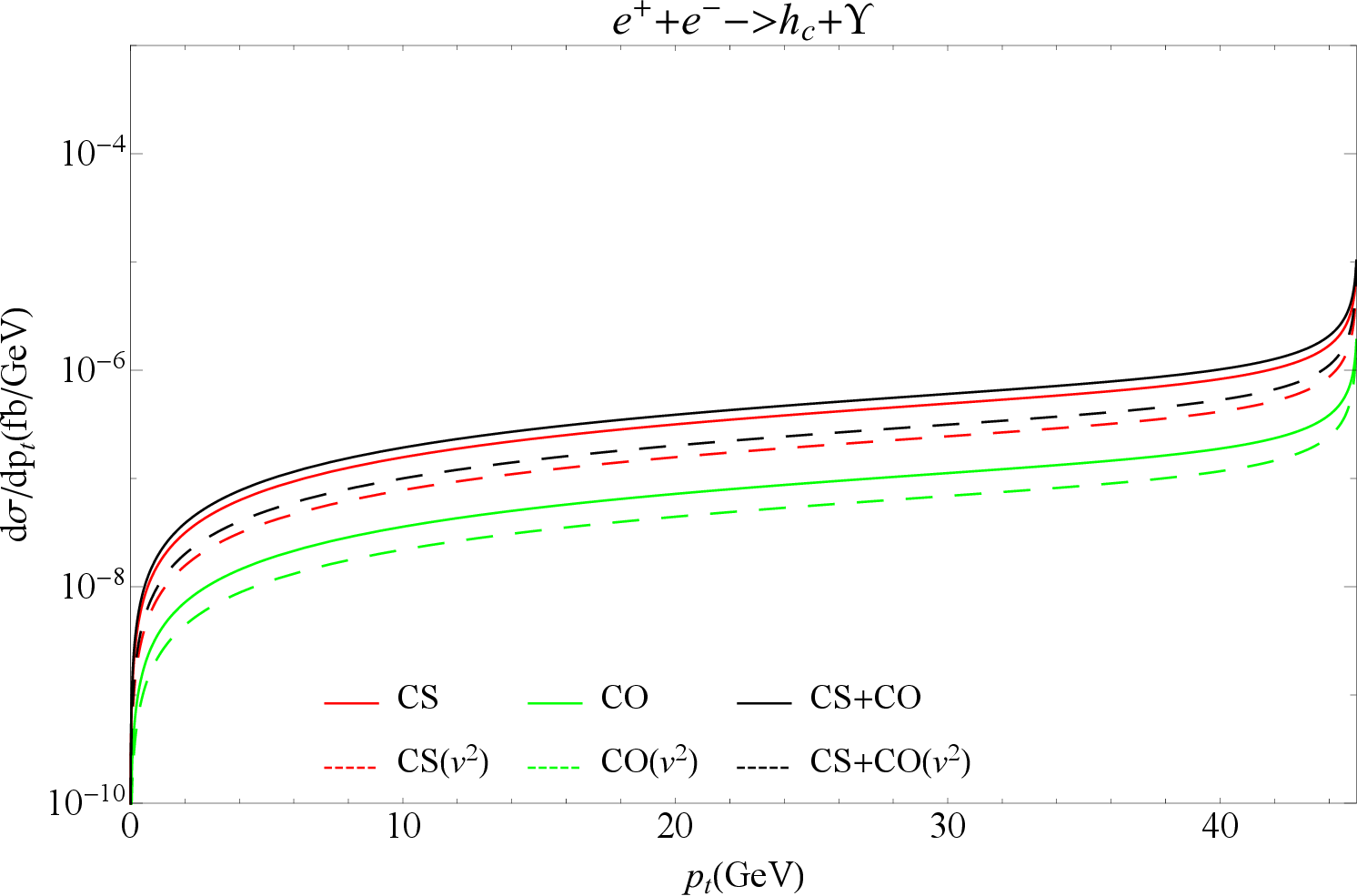}
			\includegraphics[width=0.333\textwidth]{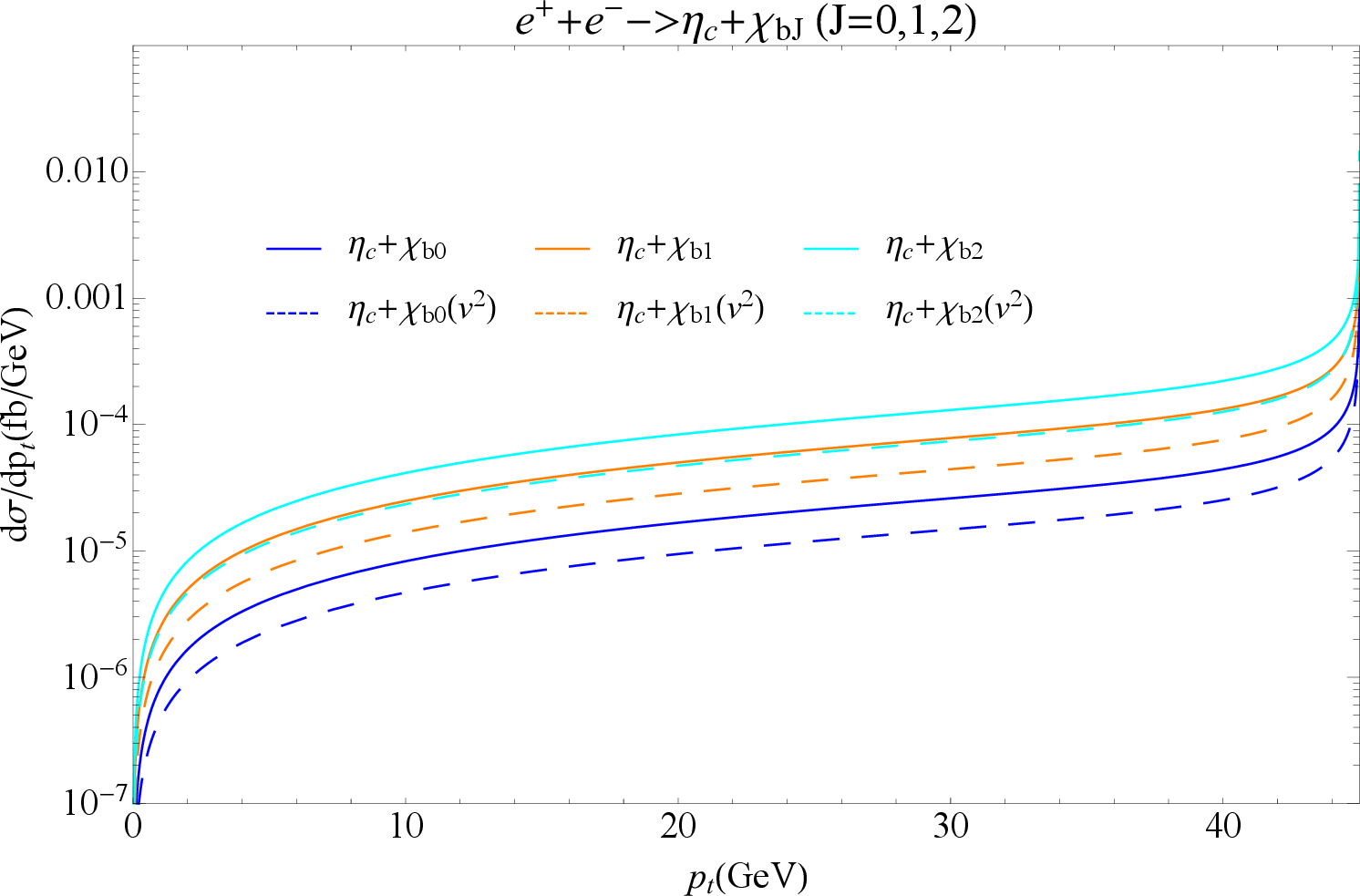}
		\end{tabular}
		\begin{tabular}{c c c }		
			\includegraphics[width=0.333\textwidth]{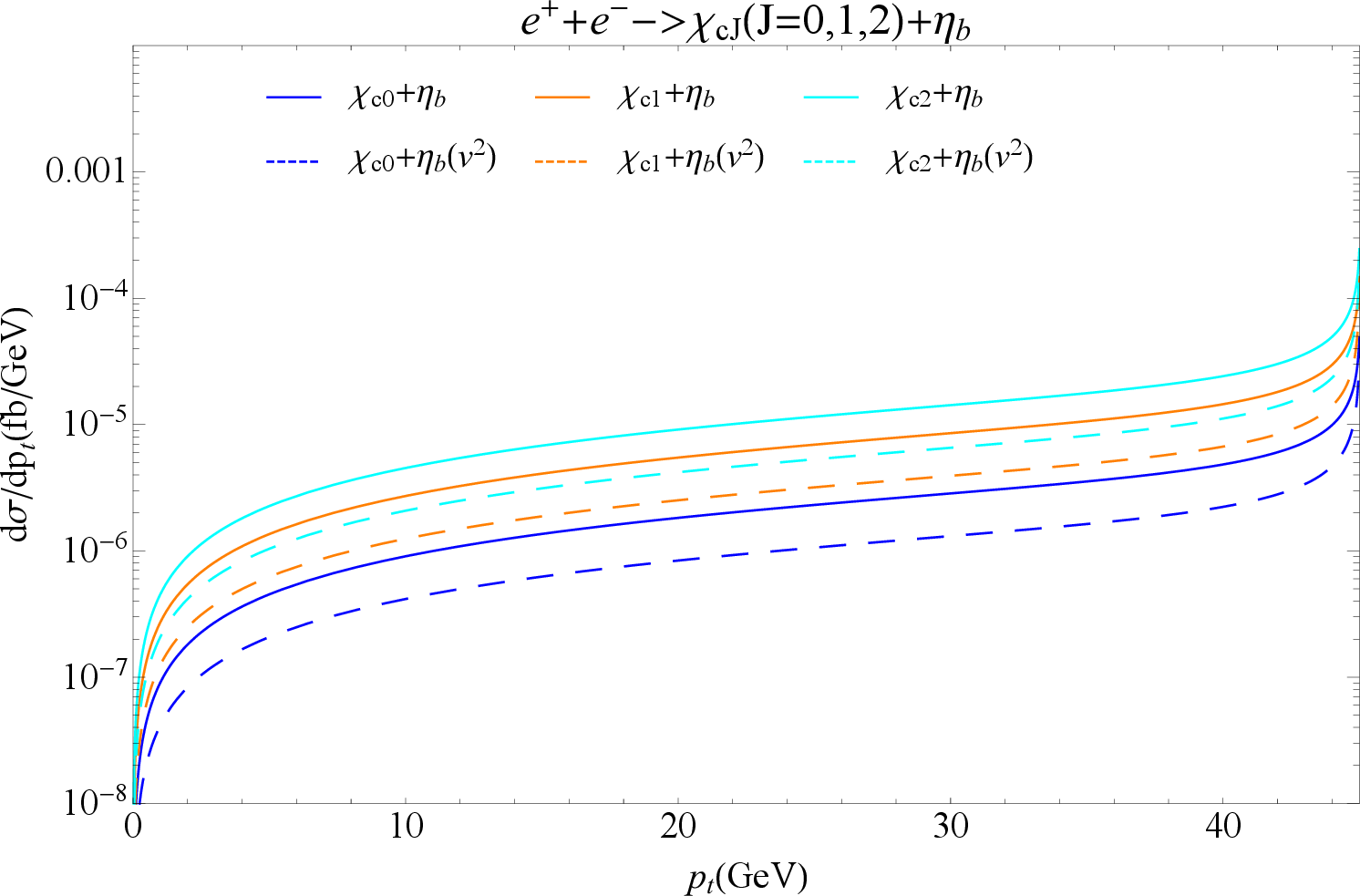}
			\includegraphics[width=0.333\textwidth]{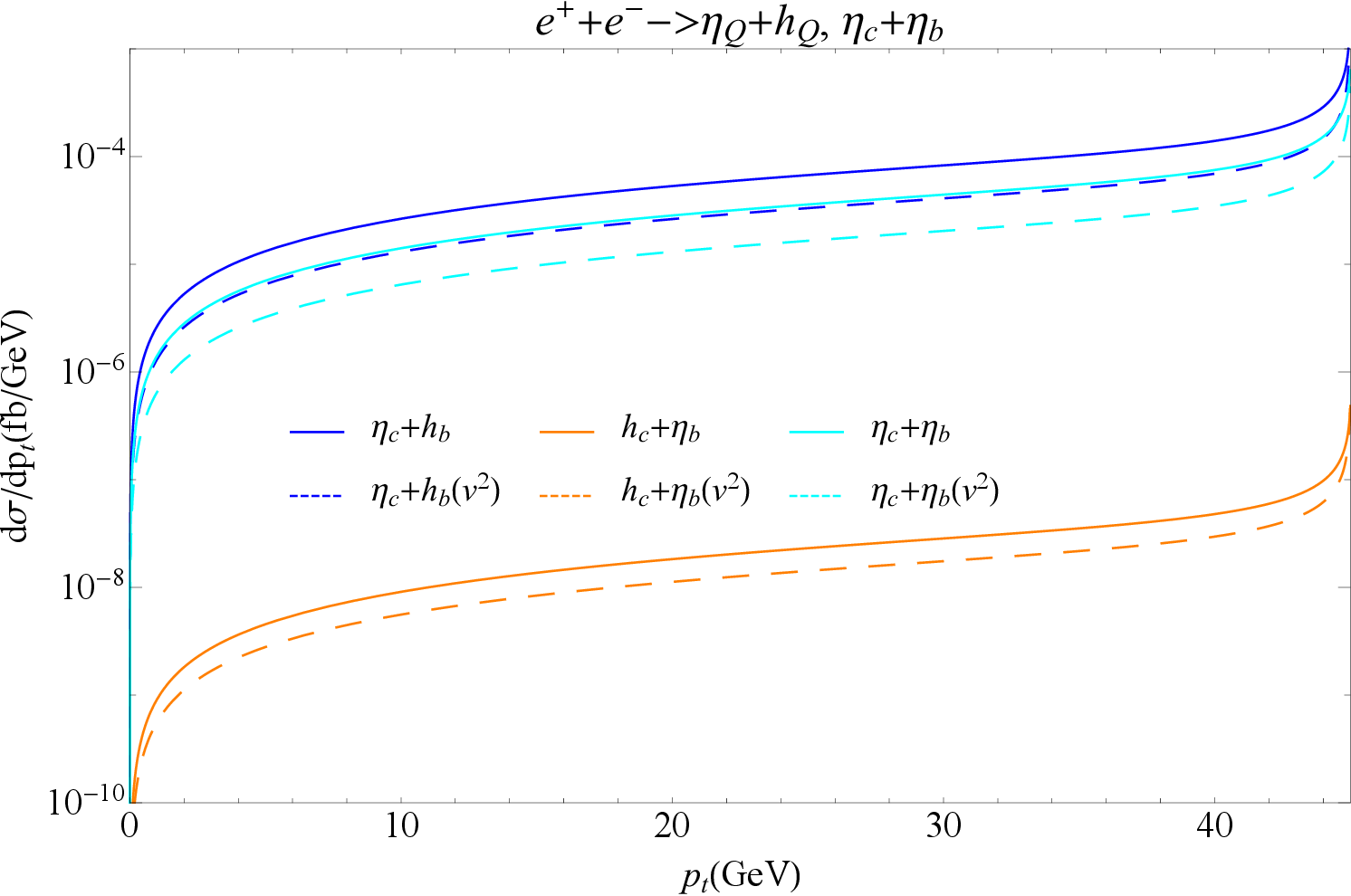}
		\end{tabular}
			\caption{(Color online) The differential cross section $d\sigma/dp_t$   at $\sqrt{s}$=$m_Z$. The solid line represents LO  and dashed line represents NLO($v^2$) result. }
		\label{ccpt}
			
		\end{figure*}
	\end{widetext}

	The differential cross section $\frac{d\sigma}{dp_t}$ can be written as below.
	\bea
	\frac{d\sigma}{dp_t}=|\frac{d\cos\theta}{dp_t}|(\frac{d\sigma}{d\cos\theta})=\frac{p_t}{|\vec{p}_3|\sqrt{\vec{p}_3^2-p_t^2}}(\frac{d\sigma}{d\cos\theta})
	\label{pt1}
	\eea
	where $\vec{p}_3$ is the momentum of the $H_1$ charmonium, we have
	\bea
	\vec{p}_3=\frac{\sqrt{\lambda[s,(2E_{q3})^2,(2E_{q4})^2]}}{2\sqrt{s}},~ \lambda[x,y,z]=x^2+y^2+z^2-2(xy+xz+yz)
	\label{pt2}
	\eea
	Combining Eqs.~(\ref{pt1}) and (\ref{pt2}), we can also get an $\mathcal{O}(v^2)$ expression:
	\bea
	\frac{d\sigma}{dp_t}=  \{ \frac{4p_ts}{[(A-4p_t^2s)A]^{1/2}}+\frac{32[m_c^2v_{c\bar{c}}^2(4m_b^2-4m_c^2+s)+m_b^2v_{b\bar{b}}^2(4m_c^2-4m_b^2+s)](A-2p_t^2s)p_ts}{[(A-4p_t^2s)A]^{3/2}} \} (\frac{d\sigma}{d\cos\theta})
	+\mathcal{O}(v^4)
	\eea
	with $A=16(m_b^2-m_c^2)^2+s(s-8m_b^2-8m_c^2)$. 
    As shown in Fig.~\ref{ccpt}, the $p_t$ distributions of the CS channels exhibit the same trend as that of the CO channels.

	\begin{table}
		\caption{Production cross sections (units:$\times 10^{-4}$ fb) at $\sqrt{s}=91.1876$ GeV. ``---'' denotes forbidden processes. LO and NLO($v^2$) denote the leading order and next-to-leading order in the $v^2$ expansion, respectively. CS and CO represent the cross sections for the CS channel and the sum of all CO channels, respectively. }
		\begin{tabular}{|c|c|c|c|c| |c|c|c|c|c|}
			\hline
			~	&\multicolumn{2}{c|}{CS} &  \multicolumn{2}{c| |}{CS+CO}	&~	&\multicolumn{2}{c|}{CS} &   \multicolumn{2}{c|}{CS+CO}\\
			\hline
			~&~LO~&NLO($v^2$) & ~LO~&NLO($v^2$)&	~&~LO~&NLO($v^2$) & ~LO~&NLO($v^2$)\\
			\hline
			\hline
			$J/\psi+\Upsilon$& 364.9&141.6& 370.5 & 144.5&	$\eta_c+\eta_b$&---  &--- &38.4 &17.6 \\
			\hline
			$J/\psi+\eta_{b}$& 18.7&9.2& 21.8 & 10.7&	$\eta_c+\Upsilon$& 0.6&0.4& 61.5 & 30.4\\
			\hline
			$J/\psi+\chi_{b0}$& 0.2&0.1& 4.1 & 2.3&	$\chi_{c0}+\Upsilon$& 0.02&0.01&3.9 &1.9 \\
			\hline
			$J/\psi+\chi_{b1}$& 1.0&0.5& 12.8 & 7.3&$\chi_{c1}+\Upsilon$& 0.1&0.07&11.7 &5.8 \\
			\hline
			$J/\psi+\chi_{b2}$& 0.3&0.2& 20.0 & 11.5	&$\chi_{c2}+\Upsilon$& 0.04&0.03&19.3 &9.5 \\
			\hline
			$J/\psi+h_{b}$& 1.0&0.5& 6.2 & 3.0&$h_c+\Upsilon$& 0.4&0.2&0.5 &0.3 \\
			\hline
			$\eta_c+h_{b}$& ---&---&71.4 &35.1&  $h_c+\eta_{b}$&--- &--- &0.02 &0.02 \\
			\hline
			$\eta_c+\chi_{b0}$&  ---&---&22.6 &12.8&  	$\chi_{c0}+\eta_{b}$&---  & ---&2.5 &1.1 \\
			\hline
			$\eta_c+\chi_{b1}$& --- &---&67.8 &38.3& 	$\chi_{c1}+\eta_{b}$& --- &---&7.4 &3.4 \\
			\hline
			$\eta_c+\chi_{b2}$& ---& ---&113.0 &63.8&  	$\chi_{c2}+\eta_b$&--- &--- &12.4 &5.7 \\
			\hline
	
		\end{tabular}
		\label{TCS1}
	\end{table}

 	\begin{widetext}
		\begin{figure*}[htbp]
		\begin{tabular}{c c c}
			\includegraphics[width=0.333\textwidth]{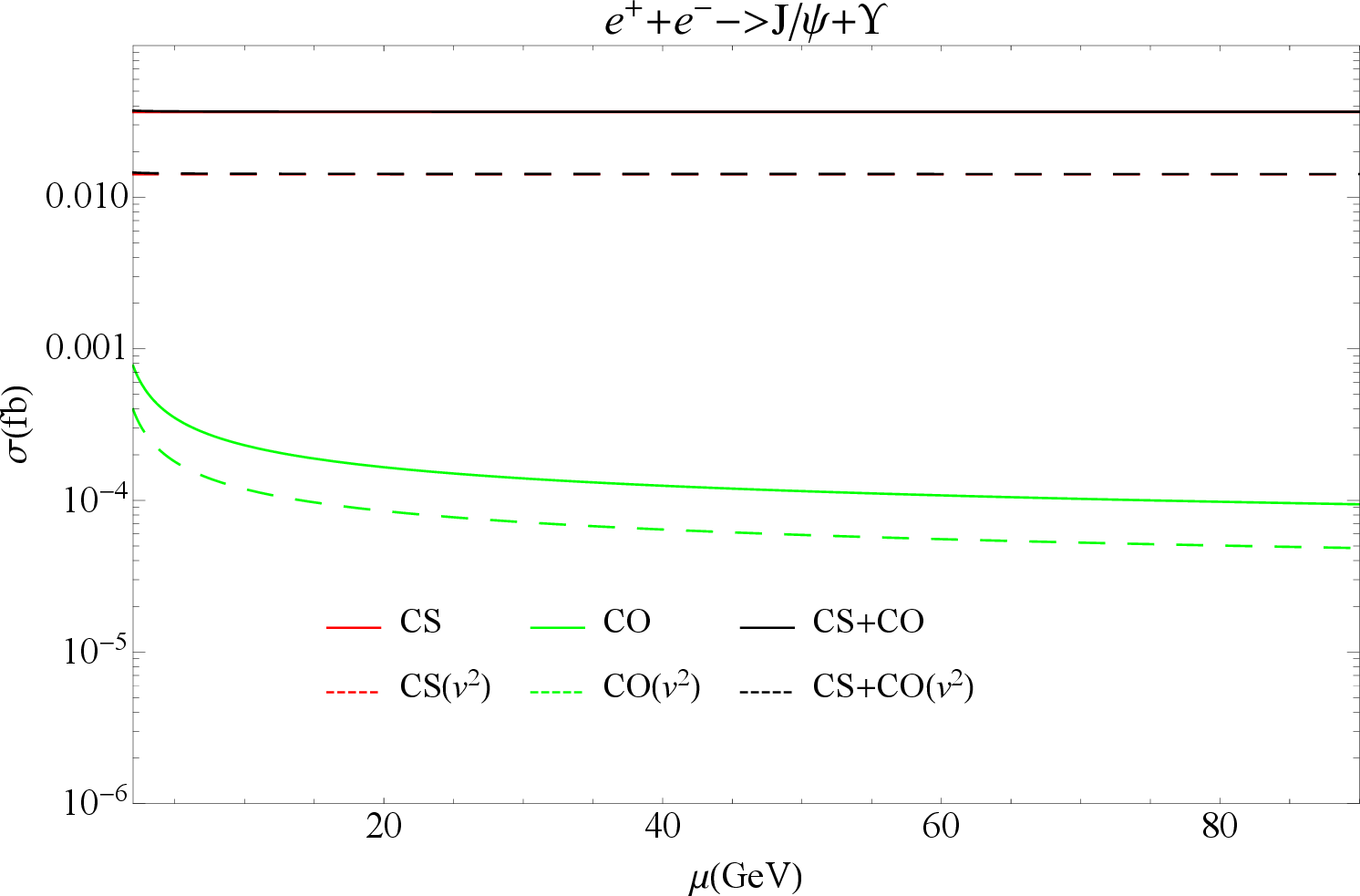}
			\includegraphics[width=0.333\textwidth]{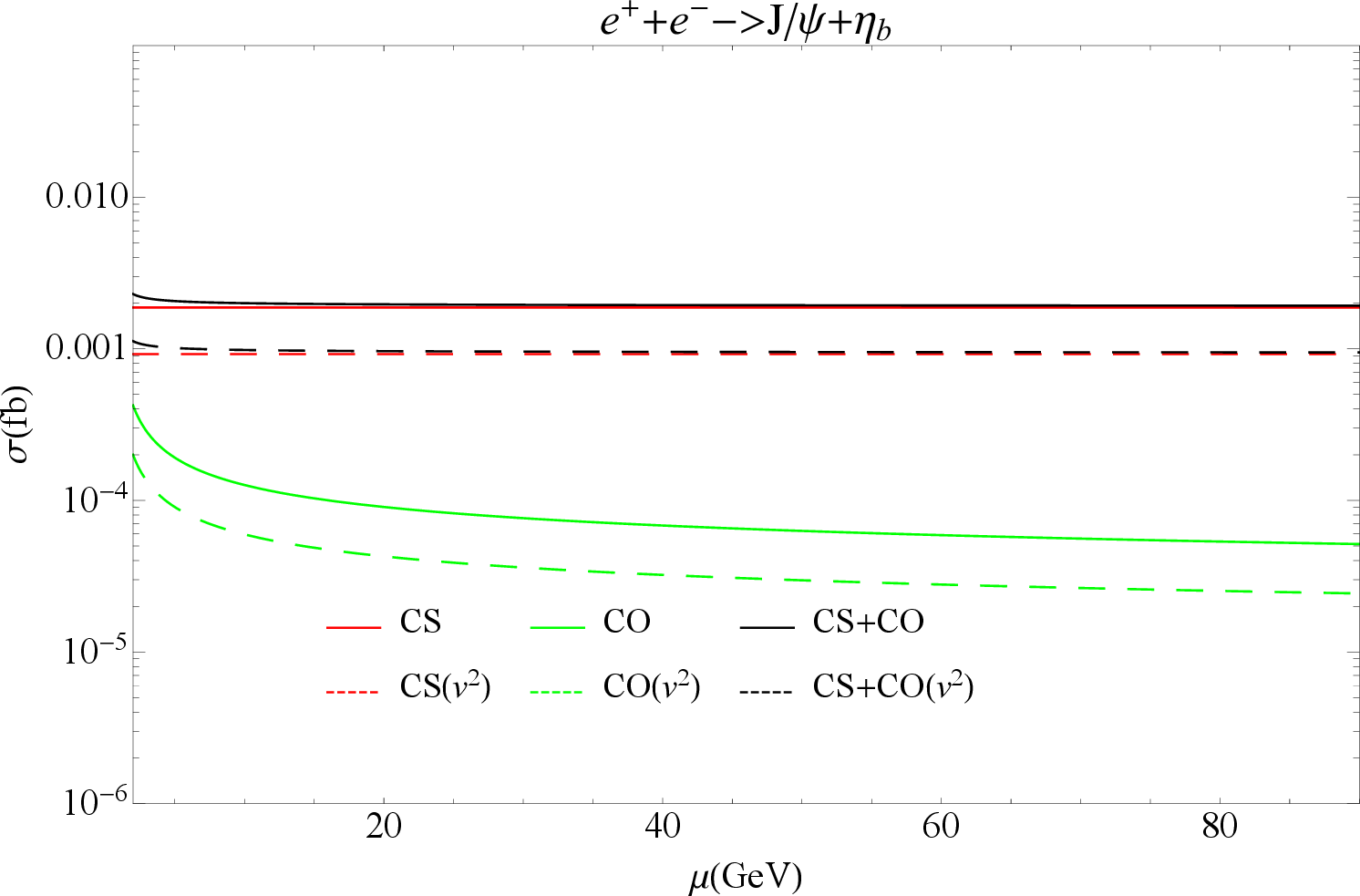}
			\includegraphics[width=0.333\textwidth]{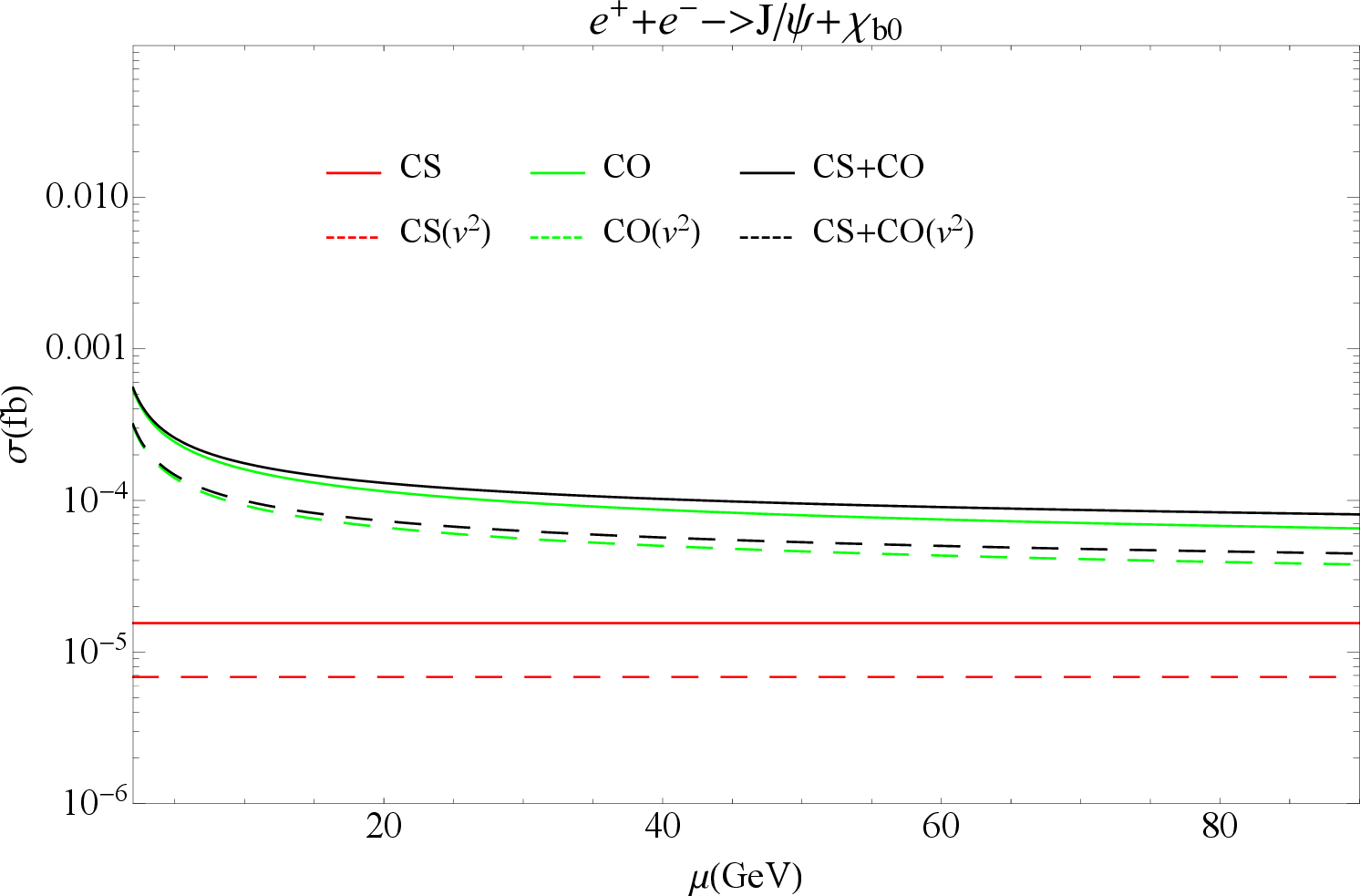}
		\end{tabular}
		\begin{tabular}{c c c}						
			\includegraphics[width=0.333\textwidth]{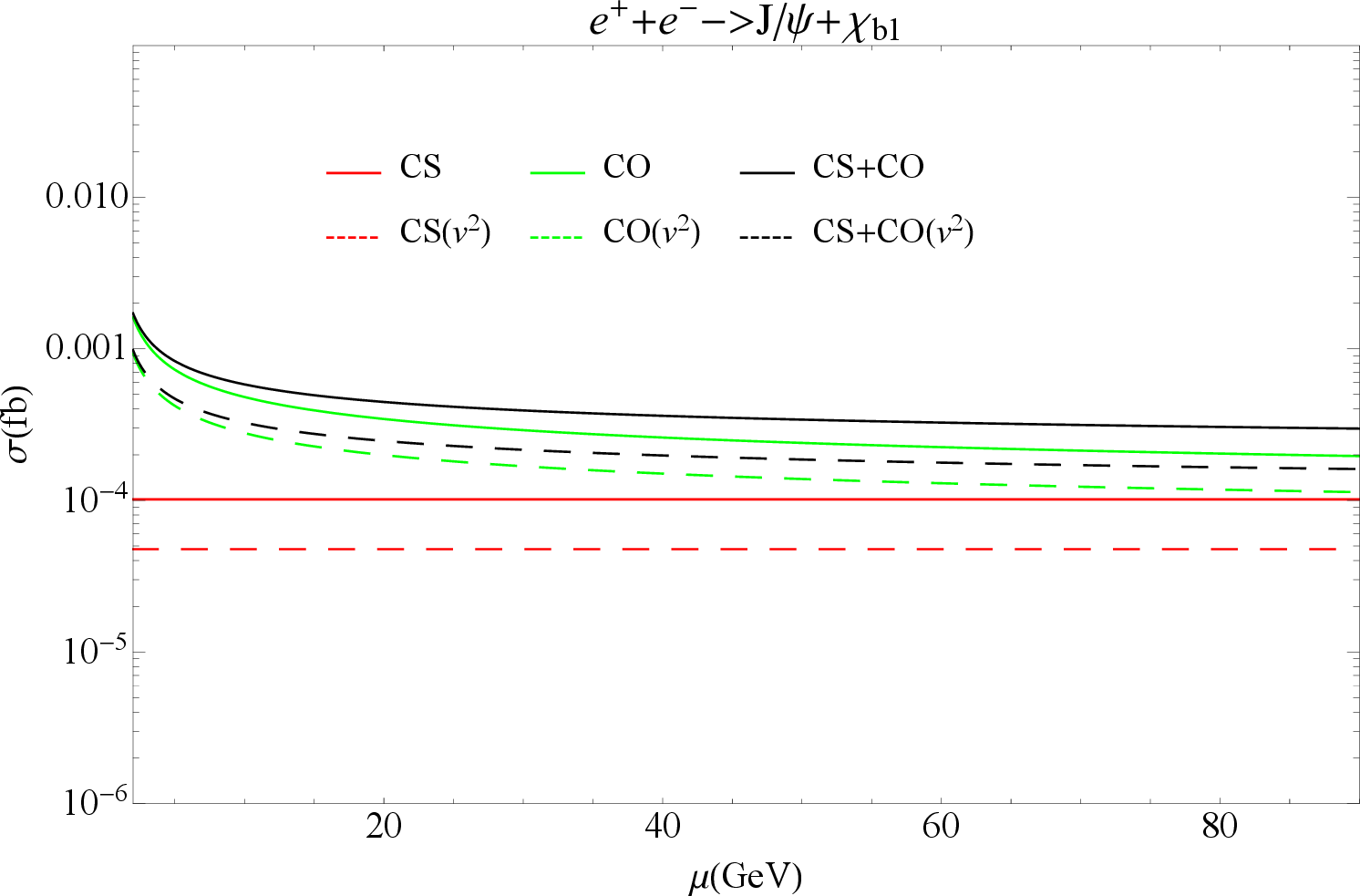}
			\includegraphics[width=0.333\textwidth]{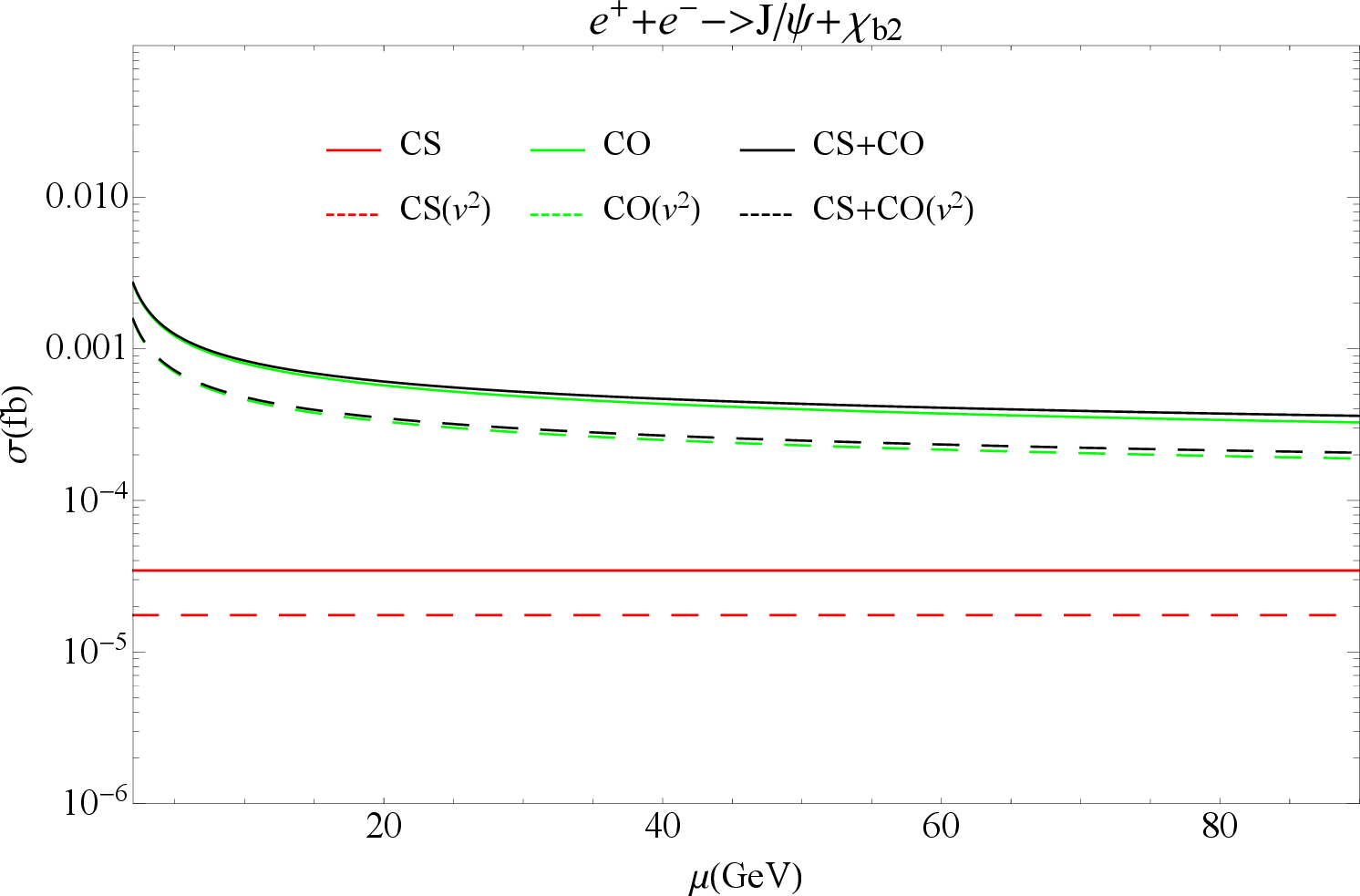}
			\includegraphics[width=0.333\textwidth]{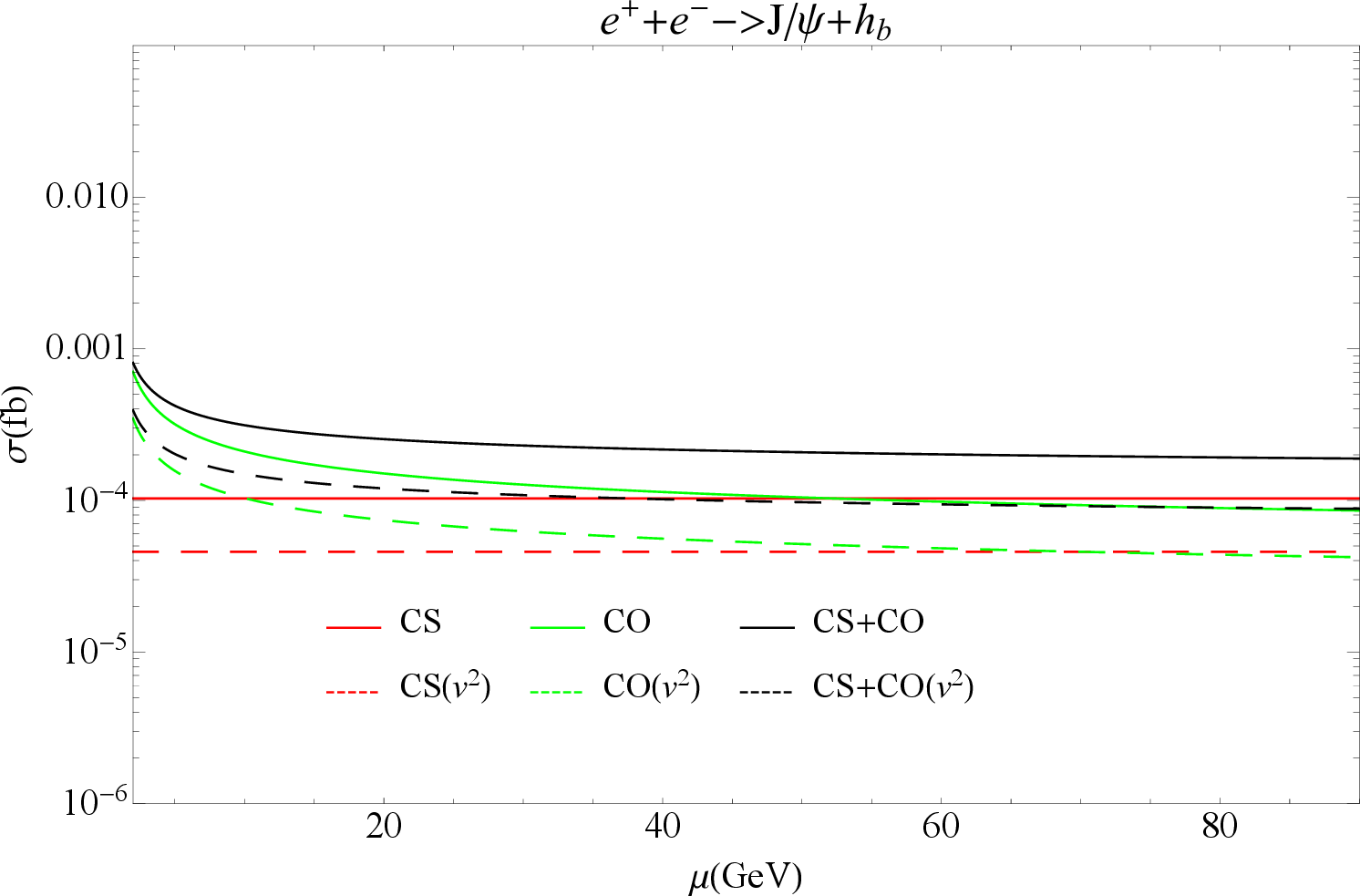}
		\end{tabular}
		
		\begin{tabular}{c c c }
			
			\includegraphics[width=0.333\textwidth]{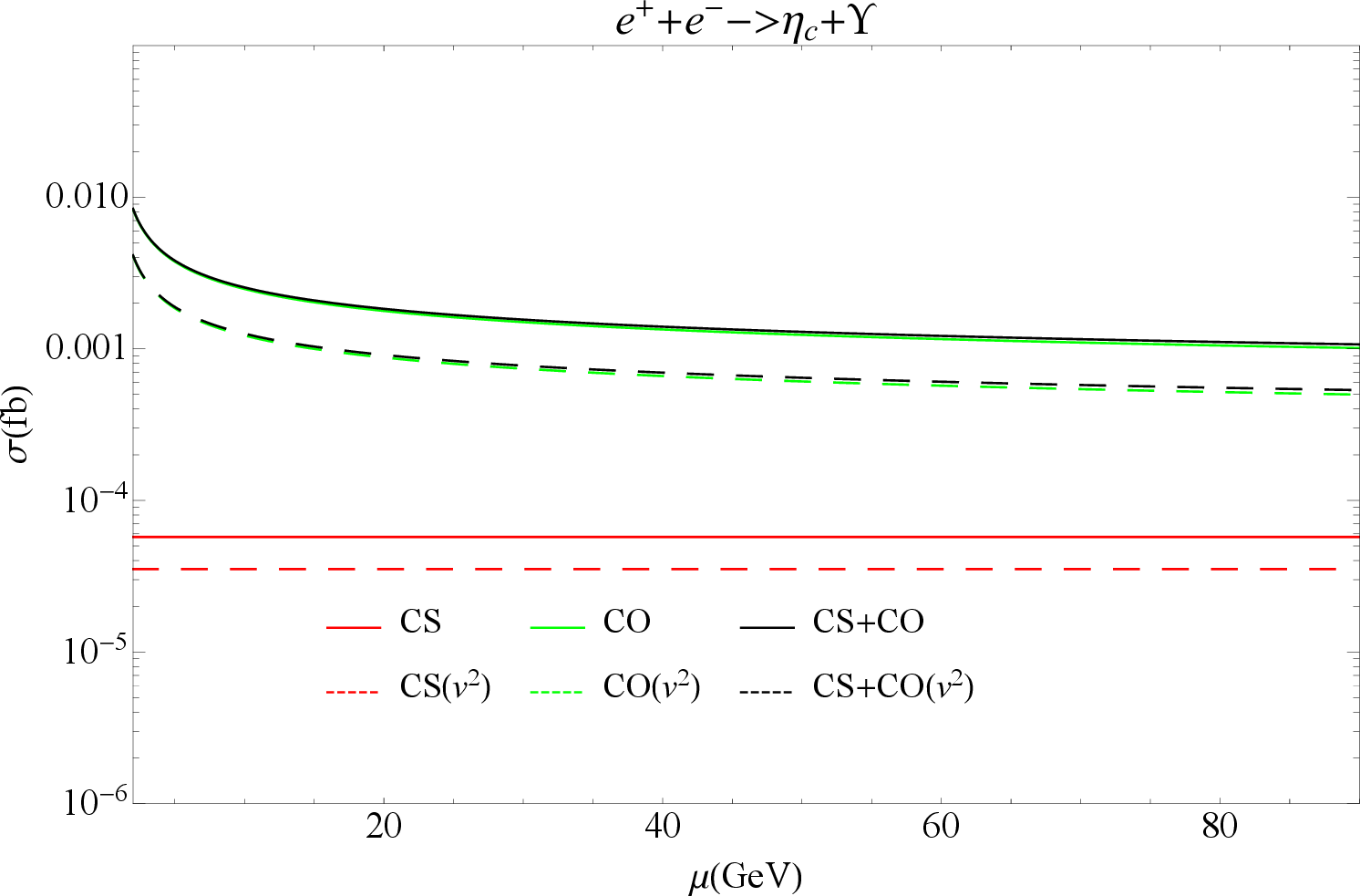}
			\includegraphics[width=0.333\textwidth]{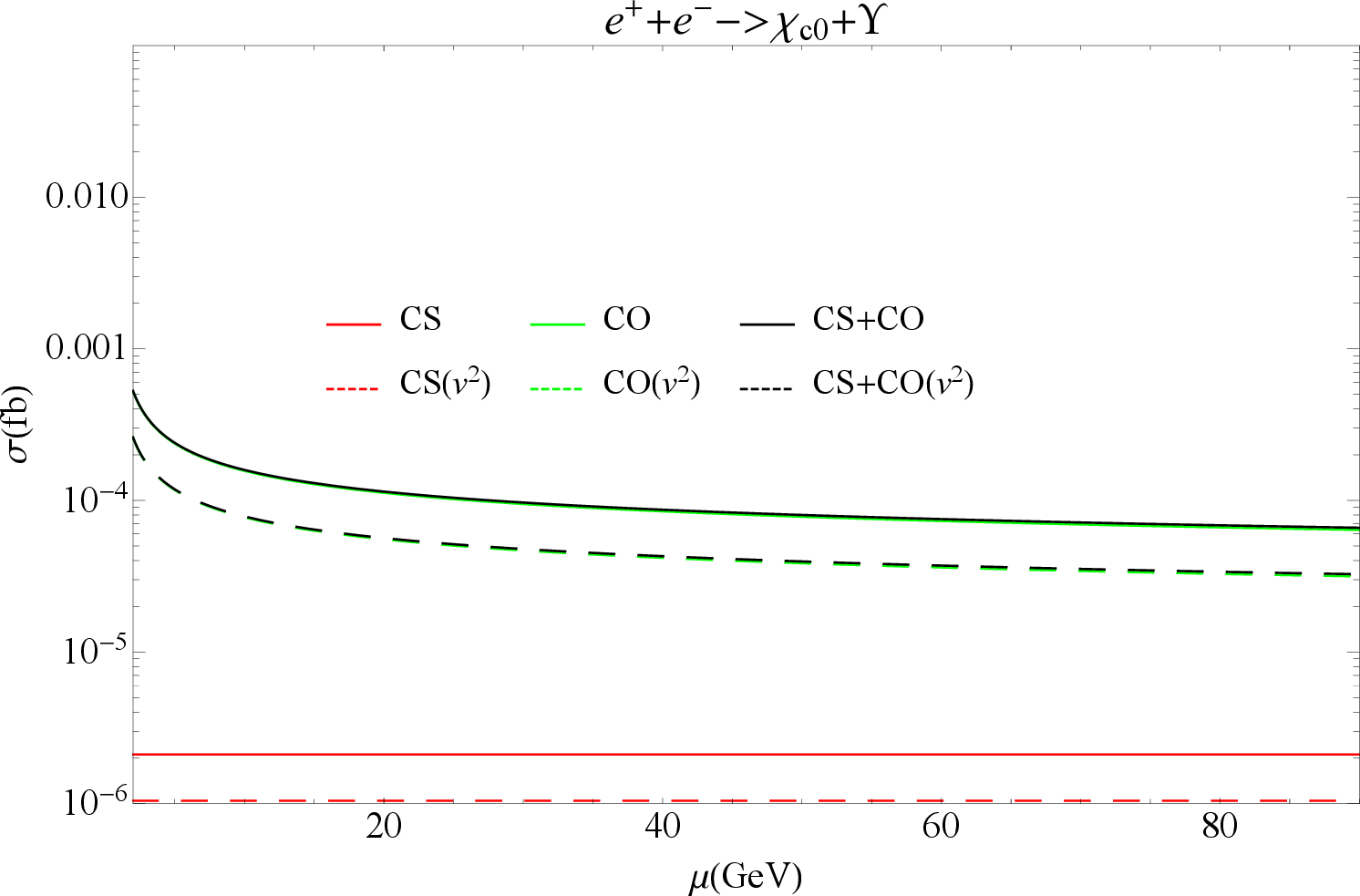}
			\includegraphics[width=0.333\textwidth]{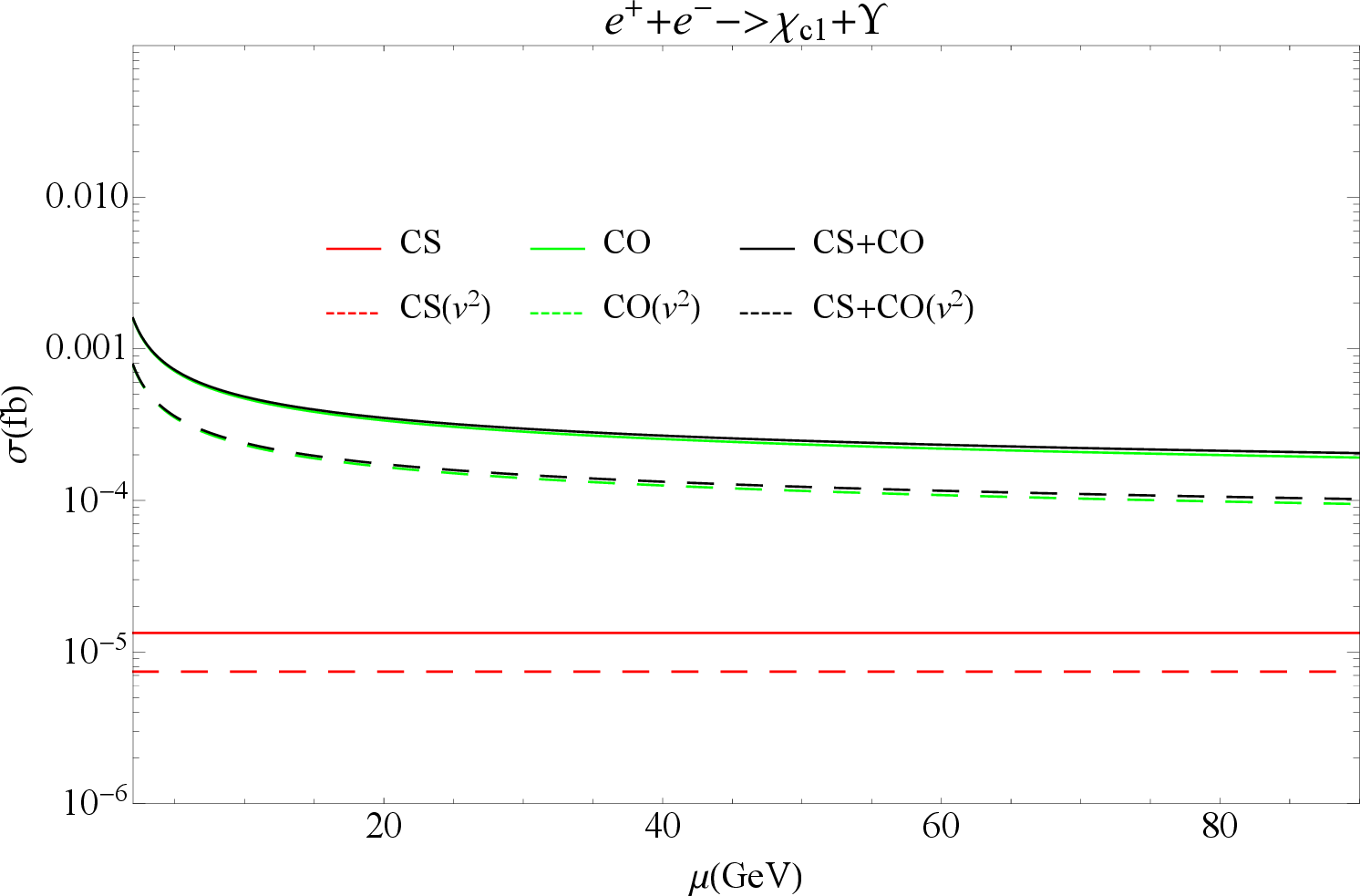}
		\end{tabular}
		\begin{tabular}{c c c }
			\includegraphics[width=0.333\textwidth]{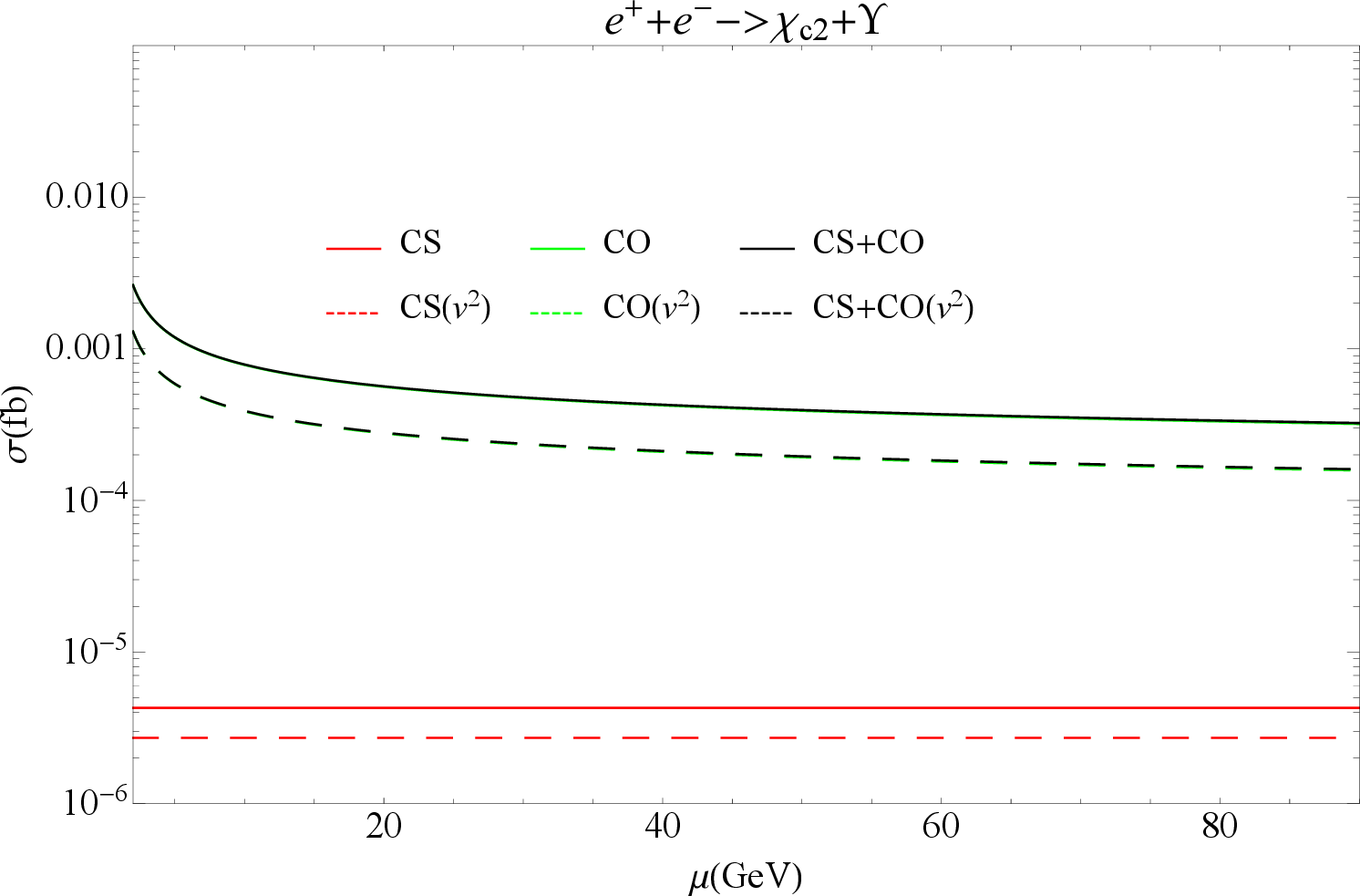}
			\includegraphics[width=0.333\textwidth]{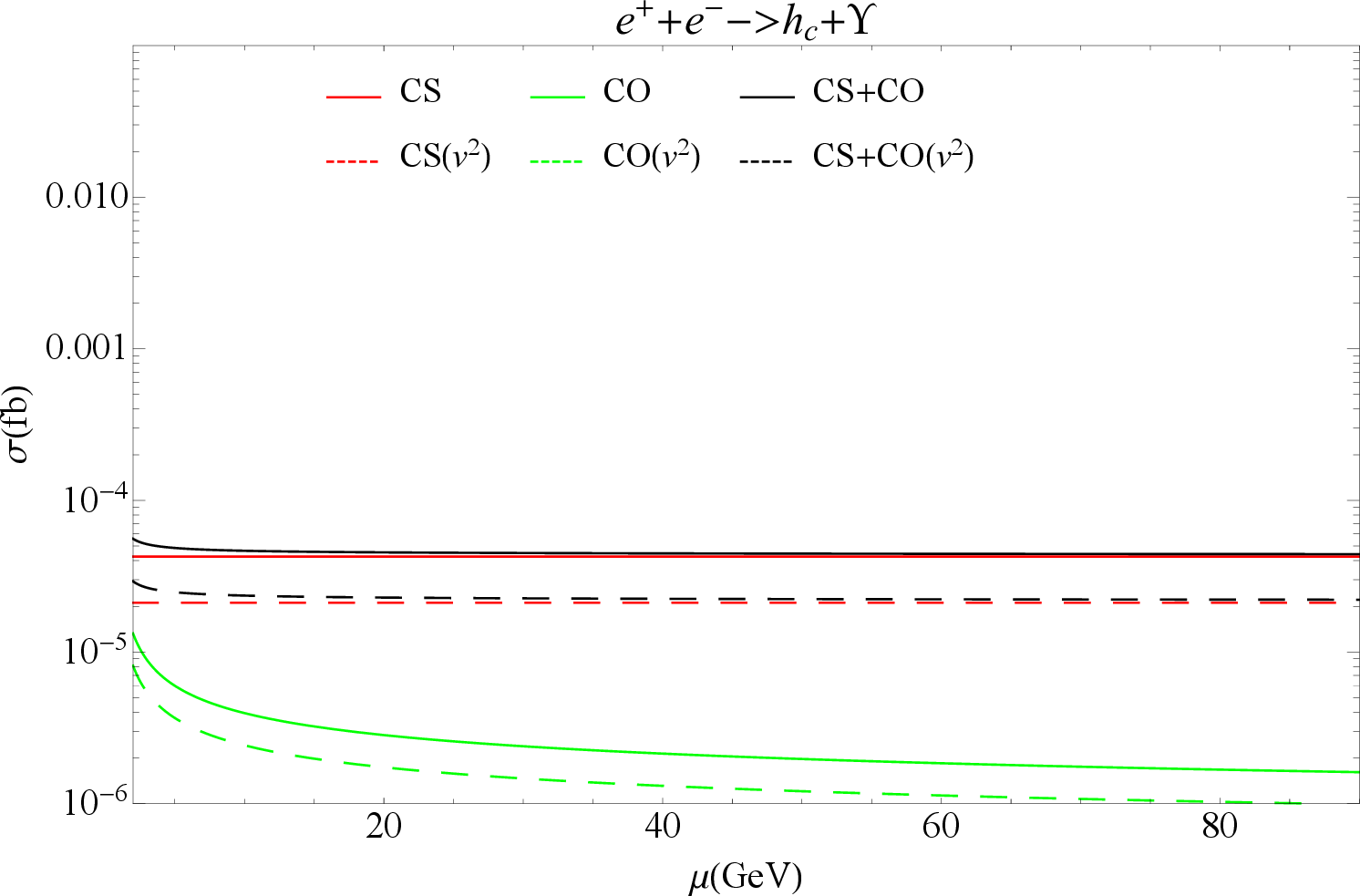}
			\includegraphics[width=0.333\textwidth]{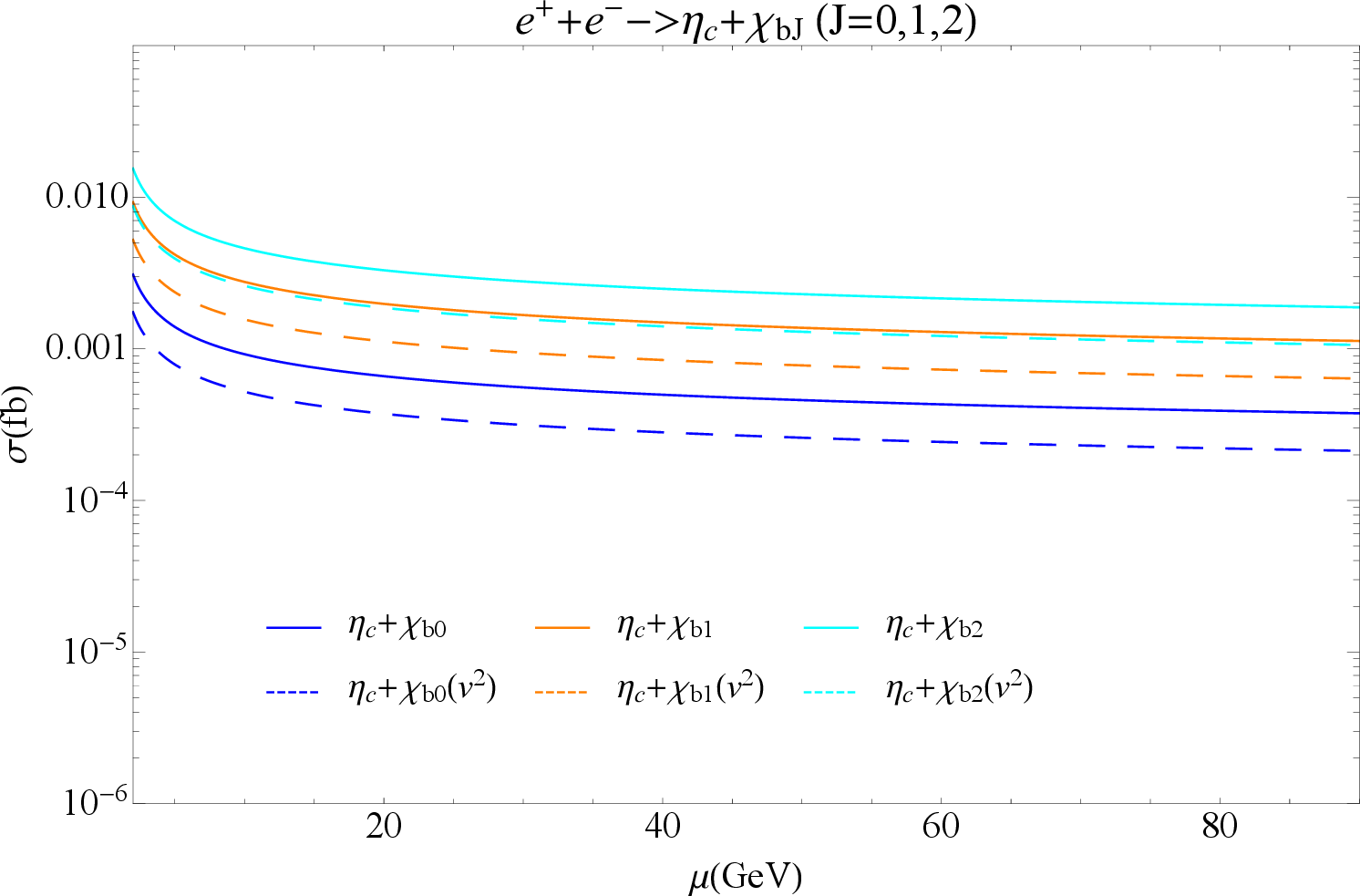}
		\end{tabular}
		\begin{tabular}{c c c }		
			\includegraphics[width=0.333\textwidth]{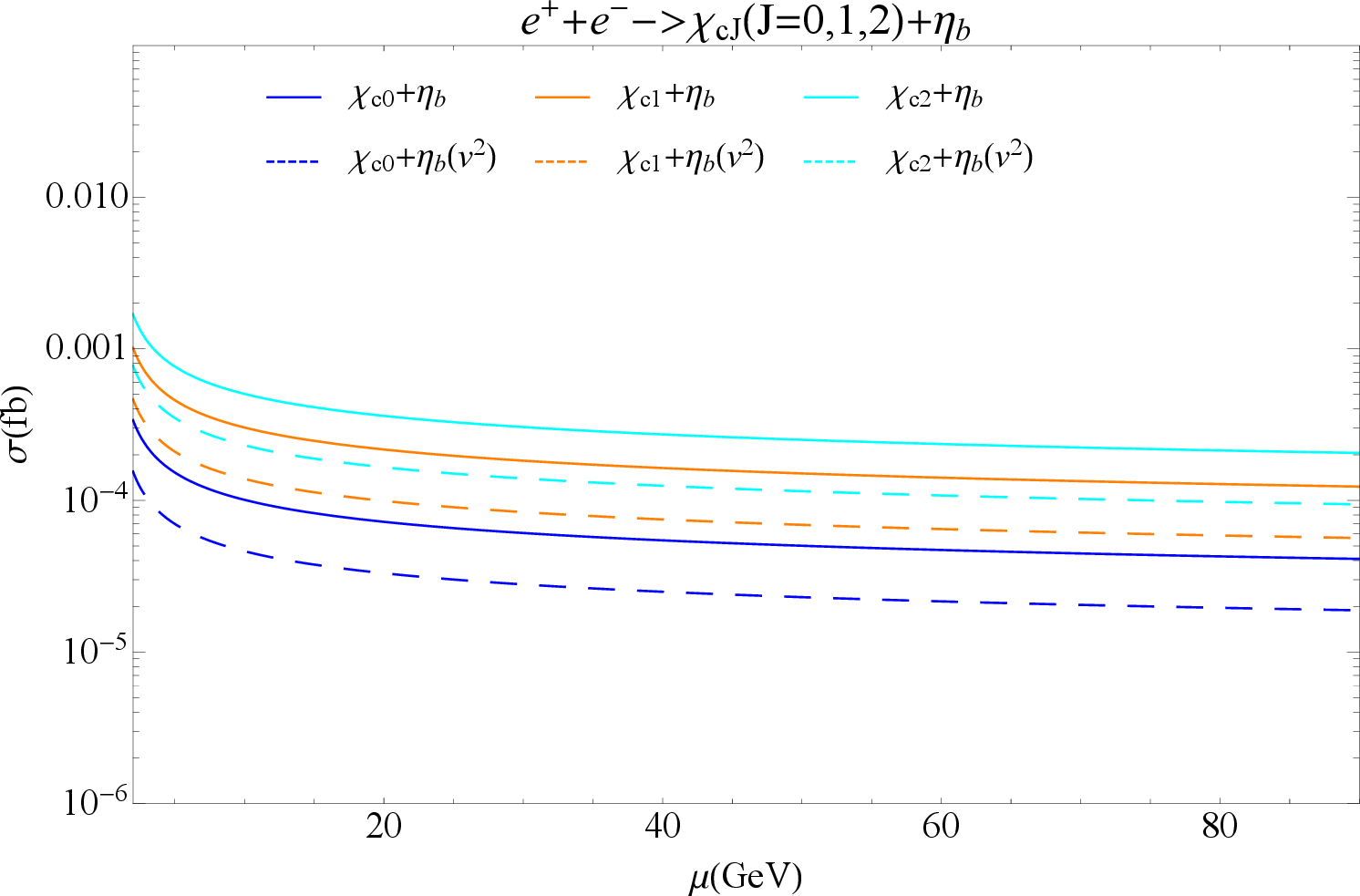}
			\includegraphics[width=0.333\textwidth]{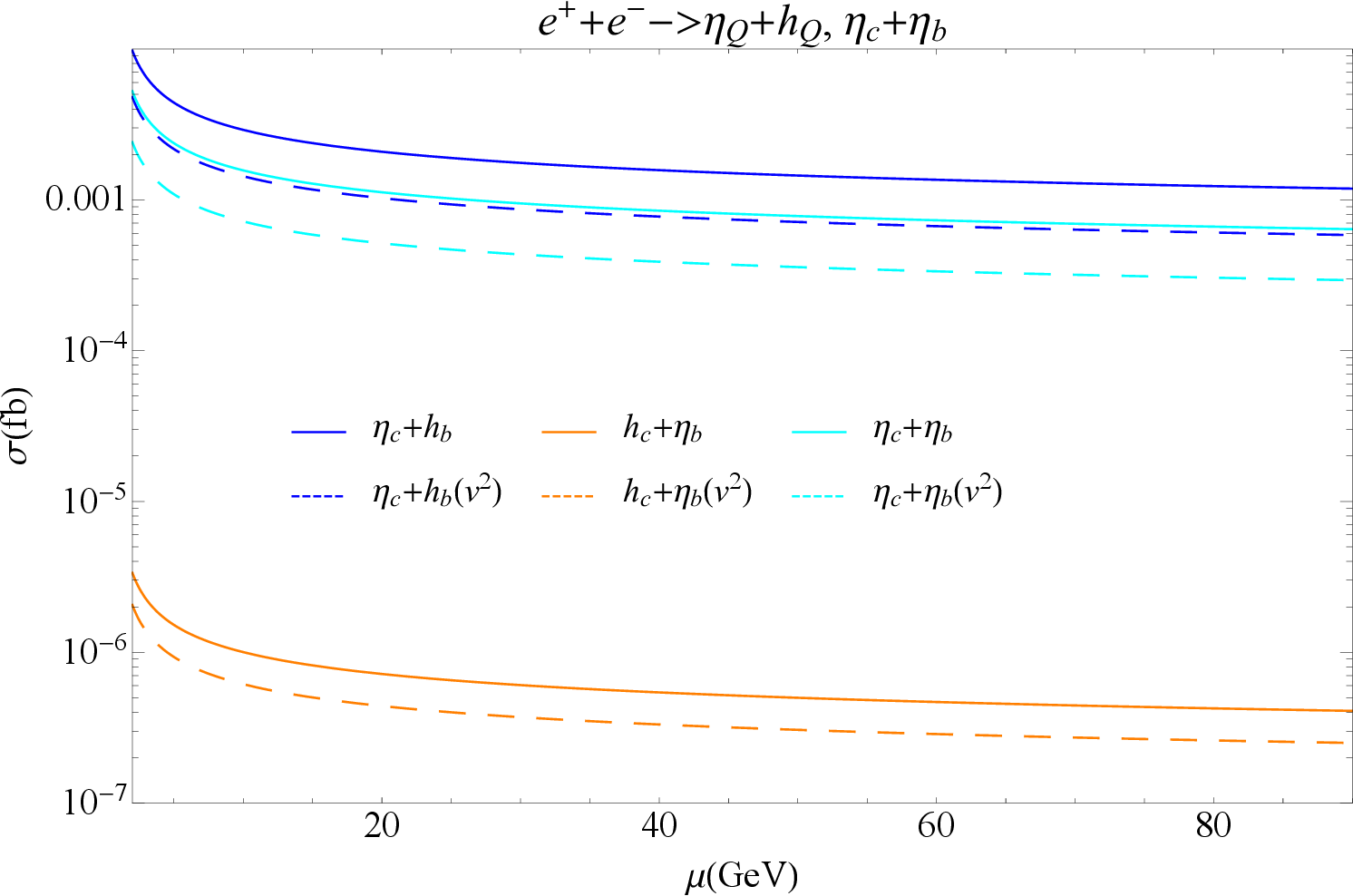}
		\end{tabular}
				\caption{(Color online) Cross section $\sigma$ versus renormalization scale $\mu$ at $\sqrt{s}$=$m_Z$. The solid line represents LO  and dashed line represents NLO($v^2$) result.   }
			\label{z0ccmu}
		\end{figure*}
	\end{widetext}

		\subsection{Uncertainties}
	
In line with previous work \cite{Wang:2025sbx}, we investigate the uncertainties arise from renormalization scale, the deviation of the collision energy from $m_Z$, heavy quark mass, and the LDMEs. 

    The uncertainties induced by the renormalization scale $\mu$ are shown in Fig.~\ref{z0ccmu}, for which we adopt the two-loop expression of the strong running coupling constant $\alpha_s$. We have $\frac{\alpha_s(\mu)}{4\pi}=\frac{1}{\beta_0L}-\frac{\beta_1\ln L}{\beta_0^3L^2}$, 
	where $L=\ln(\mu^2/\Lambda^2_{QCD})$ with $\Lambda_{QCD}\simeq338MeV$. $\beta_0=(11/3)C_A-(4/3)T_fn_f$ and $\beta_1=(34/3)C_A^2-4C_FT_fn_f-(20/3)C_AT_f$ are the one-loop and two-loop coefficients of the QCD beta function, respectively. The number of active quark flavors $n_f$ is set to 3. As observed from the figure, the renormalization scale dependence is large when $\mu<20GeV$ and becomes small as $\mu$ continues to increase.
	
	In Table~\ref{ccmz}, we show the sensitivity of the total cross sections to the collision energy around the $Z^0$ peak. Here, we estimate the uncertainties in the cross sections by assuming a $3\%$ deviation of c.m. energy at the Z factory with respect to the Z pole. This analysis method is adopted from Ref.~\cite{Sun:2013liv}.
	The total cross sections are reduced to $15-20\%$ of the peak values, a result consistent with Tables VIII, VX in Ref.~\cite{Sun:2013liv} and Tables VII, VIII in Ref.~\cite{Wang:2025sbx}.
	
	Taking a $\pm0.15 ~GeV$ variation in $m_Q$ ($m_c=(1.5\pm0.15) ~GeV, m_b=(4.7\pm0.15)~ GeV$ for $S$-wave heavy quarkonia, and $m_c=(1.75\pm0.15) ~GeV, m_b=(4.94\pm0.15)~ GeV$ for $P$-wave heavy quarkonia), the associated uncertainties from the heavy quark mass are summarized in Table~\ref{ccmc}. Regarding the total cross sections, the uncertainties are $12-15\%$ for $h_c+\Upsilon/\eta_b$, $16-22\%$ for $J/\psi+\chi_{bJ}$, $21-31\%$ for $\eta_Q+\chi_{Q'J}$ and $\chi_{cJ}+\Upsilon$, $25-37\%$ for $J/\psi+\eta_b/h_b$ and $\eta_c+h_b/\eta_b$, $27-40\%$ for $J/\psi+\Upsilon$, and $36-46\%$ for $\eta_c+\Upsilon$ production.

\begin{widetext}
	\begin{figure*}[htbp]
		\begin{tabular}{c c c }
			\includegraphics[width=0.4\textwidth]{ 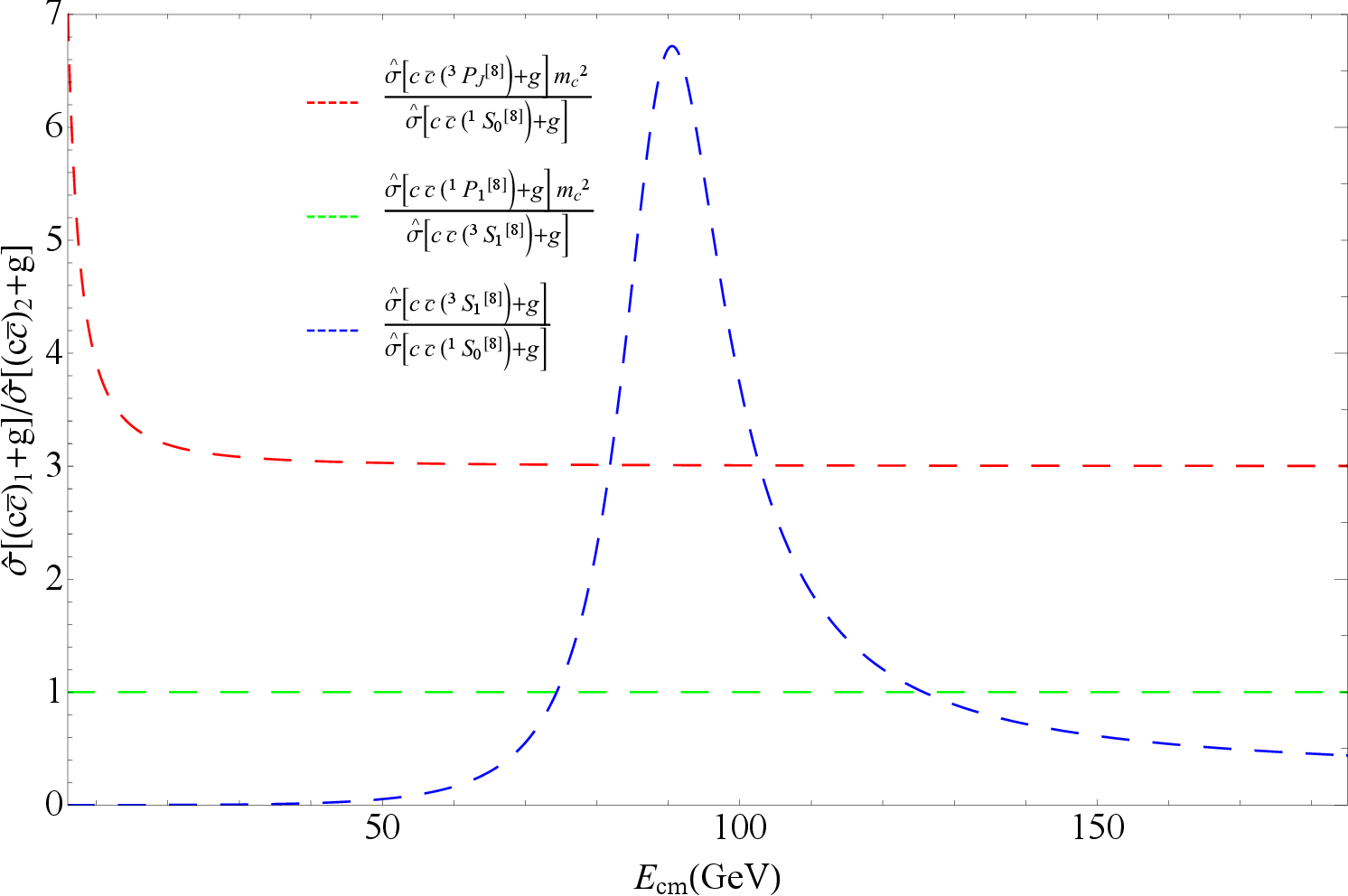} 
				\includegraphics[width=0.4\textwidth]{ 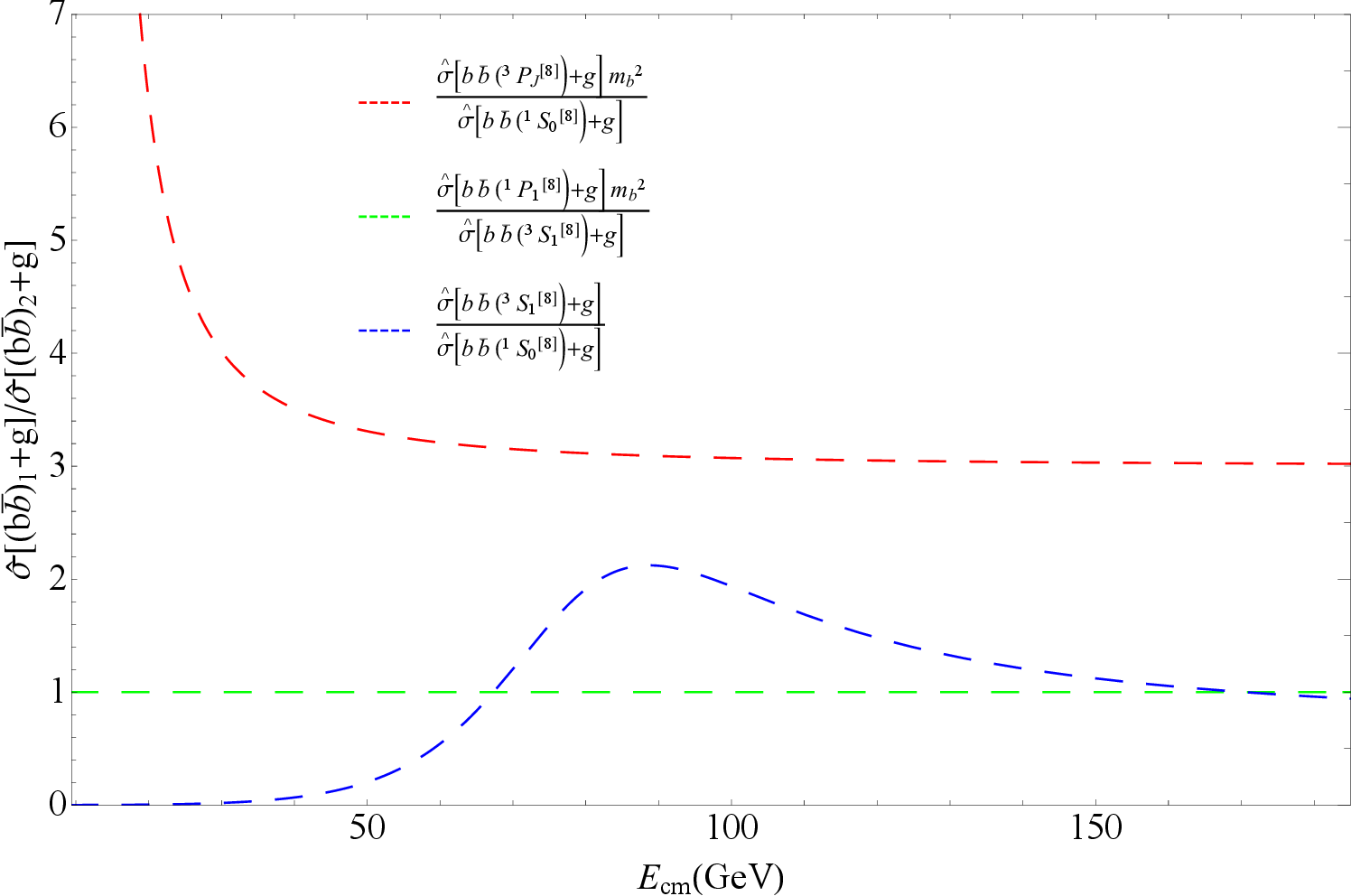} 
		\end{tabular}
		\caption{ (Color online) The ratios of the short-distance cross sections of different production channels for the charm quark pair(left) and the bottom quark pair(right). }
		\label{fraSDCS}
	\end{figure*}
\end{widetext}

\begin{widetext}
	\begin{figure*}[htbp]
		\begin{tabular}{c c}
			\includegraphics[width=0.45\textwidth]{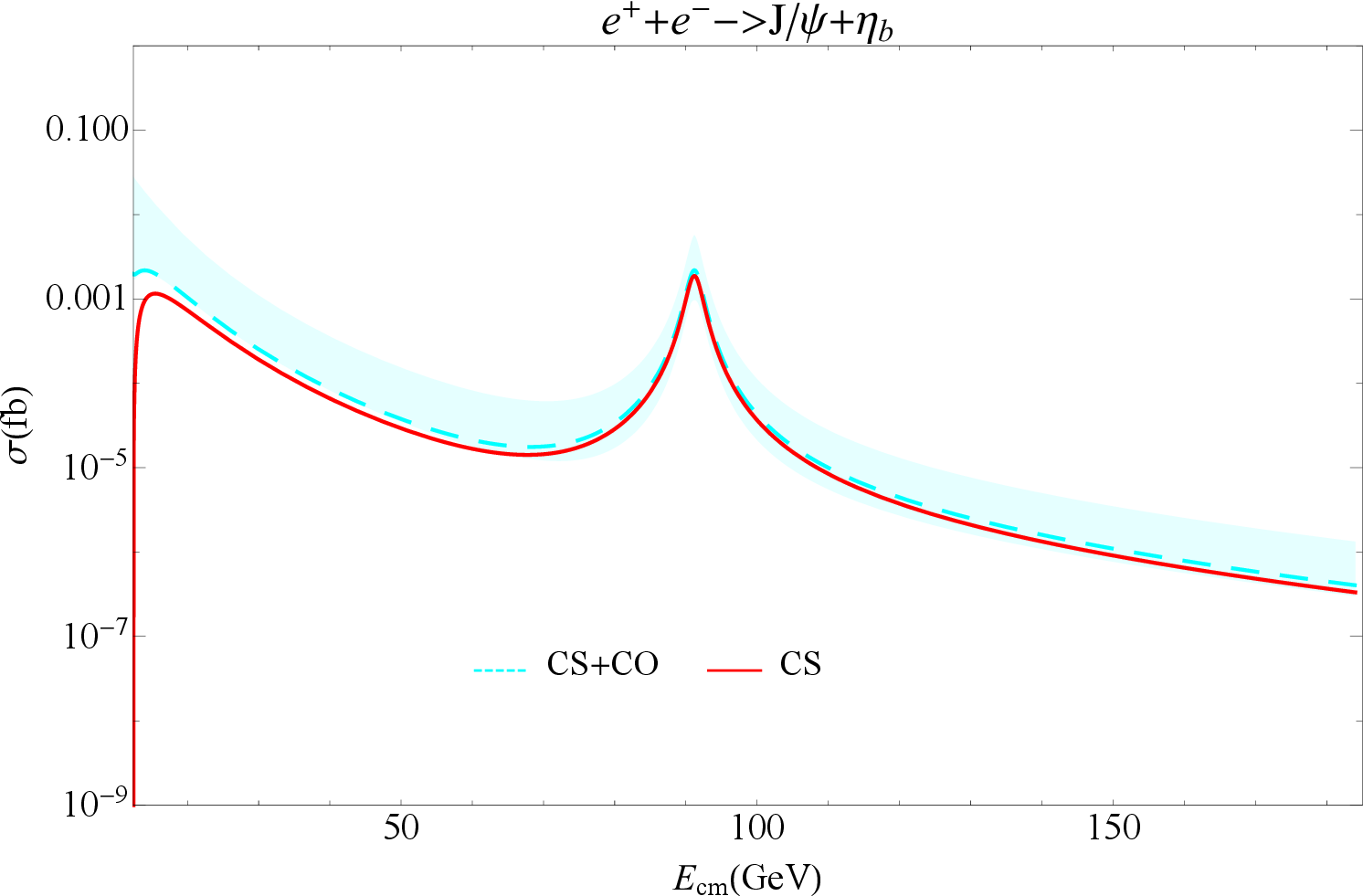}
			\includegraphics[width=0.45\textwidth]{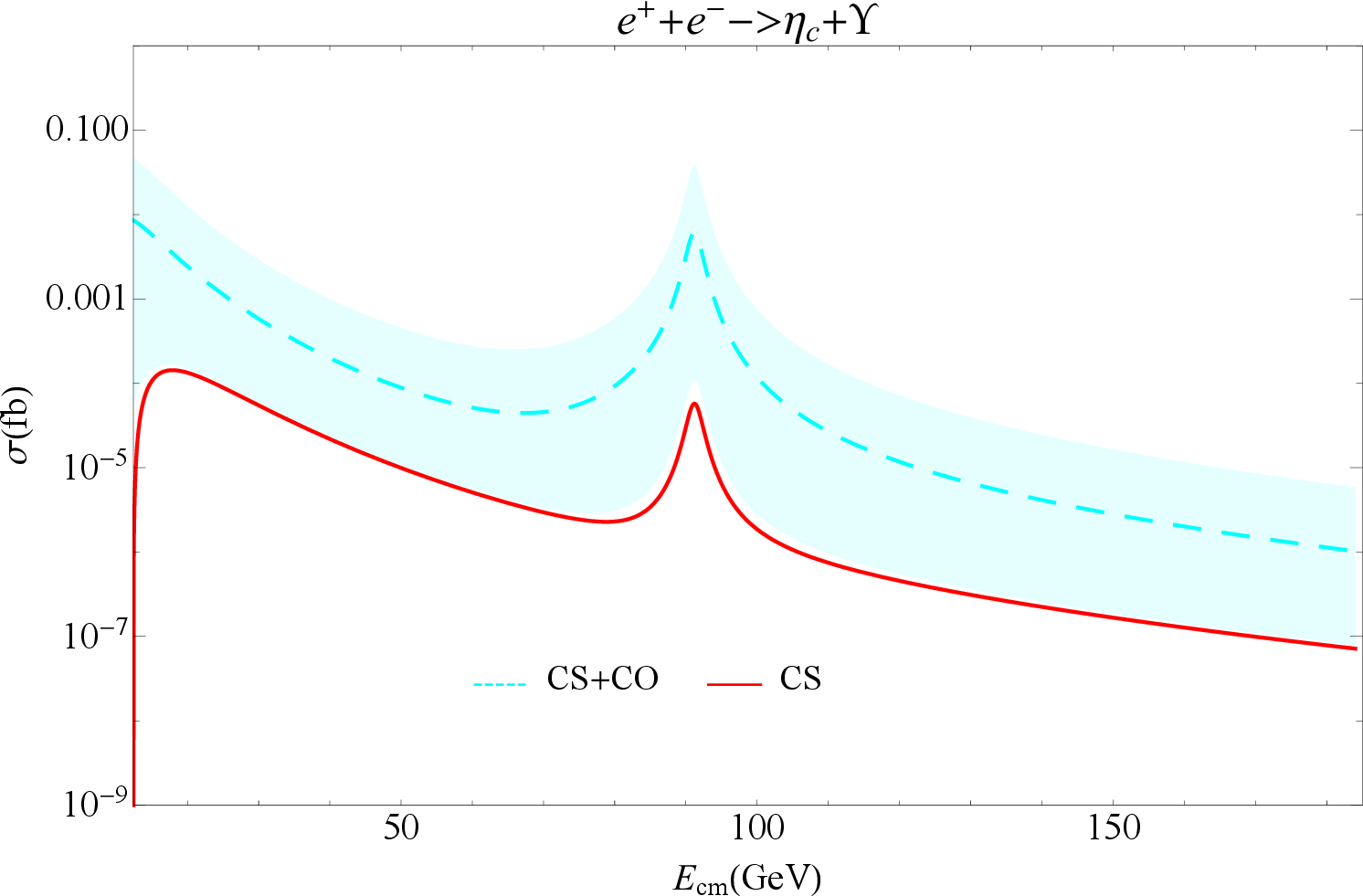}
		\end{tabular}
		\caption{
(Color online) The $J/\psi+\eta_b$ (left) and $\eta_c+\Upsilon$ (right) production including uncertainties from color-octet (CO) long-distance matrix elements (LDMEs).
The red solid line denotes color-singlet (CS) cross sections, and the cyan dashed line shows our CS+CO results.
The uncertainty band is obtained from different CO LDME combinations.
For $J/\psi+\eta_b$, LDMEs for the upper limit are taken from Ref.~\cite{Chao:2012iv,Braaten:2000cm}, and those for the lower limit from Ref.~\cite{Gong:2012ug,Sharma:2012dy}.
For $\eta_c+\Upsilon$, LDMEs for the upper and lower limit are from Ref.~\cite{Chao:2012iv,Braaten:2000cm,Chao:2012iv,Feng:2015wka}.
}
		\label{uncertldmes}
	\end{figure*}
\end{widetext}

Uncertainties arising from LDMEs are difficult to quantify due to the large uncertainties in their own values. These values are extracted from experimental data; yet significant discrepancies exist among the results from various studies\cite{Bodwin:1992qr,Braaten:1994vv,Butenschoen:2011yh,Chao:2012iv,Gong:2012ug,Bodwin:2014gia,Braaten:1999qk,Fleming:1998md,Braaten:1996pv,Beneke:1998ks,Jia:2012qx,Ma:2010vd,Zhang:2014coi,Beneke:1996yw,Shao:2014fca}, while the values adopted in the present work are relatively moderate. Furthermore, the production processes generally involve several LDMEs. For example, three relevant LDMEs are involved for $S$-wave mesons ($J/\psi$, $\eta_c$, $\Upsilon$, $\eta_b$). For these two reasons, the uncertainties arising from LDMEs are expected to be quite large. 
We illustrate this using the two processes $J/\psi+\eta_b$, $\eta_c+\Upsilon$ with the LDMEs uncertainties.
For the processes $J/\psi+\eta_b$ (dominated by CS) and $\eta_c+\Upsilon$ (dominated by CO), each involves six LDMEs. 
Considering the constant ratio relations for short-distance cross sections in the high-energy region for CO channels\cite{Chen:2013mjb,Liao:2021ifc,Wang:2025sbx} (as illustrated in Fig.~\ref{fraSDCS}), we have:
\bea
\left. \frac{\hat{\sigma}[Q\bar{Q}({}^{3}P_J^{[8]}) + g]m_Q^2}{\hat{\sigma}[Q\bar{Q}({}^{1}S_0^{[8]}) + g]}\right|_{m_Q\ll E_{\text{cm}}}=3\nonumber\\
\left. \frac{\hat{\sigma}[Q\bar{Q}({}^{1}P_1^{[8]}) + g]m_Q^2}{\hat{\sigma}[Q\bar{Q}({}^{3}S_1^{[8]}) + g]}\right|_{m_Q\ll E_{\text{cm}}}=1
\label{relations}
\eea
Consequently, the number of independent LDMEs is reduced to four. In Fig.~\ref{uncertldmes}, we plot the upper and lower variations in CO LDME values adopted from different works. We find that for both processes, the lower edge of the uncertainty band for the total cross sections approaches the CS level.

 	\begin{table}
 	\caption{Total cross sections ($units: \times10^{-4}fb$) up to $\mathcal{O}(v^2)$ for the charmonia+bottomonia production, with varying values of $E_{cm}$. ``---'' denotes forbidden processes. The ratios $R_{\pm}$, with definitions as $R_-=\frac{\sigma(E_{cm}=97\%m_Z)}{\sigma(E_{cm}=m_Z)}, R_+=\frac{\sigma(E_{cm}=103\%m_Z)}{\sigma(E_{cm}=m_Z)}$, characterize the variation of the cross sections with $E_{cm}$. 
 	}
 	\begin{tabular}{|c|c|c|c| |c|c|c| |c|c|c|}
 		\hline
 		~	&\multicolumn{3}{c| |}{CS } & \multicolumn{3}{c| |}{CO } & \multicolumn{3}{c|}{Total }\\
 		\hline
 		\hline
 		~&$97\%m_Z$&$103\%m_Z$&\Bigg(\makecell{$R_-$\\$R_+$}\Bigg)&$97\%m_Z$&$103\%m_Z$&\Bigg(\makecell{$R_-$\\$R_+$}\Bigg)&$97\%m_Z$&$103\%m_Z$&\Bigg(\makecell{$R_-$\\$R_+$}\Bigg)\\
 		\hline
 		\hline
 		$J/\psi+\Upsilon$& 128.0&113.6&\Bigg(\makecell{$90\%$\\$80\%$}\Bigg) &0.5&0.5 &\Bigg(\makecell{$18\%$\\$18\%$}\Bigg)&128.5 &114.1&\Bigg(\makecell{$89\%$\\$79\%$}\Bigg) \\
 		\hline
 		$J/\psi+\eta_b$& 1.6&1.6&\Bigg(\makecell{$18\%$\\$17\%$}\Bigg) &0.3&0.3 &\Bigg(\makecell{$18\%$\\$17\%$}\Bigg)&1.9 &~1.8~&\Bigg(\makecell{$18\%$\\$17\%$}\Bigg) \\
 		\hline
 		$J/\psi+\chi_{b0}$& 0.01&0.01&\Bigg(\makecell{$17\%$\\$17\%$}\Bigg) &0.4&0.4 &\Bigg(\makecell{$18\%$\\$18\%$}\Bigg)&0.4 &0.4&\Bigg(\makecell{$18\%$\\$18\%$}\Bigg) \\
 		\hline
 		$J/\psi+\chi_{b1}$& 0.08&0.08&\Bigg(\makecell{$18\%$\\$17\%$}\Bigg) &1.2&1.2 &\Bigg(\makecell{$18\%$\\$18\%$}\Bigg)&1.3 &1.3&\Bigg(\makecell{$18\%$\\$18\%$}\Bigg) \\
 		\hline
 		$J/\psi+\chi_{b2}$& 0.03&0.03&\Bigg(\makecell{$18\%$\\$18\%$}\Bigg) &~2.1~&2.1 &\Bigg(\makecell{$18\%$\\$18\%$}\Bigg)&2.1 &~2.1~&\Bigg(\makecell{$18\%$\\$18\%$}\Bigg) \\
 		\hline
 		$J/\psi+h_b$& 0.08&0.08&\Bigg(\makecell{$18\%$\\$17\%$}\Bigg) &~0.4~&0.4 &\Bigg(\makecell{$18\%$\\$17\%$}\Bigg)&0.5 &~0.5~&\Bigg(\makecell{$18\%$\\$17\%$}\Bigg) \\
 		\hline
 		$\eta_c+h_{b}$& --- &--- & --- &~6.2~&6.0 &\Bigg(\makecell{$18\%$\\$17\%$}\Bigg)&6.2 &6.0&\Bigg(\makecell{$18\%$\\$17\%$}\Bigg) \\
 		\hline
 		$\eta_c+\chi_{b0}$&---  & ---& ---&~2.2~&2.1 &\Bigg(\makecell{$18\%$\\$17\%$}\Bigg)&2.2 &~2.1~&\Bigg(\makecell{$18\%$\\$17\%$}\Bigg) \\
 		\hline
 		$\eta_c+\chi_{b1}$& ---& ---& ---&~6.7~&6.4 &\Bigg(\makecell{$18\%$\\$17\%$}\Bigg)&6.7 &~6.4~&\Bigg(\makecell{$18\%$\\$17\%$}\Bigg) \\
 		\hline
 		$\eta_c+\chi_{b2}$& --- & ---&  ---&~11.2~&10.7 &\Bigg(\makecell{$18\%$\\$17\%$}\Bigg)&11.2 &10.7&\Bigg(\makecell{$18\%$\\$17\%$}\Bigg) \\
 		\hline
 		$\eta_c+\eta_b$& ---& ---&--- &3.1 & 3.0 &\Bigg(\makecell{$18\%$\\$17\%$}\Bigg)&3.1 & 3.0&\Bigg(\makecell{$18\%$\\$17\%$}\Bigg) \\
 		\hline
 		$\eta_c+\Upsilon$& 0.06&0.07&\Bigg(\makecell{$18\%$\\$19\%$}\Bigg)&~5.3~&5.1 &\Bigg(\makecell{$18\%$\\$17\%$}\Bigg)&5.3 & 5.2&\Bigg(\makecell{$18\%$\\$17\%$}\Bigg) \\
 		\hline
 			$\chi_{c0}+\Upsilon$& 0.002&0.002&\Bigg(\makecell{$18\%$\\$19\%$}\Bigg)&~0.3~&0.3 &\Bigg(\makecell{$18\%$\\$17\%$}\Bigg)&0.3 & 0.3&\Bigg(\makecell{$18\%$\\$17\%$}\Bigg) \\
 		\hline
 			$\chi_{c1}+\Upsilon$& 0.01&0.01&\Bigg(\makecell{$18\%$\\$19\%$}\Bigg)&~1.0~&1.0 &\Bigg(\makecell{$18\%$\\$17\%$}\Bigg)&1.0 & 1.0&\Bigg(\makecell{$18\%$\\$17\%$}\Bigg) \\
 		\hline
 			$\chi_{c2}+\Upsilon$& 0.005&0.005&\Bigg(\makecell{$18\%$\\$19\%$}\Bigg)&~1.7~&1.6 &\Bigg(\makecell{$18\%$\\$17\%$}\Bigg)&1.7 & 1.6&\Bigg(\makecell{$18\%$\\$17\%$}\Bigg) \\
 		\hline
 			$h_{c}+\Upsilon$& 0.04&0.04&\Bigg(\makecell{$17\%$\\$17\%$}\Bigg)&~0.01~&0.01 &\Bigg(\makecell{$18\%$\\$19\%$}\Bigg)&0.05 & 0.05&\Bigg(\makecell{$18\%$\\$17\%$}\Bigg) \\
 		\hline
 			$h_c+\eta_b$& --- & ---& --- &0.003&0.003 &\Bigg(\makecell{$18\%$\\$19\%$}\Bigg)&0.003 & 0.003&\Bigg(\makecell{$18\%$\\$19\%$}\Bigg) \\
 		\hline
 			$\chi_{c0}+\eta_b$& --- &--- &--- &~0.2~&0.2 &\Bigg(\makecell{$18\%$\\$17\%$}\Bigg)&0.2 & 0.2&\Bigg(\makecell{$18\%$\\$17\%$}\Bigg) \\
 		\hline
 			$\chi_{c1}+\eta_b$& --- & ---&--- &~0.6~&0.6 &\Bigg(\makecell{$18\%$\\$17\%$}\Bigg)&0.6 & 0.6&\Bigg(\makecell{$18\%$\\$17\%$}\Bigg) \\
 		\hline
 			$\chi_{c2}+\eta_b$& --- & ---&--- &~1.0~&1.0 &\Bigg(\makecell{$18\%$\\$17\%$}\Bigg)&1.0 & 1.0&\Bigg(\makecell{$18\%$\\$17\%$}\Bigg) \\
 		\hline
 	\end{tabular}
 	\label{ccmz}
 \end{table}

\begin{table}[htbp] 
\centering 
\caption{Total cross sections (units: $\times10^{-4}\mathrm{fb}$) up to $\mathcal{O}(v^2)$ at $\sqrt{s}=91.1876\ \mathrm{GeV}$ with varying $m_Q$. ``---'' denotes forbidden processes. As noted in the previous text, the central values are $m_c = 1.5\ \mathrm{GeV}$ for S-wave charmonia and $1.75\ \mathrm{GeV}$ for P-wave charmonia, and $m_b = 4.7\ \mathrm{GeV}$ for S-wave bottomonia and $4.94\ \mathrm{GeV}$ for P-wave bottomonia. Here, $m_{c\pm}$ denotes cross sections obtained by fixing $m_b$ and shifting $m_c$ by $\pm0.15\ \mathrm{GeV}$, and $m_{b\pm}$ denotes cross sections obtained by fixing $m_c$ and shifting $m_b$ by $\pm0.15\ \mathrm{GeV}$. Each cross section has two uncertainties arising from $\pm0.15\ \mathrm{GeV}$ deviations in $m_c$ and $m_b$, and the total uncertainty is combined from these two uncertainties.}
\begin{tabular}{|c||c|c|c|c|c|c|c|c|c|c|}
	\hline
	Process & \multicolumn{5}{c|}{CS} & \multicolumn{5}{c|}{CS+CO} \\
	\hline
	\hline
	& \makecell{$m_{c-}$}&\makecell{$m_{c+}$}&\makecell{$m_{b-}$}&\makecell{$m_{b+}$}&\makecell{uncertainty\\$\Delta_{\pm}$}& \makecell{$m_{c-}$}&\makecell{$m_{c+}$}& \makecell{$m_{b-}$}&\makecell{$m_{b+}$}&\makecell{uncertainty\\$\Delta_{\pm}$}\\
	\hline
	\hline
	$J/\psi+\Upsilon$& 197.2 & 105.0 & 155.3 & 129.5 &\makecell{$+57.33 $\\$-38.55 $}&  201.0& 107.3 & 158.4 &132.3 &\makecell{$+58.20 $\\$-39.16 $} \\
	\hline
	$J/\psi+\eta_b$&12.67&6.914&9.532 &8.927 &\makecell{$+3.463 $\\$-2.324 $}& 14.62&8.046&11.06  &10.33 &\makecell{$+ 3.957$\\$-2.660 $} \\
	\hline
	$J/\psi+\chi_{b0}$&0.094&0.051&0.071 & 0.066 &\makecell{$+0.026 $\\$-0.017 $}&2.796&2.009&2.527  & 2.174 &\makecell{$+ 0.493$\\$-0.371 $} \\
	\hline
	$J/\psi+\chi_{b1}$&0.655&0.357&0.491 & 0.463 &\makecell{$+ 0.179$\\$-0.120 $}&  8.761&6.230&7.860  &6.786  &\makecell{$+1.575 $\\$-1.176 $} \\
	\hline
	$J/\psi+\chi_{b2}$&0.240&0.131&0.180 &0.170  &\makecell{$+0.066 $\\$-0.044 $}& 13.75 &9.919&12.46  &10.71 &\makecell{$+2.404 $\\$-1.814 $} \\
	\hline
	$J/\psi+h_{b}$&0.630&0.343&0.472 &0.444  &\makecell{$+0.173 $\\$-0.116 $}&4.116 &2.245&3.091  & 2.904 &\makecell{$+1.125 $\\$-0.755 $} \\
	\hline
	$\eta_c+h_{b}$&--- &---&--- & ---& ---&48.26 &26.34&36.26  & 34.06 &\makecell{$+13.19 $\\$-8.848 $} \\
	\hline
	$\eta_c+\chi_{b0}$&---&---&--- &--- &---  &16.65&10.11&13.40  &12.18 &\makecell{$ +3.941$\\$ -2.719$} \\
	\hline
	$\eta_c+\chi_{b1}$&---&---&--- &--- & --- &49.98&30.32&40.19  &36.55  &\makecell{$+11.85 $\\$-8.156 $} \\
	\hline
	$\eta_c+\chi_{b2}$&---&---&--- &--- &--- &83.26 &50.53&66.98  & 60.92 &\makecell{$+ 19.70$\\$- 13.59$} \\
	\hline
	$\eta_c+\eta_b$& ---&---&---  & --- & ---  &24.13 &13.28&18.25  &17.05   & \makecell{$+6.527 $\\$-4.388 $} \\
	\hline
	$\eta_c+\Upsilon$& 0.391&0.320&0.389 &0.319  &\makecell{$+0.054 $\\$-0.079 $}&41.14 &23.11&31.49  &29.31  &\makecell{$+10.84 $\\$-13.88 $} \\
	\hline
	$\chi_{c0}+\Upsilon$&0.010 &0.010&0.012 & 0.009 &\makecell{$+ 0.002$\\$-0.001 $}&2.486 &1.497&1.973  &1.846  &\makecell{$+0.582 $\\$-0.415 $} \\
	\hline
	$\chi_{c1}+\Upsilon$&0.081 &0.068&0.082 & 0.067 &\makecell{$+0.010 $\\$-0.009 $}& 7.503&4.530&5.967  & 5.577 &\makecell{$+1.750 $\\$-1.249 $} \\
	\hline
	$\chi_{c2}+\Upsilon$& 0.030&0.025&0.030 & 0.024 &\makecell{$+ 0.004$\\$-0.003 $}& 12.40&7.462&9.838  & 9.208 &\makecell{$+2.907 $\\$-2.073 $} \\
	\hline
	$h_c+\Upsilon$& 0.231&0.194&0.235 &0.190 &\makecell{$+0.031 $\\$-0.027 $}& 0.296&0.249&0.300  &0.244  &\makecell{$+0.039 $\\$-0.034 $} \\
	\hline
	$h_c+\eta_b$&--- &---&---& ---& --- &0.016 &0.014&0.017  &0.014  &\makecell{$+ 0.002$\\$- 0.002$} \\
	\hline
	$\chi_{c0}+\eta_b$&---&---& --- &--- &  ---& 1.483&0.884&1.170  & 1.096 &\makecell{$+ 0.353$\\$-0.251 $} \\
	\hline
	$\chi_{c1}+\eta_b$&---&---&---  & ---& ---& 4.449&2.652&3.513  &3.289  &\makecell{$+ 1.059$\\$-0.753 $} \\
	\hline
	$\chi_{c2}+\eta_b$&--- &---& ---&--- &---  & 7.415&4.419&5.855  &5.481  &\makecell{$+1.764 $\\$-1.256 $} \\
	\hline
\end{tabular}
\label{ccmc}
\end{table}

	  	\subsection{Events}

	  In Table~\ref{events}, the events are obtained for the CEPC (for its Z-factory mode with two interaction points, the designed integrated luminosity over two years is $\sim 16\ \mathrm{ab}^{-1}$\cite{CEPCStudyGroup:2018ghi}) and for the FCC-ee (which operates as a super Z factory in its first four years with a designed integrated luminosity of $150\ \mathrm{ab}^{-1}$\cite{Agapov:2022bhm}). For most production processes, event yields are dominated by the COM. As the CS mechanism alone gives nearly zero events, future precise measurements will strongly support the COM.
	  
	  Heavy quarkonia are reconstructed via their decay products and the reconstructed events can be estimated as,
	  \bea
	  N = N\left[H_1+H_2\right] \times Br_1 \times Br_2 \times \epsilon_1 \times \epsilon_2,
	  \eea
	  where $Br_i$ and $\epsilon_i(i=1,2)$ are the branching fractions and detection efficiency, respectively.  Precise measurements of the Higgs boson and the electroweak sector require the detector to identify leptons, photons, tau leptons, jets and flavours efficiently and precisely\cite{CEPCStudyGroup:2018ghi,Agapov:2022bhm}, therefore we assume the detection efficiency is 100\%.
	  Using the branching fractions of Ref.~\cite{ParticleDataGroup:2024cfk}, the reconstructed events for processes $\{J/\psi+\Upsilon, \eta_c+\chi_{b1}, \eta_c+\chi_{b2}, \mathrm{and}~\Upsilon+\eta_c\}$ are given in Table.~\ref{events2}. Here, \(J/\psi\) and $\Upsilon$ are reconstructed from leptonic decays, \(\eta_c\) is reconstructed from hadronic decays, 
	  $\chi_{bJ}$ are reconstructed via the decay chain $\chi_{bJ}\rightarrow \gamma+\Upsilon\rightarrow\gamma+l^+l^-$.
With the uncertainties combined from the heavy quark mass and renormalization scale, the total events for processes $\{J/\psi+\Upsilon, \Upsilon+\eta_c\}$ are $\{3^{+1.4}_{-0.9}~(1^{+0.5}_{-0.4}),~2^{+0.99}_{-1.77}~(1^{+0.57}_{-0.94})\}$ and $\{32^{+13.00}_{-8.81}~(13^{+5.1}_{-3.4}),~19^{+9.83}_{-16.58}~(10^{+4.81}_{-8.96})\}$ for the two colliders, respectively.

	  \begin{table}
	  	\caption{ The  events of  charmonia+bottomonia production at $\sqrt{s}=$91.1876 GeV for the CEPC and FCC-ee. In each cell, the values outside/inside the brackets are for leading order and next-to-leading order results in the $v^2$ expansions, respectively. }
	  	\begin{tabular}{|c|c|c| |c|c|c|}
	  		\hline
	  		\multicolumn{6}{|c|}{ CEPC($16ab^{-1}$)}\\
	  		\hline
	  		\hline
	  		~& CS& CS+CO& 	& CS & CS+CO  \\
	  		\hline
	  		$J/\psi+\Upsilon$& 584(227)&593(231)& 	$\eta_c+\eta_b$& 0(0)&61(28) \\
	  		\hline
	  		$J/\psi+\eta_b$& 30(15)&35(17)& 	$\eta_c+\Upsilon$& 1(1)&98(49) \\
	  		\hline
	  		$J/\psi+\chi_{b0}$& 0(0)&7(4)& 	$\chi_{c0}+\Upsilon$& 0(0)&6(3) \\
	  		\hline
	  		$J/\psi+\chi_{b1}$& 2(1)&20(12)& $\chi_{c1}+\Upsilon$& 0(0)&19(9) \\
	  		\hline
	  		$J/\psi+\chi_{b2}$& 1(0)&32(18) &$\chi_{c2}+\Upsilon$& 0(0)&31(15) \\
	  		\hline
	  		$J/\psi+h_{b}$& 2(1)&10(5)	& $h_{c}+\Upsilon$& 1(0)&1(0)\\
	  		\hline
	  		$\eta_c+h_b$& 0(0)&114(56)& $h_c+\eta_b$& 0(0)&0(0) \\
	  		\hline
	  		$\eta_c+\chi_{b0}$&0(0)&36(20)& $\chi_{c0}+\eta_b$&0(0)&4(2) \\
	  		\hline
	  		$\eta_c+\chi_{b1}$& 0(0)&108(61)& 	$\chi_{c1}+\eta_b$& 0(0)&12(5) \\
	  		\hline
	  		$\eta_c+\chi_{b2}$&0(0)&181(102)& $\chi_{c2}+\eta_b$&0(0)&20(9) \\
	  		\hline
	  		\hline
	  		\multicolumn{6}{|c|}{ FCC-ee($150ab^{-1}$)}\\
	  		\hline
	  		\hline
	  		$J/\psi+\Upsilon$& 5473(2124)&5558(2167)& 	$\eta_c+\eta_b$& 0(0)&576(264) \\
	  		\hline
	  		$J/\psi+\eta_b$& 281(138)&327(160)& 	$\eta_c+\Upsilon$& 9(5)&922(455) \\
	  		\hline
	  		$J/\psi+\chi_{b0}$& 2(1)&61(35)& 	$\chi_{c0}+\Upsilon$& 0(0)&58(29) \\
	  		\hline
	  		$J/\psi+\chi_{b1}$& 15(7)&192(109)& $\chi_{c1}+\Upsilon$& 2(1)&175(86) \\
	  		\hline
	  		$J/\psi+\chi_{b2}$& 5(3)&300(173) &$\chi_{c2}+\Upsilon$& 1(0)&289(143) \\
	  		\hline
	  		$J/\psi+h_{b}$& 15(7)&93(45)	& $h_{c}+\Upsilon$& 6(3)&8(4)\\
	  		\hline
	  		$\eta_c+h_b$& 0(0)&1070(527)& $h_c+\eta_b$& 0(0)&0(0) \\
	  		\hline
	  		$\eta_c+\chi_{b0}$&0(0)&339(191)& $\chi_{c0}+\eta_b$&0(0)&37(17) \\
	  		\hline
	  		$\eta_c+\chi_{b1}$& 0(0)&1017(574)& 	$\chi_{c1}+\eta_b$& 0(0)&111(51) \\
	  		\hline
	  		$\eta_c+\chi_{b2}$&0(0)&1694(957)& $\chi_{c2}+\eta_b$&0(0)&185(85) \\
	  		\hline
	  		
	  	\end{tabular}
	  	\label{events}
	  \end{table}
	  
	  \begin{table}
	  	\caption{ The reconstructed events of charmonia+bottomonia production from their decay products at $\sqrt{s}=$91.1876 GeV for the CEPC and FCC-ee. In each cell, the values outside/inside the brackets are for leading order and next-to-leading order results in the $v^2$ expansions, respectively. The efficiency is selected as 100\%.}
	  	\begin{tabular}{|c|c|c|c|c|c|c|c|c|}
	  		\hline
	  		$H_1+H_2$& \multicolumn{2}{c|}{ $H_1$ } &\multicolumn{2}{c|}{ $H_2$ }& \multicolumn{2}{c|}{ CEPC($16ab^{-1}$)}&	\multicolumn{2}{c|}{ FCC($150ab^{-1}$)}\\
	  		\hline
	  		&   Decay &$Br$&  Decay	&   $Br$ & CS &CS+CO&CS&CS+CO \\
	  		\hline
	  		$J/\psi+\Upsilon$&$l^+l^-(l=e,\mu)$&12\%& $l^+l^-(l=e,\mu)$&4.86\%&3(1)&3(1)&32(12)&32(13) \\
	  		\hline
	  		$\eta_c+\chi_{b1}$&hadron decay&43\%& $\gamma l^+l^-(\gamma\Upsilon)$&1.71\%&0(0)&1(0)&0(0)&7(4) \\
	  		\hline
	  		$\eta_c+\chi_{b2}$&hadron decay&43\%& $\gamma l^+l^-(\gamma\Upsilon)$&0.87\%&0(0)&1(0)&0(0)&6(4) \\
	  		\hline
	  		$\eta_c+\Upsilon$&hadron decay&43\%& $l^+l^-(l=e,\mu)$&4.86\%&0(0)&2(1)&0(0)&19(10) \\
	  		\hline
	  		
	  	\end{tabular}
	  	\label{events2}
	  \end{table}

 \section{Summary}
 \label{summary}

In this paper, we study the associated charmonium–bottomonium production in $e^+e^-$ annihilation via a $\gamma^*/Z^0$ propagator at Z factories, including both CS and CO channels up to NLO($v^2$).
For associated charmonium–bottomonium production, CO contributions through gluon fragmentation into the $^3S_1^{[8]}$ state dominate nearly all processes, not only at the $Z^0$ pole but also near the production threshold. By contrast, the CS mechanism dominates near threshold for double charmonium or double bottomonium production, owing to the non-fragmentation mechanism. These processes can therefore serve as a valuable probe to investigate COM in heavy quarkonium production across both lower and higher energy regions, and to constrain the CO LDMEs.
Besides CO contributions, we also include relativistic corrections, which are found to be significant and reduce the $\mathcal{O}(v^0)$ cross sections by roughly 50\%. Relevant uncertainties are discussed in detail.
Finally, we estimate the expected event yields for the CEPC and FCC-ee. Based on final-state particle reconstruction, the event yields for $J/\psi + \Upsilon$ and $\Upsilon + \eta_c$ are (1, 1) for the CEPC (2-year operation) and (13, 10) for the FCC-ee (4-year operation) in the Z-factory mode, respectively.
 
 Given that the one-loop QCD corrections within CSM are negligible\cite{Belov:2021ftc}, it is crucial to investigate QCD corrections within COM. This motivation is reinforced by a recent calculation of the next-to-next-to-leading-order (NNLO) QCD corrections for the processes $e^+e^- \to 2J/\psi$ and $2\Upsilon$ \cite{Chen:2025ocd}, which shows an enhancement of the LO cross section by approximately an order of magnitude. If a comparable NNLO enhancement occurs in the associated production of charmonium and bottomonium, the number of reconstructed events at future $e^+e^-$ colliders could reach $\mathcal{O}(10^2)$.

		\section{Appendix}
	\label{appdB}
	
Like previous work\cite{Wang:2025sbx}, we do not present the analytical expressions of the SDCs in this appendix for simplicity. Instead, we provide the ratios of SDCs from relativistic corrections (denoted as G) to LO SDCs (denoted as F) in the high c.m. energy limit ($m_Q^2 \ll s$) in Table~\ref{Rratio}.

	\begin{table}[ht!]
		\caption{Ratios of relativistic correction SDCs (denoted as G) to LO SDCs (denoted as F) in the high c.m. energy limit ($m_Q^2\ll s$)  for   $e^+e^-\rightarrow  (c\bar{c}) [{}^{2S_1+1}{L_1}_{J_1}^{c_1}]+(b\bar{b})[{}^{2S_2+1}{L_2}_{J_2}^{c_2}]$. In each cell, we define, $R_i\equiv G[n_i]/F[n_1+n_2]$, $c=1,8$ for CS or CO states. The first four rows are for CS channels, the second three rows for CO channals, and  the last row for the t-channel processes.
		}
		\begin{tabular}{|cll|cll|cll|}
		\hline
	  ${}^3S_1^{[1]}+{}^1S_0^{[1]}$& $R_1= -\frac{11}{6}$&$R_2=-\frac{5}{6}$ &	${}^3S_1^{[1]}+{}^3S_1^{[1]}$& $R_1= -\frac{11m_b^2+m_c^2}{6(m_b^2+m_c^2)}$ &$R_2=-\frac{m_b^2+11m_c^2}{6(m_b^2+m_c^2)}$ &
		${}^3S_1^{[1]}+{}^1P_1^{[1]}$ & $R_1= -\frac{11}{6}$& $R_2=-\frac{13}{10}$  \\
		\hline
	 ${}^3S_1^{[1]}+{}^3P_0^{[1]}$&$R_1= -\frac{11}{6}$&$R_2=-\frac{13}{10}$&
	 ${}^3S_1^{[1]}+{}^3P_1^{[1]}$&$R_1= -\frac{11}{6}$&$R_2=-\frac{11}{10}$&
		${}^3S_1^{[1]}+{}^3P_2^{[1]}$&$R_1=-\frac{11}{6}$&$R_2= -\frac{7}{10}$\\
		\hline
		 ${}^1S_0^{[1]}+{}^3S_1^{[1]}$&$R_1=-\frac{5}{6}$&$R_2= -\frac{11}{6}$&
		 ${}^1P_1^{[1]}+{}^3S_1^{[1]}$&$R_1=-\frac{13}{10}$&$R_2= -\frac{11}{6}$&
		 ${}^3P_0^{[1]}+{}^3S_1^{[1]}$ &$R_1=-\frac{13}{10}$&$R_2= -\frac{11}{6}$\\
		\hline
		${}^3P_1^{[1]}+{}^3S_1^{[1]}$&$R_1=-\frac{11}{10}$&$R_2= -\frac{11}{6}$&
		 ${}^3P_2^{[1]}+{}^3S_1^{[1]}$&$R_1=-\frac{7}{10}$&$R_2=-\frac{11}{6}$&
		&&\\
			\hline
		\hline
		 ${}^1S_0^{[8]}+{}^3S_1^{[8]}$&$R_1=-\frac{5}{6}$&$R_2= -\frac{11}{6}$&
		${}^3P_J^{[8]}+{}^3S_1^{[8]}$&$R_1=-\frac{31}{30}$&$R_2= -\frac{11}{6}$&
		${}^3S_1^{[8]}+{}^1S_0^{[8]}$&$R_1=-\frac{11}{6}$&$R_2= -\frac{5}{6}$\\
		\hline
		 ${}^3S_1^{[8]}+{}^3P_J^{[8]}$&$R_1=-\frac{11}{6}$&$R_2= -\frac{31}{30}$&
		${}^3S_1^{[8]}+{}^3S_1^{[8]}$&$R_1=-\frac{11m_b^2+m_c^2}{6(m_b^2+m_c^2)}$&$R_2= -\frac{m_b^2+11m_c^2}{6(m_b^2+m_c^2)}$&
		${}^3S_1^{[8]}+{}^1P_1^{[8]}$&$R_1=-\frac{11}{6}$&$R_2= -\frac{13}{10}$\\
		\hline
		${}^1P_1^{[8]}+{}^3S_1^{[8]}$&$R_1=-\frac{13}{10}$&$R_2= -\frac{11}{6}$&
	 & & &
	 & & \\
		\hline
		\hline
			${}^3S_1^{[1]}+{}^3S_1^{[1]}$&\multicolumn{2}{c |}{$R_1=R_2= -\frac{11}{6}$}&  &
		 &  &  & &   \\
		\hline
	\end{tabular}
	\label{Rratio}
	\end{table}
	\FloatBarrier

We find these ratios to be consistent with Ref.~\cite{Wang:2025sbx}, and they could also be verified by Refs.~\cite{Li:2013csa,Xu:2012am,Xu:2014zra}. The slightly different R ratios correspond to the ${}^3S_1^{[1/8]}+{}^3S_1^{[1/8]}$ channel, which depend on the heavy quark masses. Moreover, when we take $m_c=m_b$, these ratios become -1, which is consistent with the results for double charmonium/bottomonium production\cite{Wang:2025sbx}.

\section{Acknowledgements:} 
This work was supported by the National Natural Science Foundation of China (No. 11705078, 12575087).

\hspace{2cm}

	
%
	 
\end{document}